\newif\iffigs\figstrue
\documentstyle[12pt]{article}
\iffigs
  \input epsf
\else
\fi

\batchmode
  \newfont{\footscrfont}{rsfs10}
  \newfont{\footbbbfont}{msbm10}
  \newfont{\manfont}{manfnt}
\errorstopmode

\newif\ifscrf\scrftrue
\ifx\footscrfont\nullfont
  \scrffalse
\fi

\newif\ifamsf\amsftrue
\ifx\footbbbfont\nullfont
  \amsffalse
\fi

\def\ppnumber{\vbox{\baselineskip14pt\hbox{RU-96-98}
\hbox{hep-th/9611137}}}
\def\ppdate{November 1996}
\def\pplogo{\vbox{\kern-\headheight\kern -15pt
\halign{##&##\hfil\cr&{%\sc
\ppnumber}\cr\rule{0pt}{2.5ex}&\ppdate\cr}
}}

\makeatletter
\date{}
\def\dedicatory#1{\def\@date{\normalsize\it#1}}
\def\subjclass#1{\def\@thefnmark{}\@footnotetext{1991
    {\it Mathematics Subject Classification.} #1}}
\def\keywords#1{\def\@thefnmark{}\@footnotetext{
    {\it Key words and phrases.} #1}}

\def\ps@firstpage{\ps@empty \def\@oddhead{\hss\pplogo}%
  \let\@evenhead\@oddhead % in case an article starts on a left-hand page
}
\def\maketitle{\par
 \begingroup
 \def\thefootnote{\fnsymbol{footnote}}
 \def\@makefnmark{\hbox
 to 0pt{$^{\@thefnmark}$\hss}}
 \if@twocolumn
 \twocolumn[\@maketitle]
 \else \newpage
 \global\@topnum\z@ \@maketitle \fi\thispagestyle{firstpage}\@thanks
 \endgroup
 \setcounter{footnote}{0}
 \let\maketitle\relax
 \let\@maketitle\relax
 \gdef\@thanks{}\gdef\@author{}\gdef\@title{}\let\thanks\relax}

\def\abstract{\if@twocolumn
\section*{Abstract}
\else \small
\begin{center}
{\bf ABSTRACT}
\end{center}
\quotation
\fi}

\newtheorem{theorem}{Theorem}
\newtheorem{prop}{Proposition}

\def\thebibliography#1{\section*{References\@mkboth
 {REFERENCES}{REFERENCES}}\small\list
 {[\arabic{enumi}]}{\settowidth\labelwidth{[#1]}\leftmargin\labelwidth
 \advance\leftmargin\labelsep
 \usecounter{enumi}}
 \def\newblock{\hskip .11em plus .33em minus .07em}
 \sloppy\clubpenalty4000\widowpenalty4000
 \sfcode`\.=1000\relax}

\newif\iffn\fnfalse

\@ifundefined{reset@font}{\let\reset@font\empty}{} %needed for ancient LaTeX
\long\def\@footnotetext#1{\insert\footins{\reset@font\footnotesize
    \interlinepenalty\interfootnotelinepenalty
    \splittopskip\footnotesep
    \splitmaxdepth \dp\strutbox \floatingpenalty \@MM
    \hsize\columnwidth \@parboxrestore
   \edef\@currentlabel{\csname p@footnote\endcsname\@thefnmark}\@makefntext
    {\rule{\z@}{\footnotesep}\ignorespaces
      \fntrue#1\fnfalse\strut}}}

\makeatother

\ifamsf
  \newfont{\bigbbbfont}{msbm10 scaled\magstep2}
  \newfont{\bbbfont}{msbm10 scaled\magstep1}  % msbm12 does not exist
  \newfont{\smallbbbfont}{msbm8}
  \newfont{\tinybbbfont}{msbm6}
  \newfont{\smallfootbbbfont}{msbm7}
  \newfont{\tinyfootbbbfont}{msbm5}
  \newfont{\biggthfont}{eufm10 scaled\magstep2}
  \newfont{\gthfont}{eufm10 scaled\magstep1}  % eufm12 does not exist
  \newfont{\smallgthfont}{eufm8}
  \newfont{\tinygthfont}{eufm6}
  \newfont{\footgthfont}{eufm10}
  \newfont{\smallfootgthfont}{eufm7}
  \newfont{\tinyfootgthfont}{eufm5}
\fi

\ifscrf
  \newfont{\scrfont}{rsfs10 scaled\magstep1}  % rsfs12 does not exist
  \newfont{\smallscrfont}{rsfs7}
  \newfont{\tinyscrfont}{rsfs7}
  \newfont{\smallfootscrfont}{rsfs7}
  \newfont{\tinyfootscrfont}{rsfs7}
\fi

\ifamsf
  \newcommand{\Bbb}[1]{\iffn
      \mathchoice{\mbox{\footbbbfont #1}}{\mbox{\footbbbfont #1}}
      {\mbox{\smallfootbbbfont #1}}{\mbox{\tinyfootbbbfont #1}}\else
      \mathchoice{\mbox{\bbbfont #1}}{\mbox{\bbbfont #1}}
      {\mbox{\smallbbbfont #1}}{\mbox{\tinybbbfont #1}}\fi}
  \newcommand{\Goth}[1]{\iffn
      \mathchoice{\mbox{\footgthfont #1}}{\mbox{\footgthfont #1}}
      {\mbox{\smallfootgthfont #1}}{\mbox{\tinyfootgthfont #1}}\else
      \mathchoice{\mbox{\gthfont #1}}{\mbox{\gthfont #1}}
      {\mbox{\smallgthfont #1}}{\mbox{\tinygthfont #1}}\fi}
\else
  \def\bigbbbfont{\bf}
  \def\Bbb{\bf}
  \def\Goth{\cal}
\fi

\ifscrf
  \newcommand{\Scr}[1]{\iffn
    \mathchoice{\mbox{\footscrfont #1}}{\mbox{\footscrfont #1}}
    {\mbox{\smallfootscrfont #1}}{\mbox{\tinyfootscrfont #1}}\else
    \mathchoice{\mbox{\scrfont #1}}{\mbox{\scrfont #1}}
    {\mbox{\smallscrfont #1}}{\mbox{\tinyscrfont #1}}\fi}
\else
  \def\Scr{\cal}
\fi

\def\operatorname#1{\mathop{\rm #1}\nolimits}
\def\C{{\Bbb C}}

\def\P{{\Bbb P}}

\def\R{{\Bbb R}}
\def\Z{{\Bbb Z}}
\def\H{{\Bbb H}}

\def\Area{\operatorname{Area}}
\def\Vol{\operatorname{Vol}}

\def\Pic{\operatorname{Pic}}
\def\disc{\operatorname{disc}}

\def\Gr{\operatorname{Gr}}
\def\SO{\operatorname{SO}}
\def\Sl{\operatorname{SL}}
\def\GO{\operatorname{O{}}}
\def\SU{\operatorname{SU}}
\def\GU{\operatorname{U{}}}
\def\Sp{\operatorname{Sp}}
\def\Spin{\operatorname{Spin}}
\def\rank{\operatorname{rank}}

\def\opeq#1{\advance\lineskip#1 \advance\baselineskip#1
        \advance\lineskiplimit#1}

\def\eqalign#1{\null\,\vcenter{\opeq{2.5\jot}\mathsurround=0pt
        \everycr={}\tabskip=0pt
        \halign{\strut\hfil$\displaystyle{##}$&$\displaystyle{{}##}$\hfil
        \crcr#1\crcr}}\,\null}

\def\sm{$\sigma$-model}
\def\nlsm{non-linear \sm}

\def\CY{Calabi--Yau}

\def\cM{{\Scr M}}
\def\cA{{\Scr A}}

\def\cD{{\Scr D}}
\def\cH{{\Scr H}}
\def\cT{{\Scr T}}
\def\cL{{\Scr L}}
\def\cF{{\Scr F}}

\def\cMc{{\hfuzz=100cm\hbox to 0pt{$\;\overline{\phantom{X}}$}\cM}}
\def\barcD{{\hfuzz=100cm\hbox to 0pt{$\;\overline{\phantom{X}}$}\cD}}

\def\ff#1#2{{\textstyle\frac{#1}{#2}}}

\def\RoR{$R\leftrightarrow1/R$}

\def\spnh{\Spin(32)/\Z_2}

\def\mzx{\mbox{\manfont\char"1D}}

\ifamsf
  \def\ltimes{\mathbin{\mbox{\bbbfont\char"6E}}}
  
\else
  \def\ltimes{.}
  
\fi

\begin{document}
\setcounter{page}0
\title{\LARGE K3 Surfaces and String Duality\\[10mm]}
\author{
Paul S. Aspinwall\\[0.7cm]
\normalsize Dept.~of Physics and Astronomy,\\
\normalsize Rutgers University,\\
\normalsize Piscataway, NJ 08855\\[10mm]
}

{\hfuzz=10cm\maketitle}

\def\Large{\large}
\def\LARGE{\large\bf}

\vskip 1cm

\begin{abstract}

The primary purpose of these lecture notes is to explore the moduli
space of type IIA, type IIB, and heterotic string compactified on a K3
surface. The main tool which is invoked is that of string duality. K3
surfaces provide a fascinating arena for string compactification as
they are not trivial spaces but are sufficiently simple for one to be able
to analyze most of their properties in detail. They also make an
almost ubiquitous appearance in the common statements concerning
string duality. We review the necessary
facts concerning the classical geometry of K3 surfaces that will be
needed and then we review ``old string theory'' on K3 surfaces in
terms of conformal field theory. The type IIA string, the type IIB
string, the $E_8\times E_8$ heterotic string, and $\spnh$ heterotic
string on a K3 surface are then each analyzed in turn. 
The discussion is biased in favour of purely geometric notions
concerning the K3 surface itself.

\end{abstract}

\vfil\break

\tableofcontents

%%%%%%%%%%%%%%%%%%%%%%%%%%%%%%%%%%%%%%%%%%%%%%%%%%%%%%%%%%%%%%%%

\section{Introduction}

The notion of ``duality'' has led to something of a revolution in
string theory in the past year or two. Two theories are considered
dual to each other if they ultimately describe exactly the same
physics. In order for this to be a useful property, of course, it is best
if the two theories appear, at first sight, to be completely
unrelated. Different notions of duality abound depending on how the
two theories differ. The canonical example is that of ``S-duality''
where the coupling of one theory is inversely related to that of the
other so that one theory would be weakly coupled when the other is
strongly coupled. ``T-duality'' can be defined by similarly
considering length scales rather than coupling constants. Thus a
theory at small distances can be ``T-dual'' to another theory at large
distances.

In the quest for understanding duality many examples of dual pairs
have been postulated. The general scenario is that one takes a string
theory (or perhaps M-theory) and compactifies it on some space and
then finds a dual partner in the form of another (or perhaps the same)
string theory compactified on some other space. In this form duality
has become a subject dominated by geometrical ideas since most of the
work involved lies in analyzing the spaces on which the string theory
is compactified.

One of the spaces which has become almost omnipresent in the study of
string duality is that of the K3 surface. 
We will introduce the K3 surface in section \ref{s:clas} but let us
make a few comments here. Mathematicians have been studying the
geometry of the K3 surface as a real surface or a complex surface
for over one hundred 
years \cite{Hud:K3}. In these lectures we will always be working with
complex numbers and so ``surface'' will mean a space of complex
dimension two, or real dimension four. A curve will be complex
dimension one, etc. Physicists' interest in the K3 surface (for an early
paper see, for example, \cite{HP:K3anom}) was not sparked until Yau's
proof \cite{Yau:} of Calabi's conjecture in 1977. Since then the K3
surface has become a commonly-used ``toy model'' for compactifications
(see, for example, \cite{DNP:K3}) as it provides the second simplest
example of a Ricci-flat compact manifold after the torus.

The study of duality is best started with toy models and so the K3
surface and the torus are bound to
appear often. Another reason for the appearance of the K3 surface, as
we shall see in these lectures, is 
that the mathematics of the heterotic string appears to be
intrinsically bound to the geometry of the K3 surface. Thus, whenever
the heterotic string appears on one side of a pair of dual theories,
the K3 surface is likely to make an appearance in the analysis.

The original purpose of these lectures was to give a fairly complete
account of the way K3 surfaces appear in the subject of string
duality. For reasons outlined above, however, this is almost tantamount
to covering the entire subject of string duality. In order to make the
task manageable therefore we will have to omit some current areas of
active research. Let us first then discuss what will {\em not\/} be
covered in these lectures. Note that each of these subjects are covered
excellently by the other lecture series anyway.

Firstly we are going to largely ignore M-theory. M-theory may well
turn out to be an excellent model for understanding string theory 
or perhaps even replacing string theory. It also provides a simple
interpretation for many of the effects we will discussing.

Secondly we are going to ignore open strings and D-branes. There is no
doubt that D-branes offer a very good intuitive approach to many of
the phenomena we are going to study. It may well also be that D-branes are
absolutely necessarily for a complete understanding of the foundations
of string theory. 

Having said that M-theory and D-branes are very important we will now
do our best to not mention them. One reason for this is to able to finish
giving these lectures on time but another, perhaps more important
reason is to avoid introducing unnecessary assumptions. We want to
take a kind of ``Occam's Razor'' approach and only introduce
constructions as necessary. To many people's tastes our arguments will
become fairly cumbersome, especially when dealing with the heterotic
string, and it will certainly be true that a simpler picture could be
formulated using M-theory or D-branes in some instances. What is
important however is 
that we produce a self-consistent framework in which one may analyze
the questions we wish to pose in these lectures.

Thirdly we are going to try to avoid explicit references to solitons. Since
nonperturbative physics is absolutely central to most of the later
portions of these lectures one may view our attitude as
perverse. Indeed, one really cannot claim to understand much of the
physics in these lectures without considering the soliton effects. What we
will be focusing on, however, is the structure of moduli spaces and we
will be able to get away with ignoring solitons in this context quite
effectively. The only time that solitons will be of interest is when
they become massless.

Our main goal is to understand the type II string and the heterotic
string compactified on a K3 surface and what such models are dual
to. Of central importance will be the notion of the {\em moduli
space\/} of a given theory. 

In section \ref{s:clas} we will introduce the K3 surface itself and
describe its geometry. The facts we require from both differential
geometry and algebraic geometry are introduced. In section \ref{s:ws}
we will review the ``old'' approach to a string theory on a K3 surface
in terms of the world-sheet conformal field theory.

In section \ref{s:II} we begin our discussion of full string theory on
a K3 surface in terms of the type IIA and type IIB string. The start
of this section includes some basic facts about target-space
supergravity which are then exploited.

The heterotic string is studied in section \ref{s:het} but before that
we need to take a long detour into the study of string theories
compactified down to four dimensions. This detour comprises section
\ref{s:4d} and builds the techniques required for section \ref{s:het}.
The heterotic string on a K3 surface is a very rich and complicated
subject. The analysis is far from complete and section \ref{s:het} is
technically more difficult than the preceding sections.

Note that blocks of text beginning with a ``\mzx'' are rather
technical and may be omitted on first reading.

%%%%%%%%%%%%%%%%%%%%%%%%%%%%%%%%%%%%%%%%%%%%%%%%%%%%%%%%%%%%%%%%%%%

\section{Classical Geometry}	\label{s:clas}

In the mid 19th century the Karakorum range of mountains in Northern
Kashmir acted as a natural protection to India, then under British rule,
from the Chinese and Russians to the north. Accordingly, in 1856,
Captain T.~G.~Montgomerie was sent out to make some attempt to map the
region. From a distance of 128 miles he measured the peaks across the
horizon and gave them the names K1, K2, K3, \ldots, where the ``K''
stood simply for ``Karakorum'' \cite{M:K2}. While it later transpired
that most of these mountain peaks already had names known to the
Survey of India, the second peak retained the name Montgomerie
had assigned it.

It was not until almost a century later in 1954 that K2, the world's
second highest peak, was climbed by Achille Compagnoni and Lino
Lacedelli in an Italian expedition. This event led shortly afterwards
to the naming of an object of a quite different character. The first
occurrence of the name ``K3'' referring to an algebraic variety occurs
in print in \cite{Weil:rep}. It is explained in \cite{Weil:oev} that
the naming is after Kummer, K\"ahler and Kodaira and is inspired by
K2. It was Kummer who did much of the earliest work to 
explore the geometry of the space in question.\footnote{This may
explain the erroneous notion in some of the physics literature that
the naming is $K_3$ after ``Kummer's 
third surface'' (whatever his first two surfaces may have been). To
confuse the issue slightly there is a special kind of K3 surface known
as a ``Kummer surface'' which is introduced in section \ref{ss:orb}.}

\subsection{Definition}		\label{ss:def}

So what exactly is a K3 surface? Let us first introduce a few
definitions. For a general guide to some of the basic principles used
in these lectures we refer the reader to chapter 0 of
\cite{GH:alg}. First we define the Hodge numbers of a space $X$ as the
dimensions of 
the Dolbeault cohomology groups
\begin{equation}
  h^{p,q}(X) = \dim(H^{p,q}(X)).
\end{equation}
Next consider the {\em canonical class\/}, $K$, which we may be taken
to be defined as
\begin{equation}
  K=-c_1(T_X),
\end{equation}
where $T_X$ is the holomorphic tangent bundle of $X$. 

A K3 surface, $S$, is defined as a compact complex K\"ahler manifold of complex
dimension two, i.e., a surface, such that
\begin{equation}
  \eqalign{h^{1,0}(S) &= 0\cr K &= 0.\cr}
	\label{eq:K3d}
\end{equation}
Note that we will sometimes relax the requirement that $S$ be a
manifold.

The remarkable fact that makes K3 surfaces so special is the following
\begin{theorem}
  Any two K3 surfaces are diffeomorphic to each other.
\end{theorem}
Thus, if we can find one example of a K3 surface we may deduce all of the
topological invariants. The simplest realization is to find a simple
example as a complex surface embedded in a complex projective space,
i.e., as an {\em algebraic variety}. The obvious way to do this is to
consider the hypersurface defined by the equation
\begin{equation}
  f = x_0^n+x_1^n+x_2^n+x_3^n=0
\end{equation}
in the projective space $\P^3$ with homogeneous coordinates
$[x_0,x_1,x_2,x_3]$.

It follows from the Lefschetz hyperplane theorem \cite{GH:alg} that
$h^{1,0}$ of such a 
hypersurface will be zero. Next we need to find if we can
determine $n$ such that $K=0$. Associated to the canonical class is
the {\em canonical line bundle\/}. This is simply the holomorphic line
bundle, $L$, such that $c_1(L)=K$. From our definition of $K$ it
follows that the canonical line bundle for a manifold of dimension $d$
can be regarded as the $d$th exterior power of the holomorphic
cotangent bundle. Thus a section of the canonical line bundle can be
regarded as a holomorphic $d$-form.

The fact that $K=0$ for a K3 surface tells us that the canonical line
bundle is trivial and thus has a holomorphic section which is nowhere
zero. Consider two such sections, $s_1$ and $s_2$. The ratio $s_1/s_2$
is therefore a holomorphic function defined globally over the compact
K3 surface. From basic complex analysis it follows that $s_1/s_2$ is a
constant. We see that the K3 surface admits a globally defined, nowhere
vanishing, holomorphic
2-form, $\Omega$, which is unique up to a constant. It also follows
that $h^{2,0}(S)=1$.

Let us try to build $\Omega$ for our hypersurface of degree $n$ in
$\P^3$. First define affine coordinates in the patch $x_0\neq0$:
\begin{equation}
y_1=\frac{x_1}{x_0},\quad y_2=\frac{x_2}{x_0},\quad 
y_3=\frac{x_3}{x_0}.
\end{equation}

An obvious symmetric choice for $\Omega$ is then
\begin{equation}
  \Omega = \frac{dy_1\wedge dy_2}{\partial f/\partial y_3}
    = \frac{dy_2\wedge dy_3}{\partial f/\partial y_1}
    = \frac{dy_3\wedge dy_1}{\partial f/\partial y_2}.
\end{equation}
This is clearly nonzero and holomorphic in our patch $x_0\neq0$. We
can now consider another patch such as $x_1\neq0$. A straight forward
but rather tedious calculation then shows that $\Omega$ will only
extend into a holomorphic nonzero 2-form over this next patch if
$n=4$.

Our first example of a K3 surface is called the {\em quartic surface},
given by a hypersurface of degree 4 in $\P^3$. We could have arrived
at this same conclusion in a somewhat more abstract way by using the
{\em adjunction formula}. Consider the tangent bundle of $S$, which we
denote $T_S$, together with the normal bundle $N_S$ for the embedding
$S\subset\P^3$. One can then see that
\begin{equation}
  T_S\oplus N_S = T_{\P^3|S},
\end{equation}
where $T_{\P^3|S}$ is the restriction of the tangent bundle of the
embedding $\P^3$ to the hypersurface $S$.

Introducing the formal sum of Chern classes of a bundle $E$ (see, for example,
\cite{EGH:dg}) 
\begin{equation}
  c(E) = 1 + c_1(E) + c_2(E) +\ldots,
\end{equation}
we see that
\begin{equation}
  c(T_{\P^3|S}) = c(T_S) \wedge c(N_S).
\end{equation}

Treating a wedge product as usual multiplication from now on, it is
known that \cite{GH:alg} 
\begin{equation}
  c(T_{\P^k}) = (1+x)^{k+1},
\end{equation}
where $x$ is the fundamental generator of $H^2(\P^k,\Z)$. Since $H^2$
is dual to $H^{2(k-1)}$ which is dual to $H_{2(k-1)}$, we may also regard
$x$ as the homology class of a {\em hyperplane\/}
$\P^{k-1}\subset\P^k$ embedded in the obvious way by setting one of
the homogeneous coordinates to zero.\footnote{Throughout these
lectures we will often use the same notation for the 2-form and the
associated $2(k-1)$-cycle.}  Stated in this way one can see that
$c_1(N_S)$ is given by $nx$.

Thus we have that
\begin{equation}
  \eqalign{c(T_S) &= \frac{(1+x)^4}{1+nx}\cr
   &= 1 + (4-n)x + (6-4n+n^2)x^2.\cr}
\end{equation}
Again we see that $n=4$ is required to obtain $K=0$. Now we also have
$c_2$ which enables us to work out the Euler characteristic, $\chi(S)$,
of a K3 surface:
\begin{equation}
  \eqalign{\chi(S) &= \int_S c_2(T_S)\cr
     &= \int_{\P^3} l_S\wedge c_2(T_S)\cr
     &= \int_{\P^3} 4x.6x^2\cr
     &= 24,\cr}
\end{equation}
where $l_S$ is the 2-form which is the dual of the dual of the
hypersurface $S$ in the sense explained above. One may also show using
the Lefschetz hyperplane theorem that
\begin{equation}
  \pi_1(S)=0.
\end{equation}

\def\m#1{\makebox[10pt]{$#1$}}
We now have enough information to compute all the Hodge numbers,
$h^{p,q}$. Since $\pi_1(S)=0$, we have that the first Betti number
$b_1(S)=\dim H^1(S)=h^{1,0}+h^{0,1}$ must be zero. The Euler
characteristic then fixes $b_2(S)$ which then determines $h^{1,1}$
since we already know $h^{2,0}=1$ from above. The result is
\begin{equation}
  {\arraycolsep=2pt
  \begin{array}{*{5}{c}}
    &&\m{h^{0,0}}&& \\ &\m{h^{1,0}}&&\m{h^{0,1}}& \\
    \m{h^{2,0}}&&\m{h^{1,1}}&&\m{h^{0,2}} \\
    &\m{h^{2,1}}&&\m{h^{1,2}}& \\ &&\m{h^{2,2}}&&
  \end{array}} \;=\; 
  {\arraycolsep=2pt
  \begin{array}{*{5}{c}}
    &&\m1&& \\ &\m0&&\m0& \\ \m1&&\m{20}&&\m{1.} \\
    &\m0&&\m0& \\ &&\m1&&
  \end{array}}
\end{equation}

\subsection{Holonomy}    \label{ss:hol}

\def\Hl{{\Goth H}}
Before continuing our discussion of K3 surfaces, we will take a detour
and discuss the subject of holonomy which will be of considerable use
at many points in these lectures.

Holonomy is a natural concept in the differential geometry of a
manifold, $M$, with a vector bundle, $\pi:E\to M$, with a
connection. Consider taking a 
point, $p\in M$, and a vector in the fibre, $e_1\in\pi^{-1}(p)$. Now,
following a closed path, $\Gamma$, starting and ending at $p$,
parallel transport this vector according to the connection. When you are
done, you will have another vector $e_2\in\pi^{-1}(p)$. Write
$e_2=g_\Gamma(e_1)$, where $g_\Gamma(e_1)$ is an element of the
structure group of the bundle. The (global) {\em holonomy\/} group of $E\to M$
is defined as the group generated by all such $g_\Gamma(e_1)$ for all
closed paths $\Gamma$. The holonomy, $\Hl_M$, of a Riemannian manifold, $M$,
is defined as the holonomy of the tangent bundle equipped with the
Levi-Civita connection from the metric.

The holonomy of a Riemannian manifold of real
dimension $d$ is contained in $\GO(d)$. If it is orientable this
becomes $\SO(d)$. The study of which other holonomy
groups are possible is a very interesting question and will be of some
importance to us. We refer the reader to \cite{Besse:E,Sal:hol} for a
full discussion of the results and derivations. We require the
following:
\begin{enumerate}
\item $\Hl_M\subseteq {\rm U}(\ff d2)$ if and only if $M$ is a K\"ahler
manifold.
\item $\Hl_M\subseteq \SU(\ff d2)$ if and only if $M$ is a Ricci-flat
K\"ahler manifold.
\item $\Hl_M\subseteq \Sp(\ff d4)$ if and only if $M$ is a
hyperk\"ahler manifold. 
\item $\Hl_M\subseteq \Sp(\ff d4).\Sp(1)$ if and only if $M$ is a
quaternionic K\"ahler manifold.\footnote{The ``$.$'' denotes that we
take the direct product except that the $\Z_2$ centers of each group
are identified.}
\item A ``symmetric space'' of the form $G/H$ where $G$ and $H$ are
Lie groups has holonomy $H$.
\end{enumerate}
Actually in each case the specific representation of the group in
which the fibre transforms is also fixed. A celebrated theorem due to
Berger, with contributions from Simons \cite{Sal:hol}, then states
that {\em the only other 
possibilities not yet mentioned are $\Hl_M\cong G_2$ where the fibre
transforms as a {\bf 7} or $\Hl_M\cong {\rm Spin}(7)$ where the fibre
transforms as an {\bf 8}.}

There is a fairly clear relationship between the holonomy groups and
the invariant forms of the natural metric on $M$ in the first
cases. For the most general case we have that the form
\begin{equation}
  \sum_{i=1}^d dx^i\otimes dx^i
\end{equation}
on $\R^d$ admits $\GO(d)$ as the group of invariances.
In the complex K\"ahler case we consider the Hermitian form
\begin{equation}
  \sum_{i=1}^{d/2} d\bar z^i\otimes dz^i
\end{equation}
on $\C^{d/2}$ which admits $\GU(\ff d2)$ as the invariance group.

For the next case we consider the quaternionic numbers, $\H$. In this
case the natural form
\begin{equation}
  \sum_{i=1}^{d/4} d\zeta^i\otimes d\bar\zeta^i
\end{equation}
on $\H^{d/4}$ is preserved by $\Sp(\ff d4)$. Note that
writing quaternions as $2\times 2$ matrices in the usual way gives an
embedding $\Sp(\ff d4)\subset \SU(\ff d2)$. Thus, a hyperk\"ahler
manifold is always a Ricci-flat K\"ahler manifold. In fact, one is
free to choose one of a family of complex structures. Let us denote a
quaternion by $q=a+bI+cJ+cK$, where $a,b,c,d\in\R$, $I^2=J^2=K^2=-1$ and
$IJ=K$, $JI=-K$, etc. Given a hyperk\"ahler structure we may choose a
complex structure given by $q$, where $q^2=-1$. This implies $a=0$ and
\begin{equation}
  b^2+c^2+d^2=1.	\label{eq:qsph}
\end{equation}
Thus for a given hyperk\"ahler structure we have a whole $S^2$ of
possible complex structures. We will see that the K3 surface is
hyperk\"ahler when equipped with a Ricci-flat metric. 

Because of the fact that quaternionic numbers are not commutative, we
also have the notion of a quaternionic K\"ahler manifold in addition
to that of the hyperk\"ahler manifold. The space $\H^n$ admits an action
of $\Sp(n).\Sp(1)$ by multiplication on the right by $n\times n$
quaternionic matrices in $\Sp(n)$ and by a quaternion of unit norm on
the left. This also leads to the notion of a manifold with a kind of
quaternionic structure --- this time the ``quaternionic K\"ahler
manifold''. The main difference between this and the hyperk\"ahler
manifold is that the extra $\Sp(1)$ can act on the $S^2$ of complex
structures between patches and so destroy any global complex
structure. All that remains is an $S^2$ bundle of almost
complex structures which need have no global section. 
Thus a generic
quaternionic K\"ahler manifold will {\em not\/} admit a complex
structure. When this bundle
is trivial the situation reduces to the hyperk\"ahler case. As an
example, the space $\H\P^n$ is quaternionic K\"ahler. Note that the
case $n=1$ is somewhat redundant as this reduces to
$\Sp(1).\Sp(1)\cong SO(4)$, which gives a generic orientable Riemannian
manifold.

\subsection{Moduli space of complex structures}   \label{ss:cx}

We now want to construct the moduli space of all K3 surfaces. 
In order to determine the
moduli space it is very important to specify exactly what data defines
a particular K3 surface. By considering various data we will construct
several different moduli spaces throughout these lectures. To begin
with we want to consider the K3 surface purely as an object in
algebraic geometry and, as such, we will find the moduli space of
complex structures for K3 surfaces.

To ``measure'' the complex structure we need some relatively simple
quantity which depends on the complex structure. This will be provided
by ``periods'' which are simply integrals of the holomorphic 2-form,
$\Omega$, over integral 2-cycles within $S$. To analyze periods we
first then require an understanding of the integral 2-cycles
$H_2(S,\Z)$.

Since $b_2(S)=22$ from the previous section, we see that $H_2(S,\Z)$
is isomorphic to $\Z^{22}$ as a group.\footnote{Actually we need the
result that $H_2(S,\Z)$ is {\em torsion-free\/} to make this statement
to avoid any finite subgroups appearing. This follows from
$\pi_1(S)=0$ and the various relations between homotopy and torsion in
homology and cohomology \cite{BT:}.} In addition to this group
structure we may specify an inner product between any two elements,
$\alpha_i\in H_2(S,\Z)$, given by
\begin{equation}
  \alpha_1.\alpha_2 = \#(\alpha_1\cap\alpha_2),	\label{eq:inner}
\end{equation}
where the notation on the right refers to the oriented intersection
number, which is a natural operation on homology cycles \cite{EGH:dg}.
This abelian group structure with an inner product gives $H_2(S,\Z)$ the
structure of a lattice.

The signature of this lattice can be determined from the index theorem
for the {\em signature complex\/} \cite{EGH:dg}
\begin{equation}
  \tau = \int_S\ff13(c_1^2-2c_2) = -\ff23\chi(S) = -16.
\end{equation}
Thus the 22-dimensional lattice has signature $(3,19)$.

Poincar\'e duality tells us that given a basis $\{e_i\}$ of 2-cycles
for $H_2(S,\Z)$, for each $e_i$ we may find an $e^*_j$ such that
\begin{equation}
  e_i.e^*_j = \delta_{ij},
\end{equation}
where the set $\{e^*_j\}$ also forms a basis for $H_2(S,\Z)$. Thus
$H_2(S,\Z)$ is a {\em self-dual\/} (or {\em unimodular\/}) lattice.
Note that this also means that the lattice of integral cohomology,
$H^2(S,\Z)$, is isomorphic to the lattice of integral homology, $H_2(S,\Z)$.

The next fact we require is that the lattice $H_2(S,\Z)$ is {\em
even}. That is,
\begin{equation}
  e.e\in2\Z,\quad\forall e\in H_2(S,\Z).
\end{equation}
This is a basic topology fact for any spin manifold, i.e., for
$c_1(T_X)=0\pmod2$. We will not attempt a proof of this as
this is rather difficult (see, for example, Wu's formula in
\cite{Span:}).

The classification of even self-dual lattices is extremely
restrictive. We will use the notation $\Gamma_{m,n}$ to refer to an
even self-dual lattice of signature $(m,n)$. It is known that 
$m$ and $n$ must satisfy 
\begin{equation}
m-n=0\pmod8
\end{equation}
and that if $m>0$ and $n>0$ then $\Gamma_{m,n}$ is unique up to
isometries \cite{Serre:ca,CS:sphere}. An isometry is an automorphism of the
lattice which preserves the inner product.

In our case, one may chose a basis such that the inner product on the
basis elements forms the matrix
\begin{equation}
\left(\begin{array}{c|c|c|c|c}
  &&&&\\&&&&\\
  \makebox[50pt]{$-E_8$}&&&&\\
  &&&&\\&&&&\\ \hline
  &&&&\\&&&&\\
  &\makebox[50pt]{$-E_8$}&&&\\
  &&&&\\&&&&\\ \hline
  &&U&&\\ \hline
  &&&U&\\ \hline
  &&&&U
\end{array}\right),
\end{equation}
where $-E_8$ denotes the $8\times8$ matrix given by minus the Cartan
matrix of the Lie algebra $E_8$ and $U$ represents the ``hyperbolic
plane''
\begin{equation}
U\cong\left(\begin{array}{cc}0&1\\1&0\end{array}\right). \label{eq:U}
\end{equation}

Now we may consider periods
\begin{equation}
  \varpi_i = \int_{e_i}\Omega.
\end{equation}
We wish to phrase these periods in terms of the lattice
$\Gamma_{3,19}$ we have just discussed. 
First we will fix a specific embedding of a basis, $\{e_i\}$, of
2-cycles into the lattice $\Gamma_{3,19}$. That is, we make a specific
choice of which periods we will determine. Such a choice is called a 
``marking'' and a K3 surface, together with such a marking, is called a 
``marked K3 surface''.

There is the natural embedding
\begin{equation}
  \Gamma_{3,19}\cong H^2(S,\Z)\subset H^2(S,\R)\cong \R^{3,19}.
\end{equation}
We may now divide $\Omega\in H^2(S,\C)$ as
\begin{equation}
  \Omega = x+iy,
\end{equation}
where $x,y\in H^2(S,\R)$. Now a $(p,q)$-form integrated over $S$ is
only nonzero if $p=q=2$ \cite{GH:alg} and so
\begin{equation}
\eqalign{\int_S\Omega\wedge\Omega&=(x+iy).(x+iy)=(x.x-y.y)+2ix.y\cr
  &=0.\cr}  \label{eq:xy1}
\end{equation}
Thus, $x.x=y.y$ and $x.y=0$. We also have
\begin{equation}
\eqalign{\int_S\Omega\wedge\overline\Omega&=(x+iy).(x-iy)=(x.x+y.y)\cr
  &=\int_S\|\Omega\|^2>0.\cr}  \label{eq:xy2}
\end{equation}
The vectors $x$ and $y$ must be linearly independent over $H^2(S,\R)$
and so span a 2-plane in $H^2(S,\R)$ which we will also give the name
$\Omega$. The relations (\ref{eq:xy1}) and (\ref{eq:xy2}) determine
that this 2-plane must be space-like, i.e., any vector, $v$, within it
satisfies $v.v>0$.

Note that the 2-plane is equipped with a natural orientation but that
under complex conjugation one induces $(x,y)\to(x,-y)$ and this
orientation is reversed.

We therefore have the following picture. The choice of a complex
structure on a K3 surface determines a vector space $\R^{3,19}$ which
contains an even self-dual lattice $\Gamma_{3,19}$ and an oriented
2-plane $\Omega$. If we change the complex structure on the K3 surface
we expect the periods to change and so the plane $\Omega$ will rotate
with respect to the lattice $\Gamma_{3,19}$.

We almost have enough technology now to build our moduli space of
complex structures on a K3 surface. Before we can give the result
however we need to worry about special things that can happen within
the moduli space. A K3 surface which gives a 2-plane, $\Omega$, which
very nearly contains a light-like direction, will have periods which
are only just acceptable and so this K3 surface will be near the
boundary of our moduli space. As we approach the boundary we expect the
K3 surfaces to degenerate in some way. Aside from this obvious
behaviour we need to worry that some points away from this natural
boundary may also correspond to K3 surfaces which have degenerated in 
some way. It turns out that there are such points in the moduli space
and these will be of particular interest to us in these lectures. They will
correspond to {\em orbifolds\/}, as we will explain in detail
in section \ref{ss:orb}. For now, however, we need to include the
orbifolds in our moduli 
space to be able to state the form of the moduli space.

The last result we require is the following
\begin{theorem}[Torelli]
  The moduli space of complex structures on a marked K3 surface
(including orbifold points) is given by the space of possible periods.
\end{theorem}
For an account of the origin of this theorem we refer to \cite{BPV:}.
Thus, the moduli space of complex structures on a marked K3 surface is given
by the space of all possible oriented 2-planes in $\R^{3,19}$ with
respect to a fixed lattice $\Gamma_{3,19}$.

Consider this space of oriented 2-planes in
$\R^{3,19}$. Such a space is called a Grassmannian, which we denote
$\Gr^+(\Omega,\R^{3,19})$, and we have
\begin{equation}
  \Gr^+(\Omega,\R^{3,19}) \cong \frac{\GO^+(3,19)}{(\GO(2)\times
\GO(1,19))^+}.
	\label{eq:Tcx}
\end{equation}
This may be deduced as follows. In order to build the Grassmannian of
2-planes in $\R^{3,19}$, first consider all rotations, $\GO(3,19)$, of
$\R^{3,19}$. Of these, we do not care about internal rotations within
the 2-plane, $\GO(2)$, or rotations normal to it, $\GO(1,19)$. For
oriented 2-planes we consider only the index 2 subgroup which
preserves orientation on the space-like directions. We use the ``$+$''
superscripts to denote this.

This Grassmannian builds the moduli space of marked K3 surfaces. We now
want to remove the effects of the marking. There are diffeomorphisms
of the K3 surface, which we want to regard as trivial as far as our
moduli space is concerned, but which have a nontrivial action on
the lattice $H^2(S,\Z)$.  
Clearly any diffeomorphism induces an isometry of $H^2(S,\Z)$,
preserving the inner product. We denote the full group of such
isometries as $\GO(\Gamma_{3,19})$.\footnote{Sometimes the less
precise notation $\GO(3,19;\Z)$ is used.}
Our moduli
space of marked K3 surfaces can be viewed as a kind of Teichm\"uller
space, and the image of the diffeomorphisms in $\GO(\Gamma_{3,19})$
can be viewed as the modular group. The moduli space is the quotient
of the Teichm\"uller space by the modular group. 

What is this modular group? It was shown in \cite{Borcea:d,Mat:d} that
any element of $\GO^+(\Gamma_{3,19})$ can be induced from a
diffeomorphism of the K3 surface. It was shown further in
\cite{Don:inv} that any element of $\GO(\Gamma_{3,19})$ which is not in
$\GO^+(\Gamma_{3,19})$ {\em cannot\/} be induced by a
diffeomorphism. Thus our modular group is precisely $\GO^+(\Gamma_{3,19})$.

Treating (\ref{eq:Tcx}) as a
right coset we will act on the left for the action of the modular
group. The result is that the moduli space of complex structures on a
K3 surface (including orbifold points) is
\begin{equation}
  \cM_c \cong \GO^+(\Gamma_{3,19})\backslash\GO^+(3,19)/(\GO(2)\times
\GO(1,19))^+.	\label{eq:Mcx}
\end{equation}

When dealing with $\cM_c$ it is important to realize that
$\GO^+(\Gamma_{3,19})$ has an ergodic action on the Teichm\"uller space
and thus $\cM_c$ is actually not Hausdorff. Such unpleasant
behaviour is sometimes seen in string theory in fairly pathological
circumstances \cite{Moore:nH} but it seems reasonable to expect that
under reasonable conditions we should see a fairly well-behaved moduli
space. As we shall see, the moduli space $\cM_c$ does not appear to
make any natural appearance in string theory and the related
moduli spaces which do appear will actually be Hausdorff.

\subsection{Einstein metrics}   \label{ss:Em}

The first modification we will consider is that of considering the
moduli space of Einstein metrics on a K3 surface. We will always
assume that the metric is K\"ahler.

An Einstein metric is a (pseudo-)Riemannian metric on a
(pseudo-)Riemannian manifold whose Ricci curvature is proportional to
the metric. 
Actually, for a K3 surface, this condition implies that
the metric is Ricci-flat \cite{Hit:c4}. We may thus use the terms
``Einstein'' and ``Ricci-flat'' interchangeably in our discussion of
K3 surfaces.

\def\cHs#1{\cH^{\;\>#1}}
The Hodge star will play an essential r\^ole in the discussion of the
desired moduli space. Recall that \cite{EGH:dg,Besse:E}
\begin{equation}
   \alpha\wedge*\beta = (\alpha,\beta)\omega_g,
\end{equation}
where $\alpha$ and $\beta$ are $p$-forms, $\omega_g$ is the volume
form and $(\alpha,\beta)$ is given by
\begin{equation}
  (\alpha,\beta) = p!\int \sum_{i_1i_2j_1j_2\ldots}g^{i_1i_2}g^{j_1j_2}
  \alpha_{i_1j_1\ldots}\beta_{i_2j_2\ldots}\,dx_1dx_2\ldots,
\end{equation}
in local coordinates. In particular, if $\alpha$ is self-dual, in the
sense $\alpha=*\alpha$, then $\alpha.\alpha>0$ in the notation of
section \ref{ss:cx}. Similarly an anti-self-dual 2-form will obey
$\alpha.\alpha<0$. On our K3 surface $S$ we may decompose 
\begin{equation}
  H^2(S,\R) = \cHs+\oplus\cHs-,
\end{equation}
where $\cHs\pm$ represents the cohomology of the space of
(anti-)self-dual 2-forms. We then see that 
\begin{equation}
  \dim\cHs+=3,\quad\dim\cHs-=19,
\end{equation}
from section \ref{ss:cx}.

The curvature acts naturally on the bundle of (anti)-self-dual
2-forms. By standard methods (see, for example, \cite{Besse:E}) one
may show that the curvature of the bundle of self-dual 2-forms is
actually zero when the manifold in question is a K3 surface. This is
one way of seeing directly the action of the $\SU(2)$ holonomy of
section \ref{ss:hol}. Since a K3 is simply-connected, this
shows that the bundle $\cHs+$ is trivial and thus has 3 linearly
independent sections.

Consider a local orthonormal frame of the cotangent bundle
$\{e_1,e_2,e_3,e_4\}$. We may write the three sections of $\cHs+$ as
\begin{equation}
  \eqalign{s_1 &= e_1\wedge e_2+ e_3\wedge e_4\cr
	s_2 &= e_1\wedge e_3+ e_4\wedge e_2\cr
	s_3 &= e_1\wedge e_4+ e_2\wedge e_3.\cr}
\end{equation}
Clearly an $\SO(4)$ rotation of the cotangent directions produces an
$\SO(3)$ rotation of $\cHs+$. That is, a rotation within $\cHs+$ is
induced by a reparametrization of the underlying K3 surface. One
should note that the orientation of $\cHs+$ is fixed.

Let us denote by $\Sigma$ the space $\cHs+$ viewed as a subspace of
$H^2(S,\R)$. 
Putting $dz_1=e_1+ie_2$ and $dz_2=e_3+ie_4$ we obtain a K\"ahler form
equal to $s_1$, and the holomorphic 2-form $dz_1\wedge dz_2$ is given
by $s_2+is_3$. This shows that $\Sigma$ is spanned by the 2-plane
$\Omega$ of section \ref{ss:cx} together with the direction in
$H^2(S,\R)$ given by the K\"ahler form.

This fits in very nicely with Yau's theorem \cite{Yau:} which
states that for any manifold $M$ with $K=0$, and a fixed complex
structure, given a cohomology class of the K\"ahler form, there
exists a unique Ricci-flat metric. Thus, we fix the complex structure
by specifying the 2-plane, $\Omega$, and then choose a K\"ahler form, $J$,
by specifying another direction in $H^2(S,\R)$. Clearly this third
direction is space-like, since 
\begin{equation}
  \Vol(S) = \int_S J\wedge J > 0,  \label{eq:volK3}
\end{equation}
and it is perpendicular to $\Omega$ as the K\"ahler form is of type
$(1,1)$. Thus $\Sigma$, spanned by $\Omega$ and $J$, is space-like.

The beauty of Yau's theorem is that we need never concern ourselves
with the explicit form of the Einstein metric on the K3 surface. Once
we have fixed $\Omega$ and $J$, we know that a unique metric
exists. Traditionalists may find it rather unsatisfactory that we do
not write the metric down --- indeed no explicit metric on a K3
surface has ever been determined to date --- but one of the lessons we
appear to have learnt from the analysis of \CY\ manifolds in string
theory is that knowledge of the metric is relatively unimportant.

As far as our moduli space is concerned, one aspect of the above analysis
which is important is that rotations within
the 3-plane, $\Sigma$, may affect what we consider to be the K\"ahler
form and complex structure but they do not affect the underlying
Riemannian metric. We see that a K3 surface viewed as a Riemannian
manifold may admit a whole family of complex structures. Actually this
family is parametrized by the sphere, $S^2$, of ways in which $\Sigma$
is divided into $\Omega$ and $J$.

This property comes from the fact that a K3 surface admits a 
hyperk\"ahler structure. This is obvious from section \ref{ss:hol} as a
K3 is Ricci-flat and K\"ahler and thus has holonomy $\SU(2)$, and 
$\SU(2)\cong \Sp(1)$.
The sphere (\ref{eq:qsph}) of possible complex structures is exactly
the $S^2$ degree of freedom of 
rotating within the 3-plane $\Sigma$ we found above.

Some care should be taken to show that the maps involved are surjective
\cite{Tod:inv,Mor:Katata} but we end up with the following
\cite{Kobayashi-Todorov}
\begin{theorem}
  The moduli space of Einstein metrics, $\cM_E$, for a K3 surface
(including orbifold points) is given by the Grassmannian of oriented
3-planes within the space $\R^{3,19}$ modulo the effects of
diffeomorphisms acting on the lattice $H^2(S,\Z)$.
\end{theorem}
In other words, we have a relation similar\footnote{Note that the
orientation problem makes this look more unlike (\ref{eq:Mcx}) than it
needs to! We encourage the reader to not concern themselves with these
orientation issues, at least on first reading.} to (\ref{eq:Mcx}):
\begin{equation}
\cM_E \cong \GO^+(\Gamma_{3,19})\backslash \GO^+(3,19)/(\SO(3)\times
\GO(19))\times \R_+,
\end{equation}
where the $\R_+$ factor denotes the volume of the K3 surface given by
(\ref{eq:volK3}).

This is actually isomorphic to the space
\begin{equation}
\cM_E \cong \GO(\Gamma_{3,19})\backslash \GO(3,19)/(\GO(3)\times
\GO(19))\times \R_+,		\label{eq:ME}
\end{equation}
since the extra generator required, $-1\in \GO(3,19)$, to elevate
$\GO^+(3,19)$ to $\GO(3,19)$, is in the
center and so taking the left-right coset makes no difference.

Note that (\ref{eq:ME}) is actually a Hausdorff space as the discrete
group $\GO(\Gamma_{3,19})$ has a properly discontinuous action
\cite{Allan:}. Thus we see that the global description of this space of
Einstein metrics on a K3 surface is much better behaved than the
moduli space of complex structures discussed earlier.

\subsection{Algebraic K3 Surfaces}  \label{ss:alg}

In section \ref{ss:Em} we enlarged the moduli space of complex
structures of section \ref{ss:cx} and we found a space with nice
properties. In this section we are going to find another nice moduli
space by going in the opposite direction. That is, we will restrict
the moduli space of complex structures by putting 
constraints on the K3 surface. We are going to consider {\em
algebraic\/} K3 surfaces, i.e., K3 surfaces that may be embedded
by algebraic (holomorphic) equations in homogeneous coordinates into
$\P^N$ for some $N$. 

The analysis is done in terms of algebraic curves, that is, Riemann
surfaces which have been holomorphically embedded into our K3 surface
$S$. Clearly such a curve, $C$, may be regarded as a homology cycle
and thus an element of $H_2(S,\Z)$. By the ``dual of the dual''
construction of section \ref{ss:def} we may also regard it as an
element $C\in H^2(S,\Z)$. The fact that $C$ is holomorphically embedded
may also be used to show that $C\in H^{1,1}(S)$ \cite{GH:alg}. We then
have $C\in\Pic(S)$, where we define
\begin{equation}
  \Pic(S) = H^2(S,\Z)\cap H^{1,1}(S),
\end{equation}
which is called the ``Picard group'', or ``Picard lattice'', of
$S$. We also define the ``Picard number'', $\rho(S)$, as the rank of
the Picard lattice. Any element of the Picard group, $e\in\Pic(S)$,
corresponds to a line bundle, $L$, such that $c_1(L)=e$
\cite{GH:alg}. Thus the Picard group may be regarded as the group of
line bundles on $S$, where the group composition is the Whitney
product.

As the complex structure of $S$ is varied, the Picard group
changes. This is because an element of $H^2(S,\Z)$ that was regarded
as having type purely $(1,1)$ may pick up parts of type $(2,0)$ or $(0,2)$ as
we vary the complex structure. A completely generic K3 surface will
have completely trivial Picard group, i.e., $\rho=0$.

The fact that $S$ contains a curve $C$ is therefore a restriction on
the complex structure of $S$. An algebraic K3 surface similarly has
its deformations restricted as the embedding in $\P^N$ will imply the
existence of one or more curves. As an example let us return to the
case where $S$ is a quartic surface
\begin{equation}
  f=x_0^4+x_1^4+x_2^4+x_3^4=0,	\label{eq:FerK3}
\end{equation}
in $\P^3$. A hyperplane $\P^2\subset\P^3$ will cut $f=0$ along a
curve and so shows the existence of an algebraic curve $C$. Taking
various hyperplanes will produce various curves but they are all
homologous as 2-cycles and thus define a unique element of the Picard
lattice. Thus the Picard number of this quartic surface is at least
one. Actually for the ``Fermat'' form of $f$ given in (\ref{eq:FerK3})
it turns out\footnote{I thank M.~Gross for explaining this to me.}
that $\rho=20$ (which is clearly the maximum value we may have since
$\dim H^{1,1}(S)=20$). A quartic surface need not be in the special
form (\ref{eq:FerK3}). We may consider a more general
\begin{equation}
  f=\sum_{i+j+k+l=4}a_{ijkl}\,x_0^ix_1^jx_2^kx_3^l,  \label{eq:genQ}
\end{equation}
for arbitrary $a_{ijkl}\in\C$. Generically we expect no elements of
the Picard lattice other than those generated by $C$ and so
$\rho(S)=1$.

Now consider the moduli space of complex structures on a quartic
surface. Since $C$ is of type $(1,1)$, the 2-plane $\Omega$ of section
\ref{ss:cx} must remain orthogonal to this direction. 

We may determine $C.C$ by taking two hyperplane sections and finding
the number of points of intersection. The intersection of two
hyperplanes in $\P^3$ is clearly $\P^1$ and so the intersection $C\cap
C$ is given by a quartic in $\P^1$, which is 4 points. Thus $C.C=4$ and
$C$ spans a space-like direction.

Our moduli space will
be similar to that of (\ref{eq:Mcx}) except that we may remove the
direction generated by $C$ from consideration. Thus our 2-plane is now
embedded in $\R^{2,19}$ and the discrete group is generated by the
lattice $\Lambda_C = \Gamma_{3,19}\cap C^\perp$. Note that we do not
denote $\Lambda_C$ as $\Gamma_{2,19}$ as it is {\em not\/}
even-self-dual. The moduli space in question is then
\begin{equation}
\cM_{\rm Quartic} \cong \GO(\Lambda_C)\backslash \GO(2,19)/(\GO(2)\times
\GO(19)).		\label{eq:Mquar}
\end{equation}
This is Hausdorff. Note also that it is a space of complex
dimension 19 and that a simple analysis of (\ref{eq:genQ}) shows that
$f=0$ has 19 deformations of $a_{ijkl}$ modulo reparametrizations of the
embedding $\P^3$. Thus embedding this K3 surface in $\P^3$ has brought
about a better-behaved moduli space of complex structures but we have
``lost'' one deformation as (\ref{eq:Mcx}) has complex dimension 20.

One may consider a more elaborate embedding such as a hypersurface in
$\P^2\times\P^1$ given by an equation of bidegree $(3,2)$, i.e.,
\begin{equation}
  f = \sum_{{a_0+a_1+a_2=3\atop b_0+b_1=2}} A_{a_0a_1a_2,b_0b_1}\,
      x_0^{a_0}x_1^{a_1}x_2^{a_2}y_0^{b_0}y_1^{b_1}.
\end{equation}
This is an algebraic K3 since $\P^2\times\P^1$ itself may be embedded
in $\P^5$ (see, for example, \cite{GH:alg}). Taking a hyperplane
$\P^1\times\P^1 
\subset\P^2\times\P^1$ one cuts out a curve $C_1$. Taking the hyperplane 
$\P^2\times\{p\}$ for some $p\in\P^1$ cuts out $C_2$. By the same
method we used for the quartic above we find the intersection matrix
\begin{equation}
  \left(\begin{array}{cc}2&3\\3&0\end{array}\right).
\end{equation}
Thus, denoting $\Lambda_{C_1C_2}$ by the sublattice of $\Gamma_{3,19}$
orthogonal to $C_1$ and $C_2$ we have
\begin{equation}
\cM \cong \GO(\Lambda_{C_1C_2})\backslash \GO(2,18)/(\GO(2)\times
\GO(18)).		
\end{equation}
This moduli space has dimension 18 and this algebraic K3 surface has
$\rho(S)=2$ generically. In general it is easy to see that the
dimension of the moduli space plus the generic Picard number will
equal 20. Note that the Picard lattice will have signature
$(1,\rho-1)$. This follows as it is orthogonal to $\Omega$
inside $\R^{3,19}$ and thus has at most one space-like direction but
the natural K\"ahler form inherited from the ambient $\P^N$ is itself
in the Picard lattice and so there must be at least one space-like
direction. Thus the moduli space of complex structures on an algebraic
K3 surface will always be of the form 
\begin{equation}
\cM_{\rm Alg} \cong \GO(\Lambda)\backslash \GO(2,20-\rho)/(\GO(2)\times
\GO(20-\rho)),		\label{eq:Malg}
\end{equation}
where $\Lambda$ is the sublattice of $H^2(S,\Z)$ orthogonal to the Picard
lattice. This lattice is often referred to as the {\em transcendental
lattice\/} of $S$. Note that this lattice is rarely self-dual.

\subsection{Orbifolds and blow-ups}  \label{ss:orb}

When discussing the moduli spaces above we have had to be careful to
note that we may be including K3 surfaces which are not manifolds but,
rather, {\em orbifolds}. The term ``orbifold'' was introduced many
years ago by W.~Thurston after their first appearance in the
mathematics literature in \cite{Sat:V} (where they were referred to as
``V-manifolds''). 
The general idea is to slightly enlarge the concept of a manifold to
objects which contain singularities produced by quotients.
They have subsequently played a celebrated r\^ole in
string theory after they made their entry into the subject in
\cite{DHVW:}. It is probably worthwhile noting that the definition
of an orbifold is slightly different in mathematics and physics. We
will adopt the mathematics definition which, for our purposes, we take
to be defined as follows:
\begin{quotation}\em
  An orbifold is a space which admits an open covering, $\{U_i\}$,
such that each patch is diffeomorphic to $\R^n/G_i$.
\end{quotation}
The $G_i$'s are discrete groups (which may be trivial) which can be
taken to fix the origin of $\R^n$. The physics definition however is
more global and defines an orbifold to be a space of the form $M/G$ where $M$
is a manifold and $G$ is a discrete group. A physicist's orbifold
is a special case of the 
orbifolds we consider here. Which definition is applicable to string
theory is arguable. Most of the vast amount of analysis that has been
done over the last 10 years on orbifolds has relied on the global form
$M/G$. Having said that, one of the appeals of orbifolds is that
string theory is 
(generically) well-behaved on such an object, and this behaviour only
appears to require the local condition.

The definition of an orbifold can be extended to the notion of a
complex orbifold where each patch is biholomorphic to $\C^n/G_i$ and
the induced transition functions are holomorphic. Since we may define
a metric on $\C^n/G_i$ by the natural inherited metric on $\C^n$, we
have a notion of a metric on an orbifold. In fact, it is not hard to
extend the notion of a K\"ahler-Einstein metric to include orbifolds
\cite{Kob:orb}. Similarly, a complex orbifold may be embeddable in
$\P^N$ and can 
thus be viewed as an algebraic variety. In this case the notion of
canonical class is still valid. Thus, there is nothing to stop the
definition of the K3 surface being extended to include orbifolds. As
we will see in this section, such K3 surfaces lie naturally at the
boundary of the moduli space of K3 manifolds.

An example of such a K3 orbifold is the following, which is often the
first K3 surface that string theorists encounter. Take the 4-torus
defined as a complex manifold of dimension two by dividing the complex
plane $\C^2$ with affine coordinates $(z_1,z_2)$ by the group $\Z^4$
generated by
\begin{equation}
  z_k\mapsto z_k+1,\quad z_k\mapsto z_k+i,\quad k=1,2.
\end{equation}
Then consider the $\Z_2$ group of isometries generated by
$(z_1,z_2)\mapsto(-z_1,-z_2)$. It is not hard to see that this $\Z_2$
generator fixes 16 points: $(0,0)$, $(0,\ff12)$, $(0,\ff12i)$, 
$(0,\ff12+\ff12i)$, $(\ff12,0)$, \ldots,
$(\ff12+\ff12i,\ff12+\ff12i)$. Thus we have an orbifold, which we will
denote $S_0$.

Since the $\Z_2$-action respects the complex structure and leaves the
K\"ahler form invariant we expect $S_0$ to be a complex K\"ahler
orbifold. Also, a moment's thought shows that any of the
non-contractable loops of the 4-torus may be shrunk to a point after
the $\Z_2$-identification is made. Thus $\pi_1(S_0)=0$. Also, the
holomorphic 2-form $dz_1\wedge dz_2$ is invariant. We thus expect
$K=0$ for $S_0$. All said, the orbifold $S_0$ has every right to be
called a K3 surface. 

We now want to see what the relation of this orbifold $S_0$ might be
to the general class of K3 manifolds. To do this, we are going to
modify $S_0$ to make it smooth. This process is known as
``blowing-up''. This procedure is completely local and so we may
restrict attention to a patch within $S_0$. Clearly the patch of
interest is $\C^2/\Z_2$.

The space $\C^2/\Z_2$ can be written algebraically by embedding it in
$\C^3$ as follows. Let $(x_0,x_1,x_2)$ denote the coordinates of $\C^3$ and
consider the hypersurface $A$ given by
\begin{equation}
  f = x_0x_1-x_2^2=0. 	\label{eq:Z2,1}
\end{equation}
A hypersurface is smooth if and only if $\partial f/\partial x_0=\ldots
=\partial f/\partial x_2=f=0$ has no solution. Thus $f=0$ is smooth
everywhere except at the origin where it is singular. We can
parameterize $f=0$ by putting $x_0=\xi^2$, $x_1=\eta^2$, and
$x_2=\xi\eta$. Clearly then $(\xi,\eta)$ and $(-\xi,-\eta)$ denote the
same point. This is the only identification and so $f=0$ in $\C^3$
really is the orbifold $\C^2/\Z_2$ we require.

Consider now the following subspace of $\C^3\times\P^2$:
\begin{equation}
  \{(x_0,x_1,x_2),[s_0,s_1,s_2]\in\C^3\times\P^2;\;x_is_j=x_js_i,\;
\forall i,j\}.	\label{eq:O2(-1)}
\end{equation}
This space may be viewed in two ways --- either by trying to project
it onto the $\C^3$ or the $\P^2$. Fixing a point in $\P^2$ determines
a line (i.e., $\C^1$) in $\C^3$. Thus, (\ref{eq:O2(-1)}) determines a
line bundle on $\P^2$. One may determine $c_1=-H$ for this bundle,
where $H$ is the hyperplane class. We may thus denote this bundle
${\cal O}_{\P^2}(-1)$. Alternatively one may fix a point in $\C^3$. If
this is not the point $(0,0,0)$, this determines a point in $\P^2$. At
$(0,0,0)$ however we have the entire $\P^2$. Thus ${\cal
O}_{\P^2}(-1)$ can be identified pointwise with $\C^3$ except that the
origin in $\C^3$ has been replaced by $\P^2$. The space
(\ref{eq:O2(-1)}) is thus referred to as a {\em blow-up of\/} $\C^3$
{\em at
the origin}. The fact the ${\cal O}_{\P^2}(-1)$ and $\C^3$ are
generically isomorphic as complex spaces in this way away from some
subset means that these spaces are {\em birationally equivalent\/}
\cite{GH:alg}. A space $X$ blown-up at a point will be denoted
$\widetilde X$ and the birational map between them will usually be
written
\begin{equation}
  \pi:\widetilde X\to X.
\end{equation}
That is, $\pi$ represents the {\em blow-down\/} of $\widetilde X$. The
$\P^2$ which has grown out of the origin is called the {\em
exceptional divisor}.

Now let us consider what happens to the hypersurface $A$ given by
(\ref{eq:Z2,1}) 
in $\C^3$ as we blow up the origin. We will consider the {\em proper
transform}, $\widetilde A\subset \widetilde X$. If $X$ is blown-up at
the point $p\in X$ then $\widetilde A$ is defined as the closure of
the point set $\pi^{-1}(A\setminus p)$ in $\widetilde X$.

Consider following a path in $A$ towards the origin. In the blow-up,
the point we land on in the exceptional $\P^2$ in the blow-up depends
on the angle at which we approached the origin. Clearly the line given by
$(x_0t,x_1t,x_2t)$, $t\in\C$, $x_0x_1-x_2^2=0$, will land on the point
$[s_0,s_1,s_2]\in\P^2$ where again $s_0s_1-s_2^2=0$. Thus the point
set that provides the closure away from the origin is a quadric
$s_0s_1-s_2^2=0$ in $\P^2$. It is easy to show that this curve has
$\chi=2$ and is thus a sphere, or $\P^1$.

\iffigs
\begin{figure}[t]
  \centerline{\epsfxsize=11cm\epsfbox{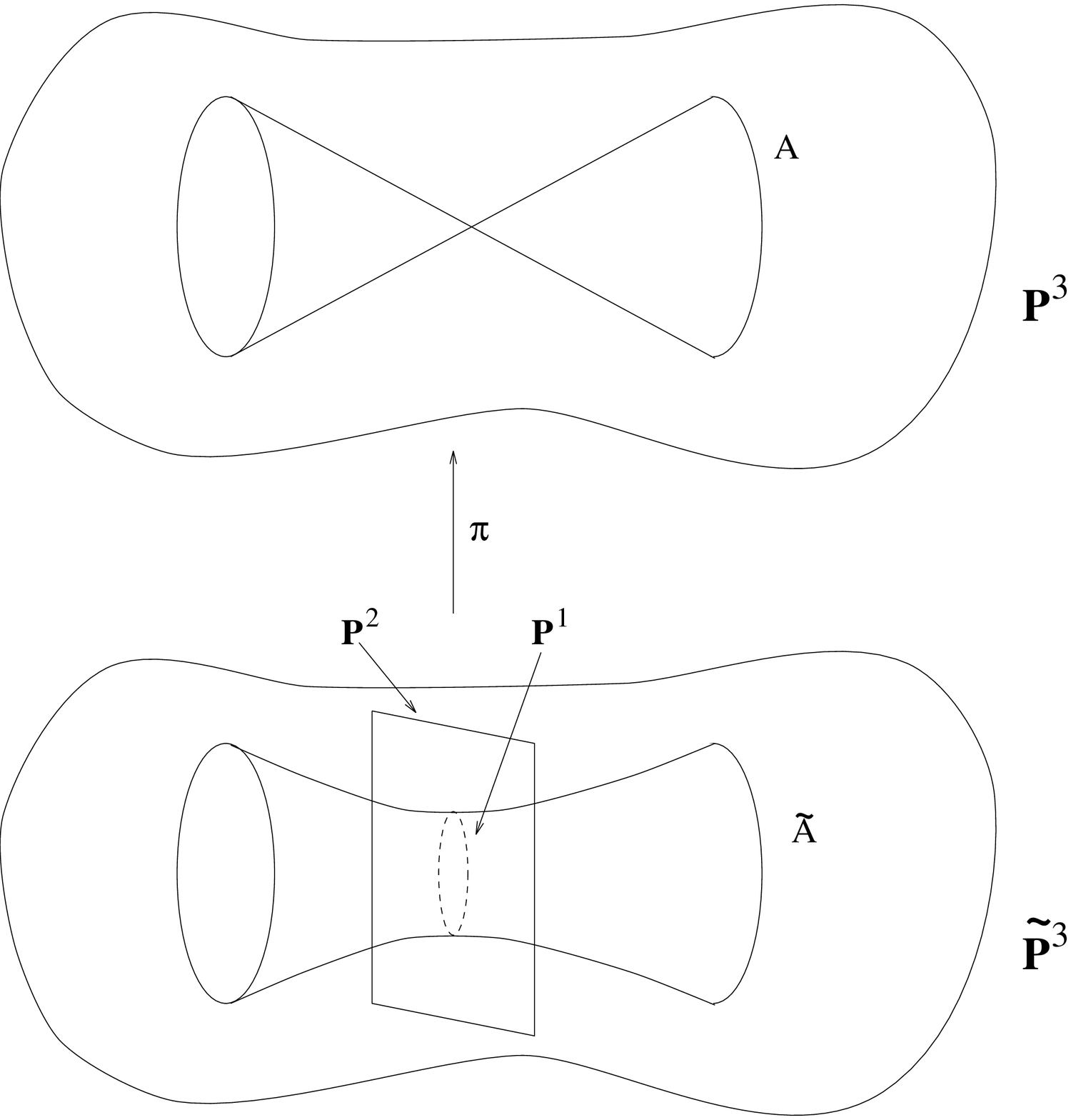}}
  \caption{Blowing up $A\protect\cong\C^2/\Z_2$.}
  \label{fig:bup1}
\end{figure}
\fi

We have thus shown that when the origin is blown-up for $A\subset\C^3$,
the proper transform of $A$ replaces the old origin, i.e., the
singularity, by a $\P^1$. Within the context of blowing up $A$, this
$\P^1$ is viewed as the exceptional divisor and we denote it $E$. What
is more, this 
resulting space, $\widetilde A$, is now smooth. We show this process
in figure \ref{fig:bup1}.

Carefully looking at the coordinate patches in $\widetilde A$ around
$E$, we can work out the normal bundle for $E\subset \widetilde
A$. The result is that this line bundle is equal to ${\cal
O}_{\P^1}(-2)$. We will refer to such a rational (i.e., genus zero)
curve in a complex surface as a ``$(-2)$-curve''.

Let us move our attention for a second to the general subject of
complex surfaces with $K=0$ and consider algebraic curves within
them. Consider a curve $C$ of genus $g$. The self-intersection of a
curve may be found by deforming the curve to another one, homologically
equivalent, and counting the numbers of points of intersection (with
orientation giving signs). In
other words, we count the number of points which remain fixed under
the deformation. Suppose we may deform and keep the curve
algebraic. Then an infinitesimal such deformation may be considered as
a section of the normal bundle of $C$ and the self-intersection is the
number of zeros, i.e., the value of $c_1$ of the normal bundle
integrated over $C$. Thus $c_1(N)$ gives the self-intersection, where
$N$ is the normal bundle. Note that two algebraic curves which
intersect transversely always have positive intersection since the
complex structure fixes the orientation. Thus this can only be carried
out when $c_1(N)>0$. We may extend the concept however when $c_1(N)<0$
to the idea of negative self-intersection. In this case we see that
$C$ cannot be deformed to a nearby algebraic curve.

The adjunction formula tells us the sum $c_1(N)+c_1(T)$, where $T$ is
the tangent bundle of $C$, must give the first Chern class of the
embedding surface restricted to $C$. Thus, if $K=0$, we have
\begin{equation}
  C.C = 2(g-1). \label{eq:aji}
\end{equation}
That is, any rational curve in a K3 surface must be a $(-2)$-curve.
Note that (\ref{eq:aji}) provides a proof of our assertion in
section \ref{ss:cx} that the self-intersection of a cycle is always an
even number --- at least in the case that the cycle is a smooth
algebraic curve.

Actually, if we blow-up all 16 fixed points of our original orbifold
$S_0$ we obtain a smooth K3 surface. To see this we need only show
that the blow-up we have done does not affect $K=0$. One can show that
this is indeed the case so long as the exceptional divisor satisfies
(\ref{eq:aji}), i.e., it is a $(-2)$-curve \cite{GH:alg}. A smooth K3
surface obtained as the blow-up of $T^4/\Z_2$ at all 16 fixed points
is called a {\em Kummer surface}. Clearly the Picard number of a
Kummer surface is at least 16. A Kummer surface need not be algebraic,
just as the original $T^4$ need not be algebraic.

Now we have enough information to find how the orbifolds, such as
$S_0$, fit into the moduli space of Einstein metrics on a K3
surface. If we blow down a $(-2)$-curve in a K3 surface we obtain,
locally, a $\C^2/\Z_2$ quotient singularity. The above description
appeared somewhat discontinuous but we may consider doing such a process
gradually as follows. Denote the $(-2)$-curve as $E$. The size of $E$
is given by the integral of the K\"ahler form over $E$, that is,
$J.E$. Keeping the complex structure of the K3 surface fixed we may
maintain $E$ in the Picard lattice but we may move $J$ so that it
becomes orthogonal to $E$. Thus $E$ has shrunk down to a point --- we
have done the blow-down.

We have shown that any rational curve in a K3 surface is an element of
the Picard lattice with $C.C=-2$. Actually the converse is true
\cite{BPV:}. That is, given an element of the Picard lattice, $e$,
such that $e.e=-2$, then either $e$ or $-e$ gives the class of a
rational curve in the K3 surface. This will help us prove the following.
Let us define the {\em roots\/} of $\Gamma_{3,19}$ as
$\{\alpha\in\Gamma_{3,19};\;\alpha.\alpha=-2\}$. 
\begin{theorem}
A point in the moduli space of Einstein metrics on a K3 surface
corresponds to an orbifold if and only if the 3-plane, $\Sigma$, is
orthogonal to a root of $\Gamma_{3,19}$.
\end{theorem}
If we take a root which is perpendicular to $\Sigma$, then it must be
perpendicular to $\Omega$ and thus in the Picard lattice. It follows
that this root (or minus the root, which is also perpendicular to
$\Sigma$) gives a rational curve in the K3. Then, since $J$ is also
perpendicular to this root, the rational curve has zero size and the
K3 surface must be singular. 
Note also that any higher genus curve cannot be shrunk down in this
way as it would be a space-like or light-like direction in the Picard
lattice and could thus not be orthogonal to $\Sigma$.
What remains to be shown is that the
resulting singular K3 surface is always an orbifold.

For a given point in the moduli space of Einstein metrics consider the
set of roots orthogonal to $\Sigma$. Suppose this set can be divided
into two mutually orthogonal sets. These would correspond to two sets
of curves which did not intersect and thus would be blown down to two
(or more) separate isolated points. Since the orbifold condition is
local we may confine our attention to the case when this doesn't
happen. The term ``root'' is borrowed from Lie group theory and we may
analyze our situation in the corresponding way. We may
choose the ``simple roots'' in our set which will span the root
lattice in the usual way (see, for example, \cite{FulHar:rep}). Now
consider the intersection matrix of the simple roots. It must have
$-2$'s down the diagonal and be negative definite (as it is orthogonal
to $\Sigma$). This is entirely analogous to the classification of
simply-laced Lie algebras and immediately tells us that there is an
$A$-$D$-$E$ classification of such events.

We have already considered the $A_1$ case above. This was the case of
a single isolated $(-2)$-curve which shrinks down to a point giving
locally a $\C^2/\Z_2$ quotient singularity. To proceed further let us
try another example. The next simplest situation is that of a
$\C^2/\Z_3$ quotient singularity given by
$(\xi,\eta)\mapsto(\omega\xi,\omega^2\eta)$, where $\omega$ is a cube
root of unity. As before we may rewrite this as a subspace of $\C^3$
as
\begin{equation}
  f=x_0x_1-x_2^3=0.
\end{equation}
The argument is very similar to the $\C^2/\Z_2$ case. The difference
is that now
as we follow a line into the blown-up singularity and consider the path
$(x_0t,x_1t,x_2t)$ within $x_0x_1-x_2^3$, the $x_2^3$ term becomes
irrelevant and the closure of the point set becomes $s_0s_1=0$ within
$\P^2$. This consists of two rational curves ($s_0=0$ and $s_1=0$)
intersecting transversely (at $[s_0,s_1,s_2]=[0,0,1]$). Thus, when we
blow-up $\C^2/\Z_3$, we obtain as an exceptional divisor two
$(-2)$-curves crossing at one point. Clearly this is the $A_2$ case.

Now consider the general case of a cyclic quotient $\C^2/\Z_n$. This
is given by $f=x_0x_1-x_2^n=0$. At first sight the discussion above
for the case $n=3$ would appear to be exactly the same for any value
of $n>3$ but actually we need to be careful that after the blow-up we
really have completely resolved the singularity. Consider the
coordinate patch in ${\cal O}_{\P^2}(-1)$ written as (\ref{eq:O2(-1)})
where $s_2\neq0$. We may use $y_0=s_0/s_2$, $y_1=s_1/s_2$ as good
affine coordinates on the base $\P^2$ and $y_2=x_2$ as a good
coordinate in the fibre. Since $x_0=y_0y_2$ and $x_1=y_1y_2$, our
hypersurface becomes 
\begin{equation}
  y_2^2(y_0y_1-y_2^{n-2})=0.
\end{equation}
Now $y_2=0$ is the equation for the exceptional divisor
$\P^2\subset\P^3$ in our patch. We are interested in the proper
transform of our surface and thus we do not want to include the full
$\P^2$ in our solution --- just the intersection with our
surface. Thus we throw this solution away and are left with
\begin{equation}
  y_0y_1-y_2^{n-2}=0.
\end{equation}
If $n=2$ or 3 this is smooth. If $n>3$ however we have a singularity
at $y_0=y_1=y_2=0$. This point is at $s_0=s_1=0$ which is precisely
where the two $\P^1$'s produced by the blow-up intersect. One may
check that the other patches contain no singularities.

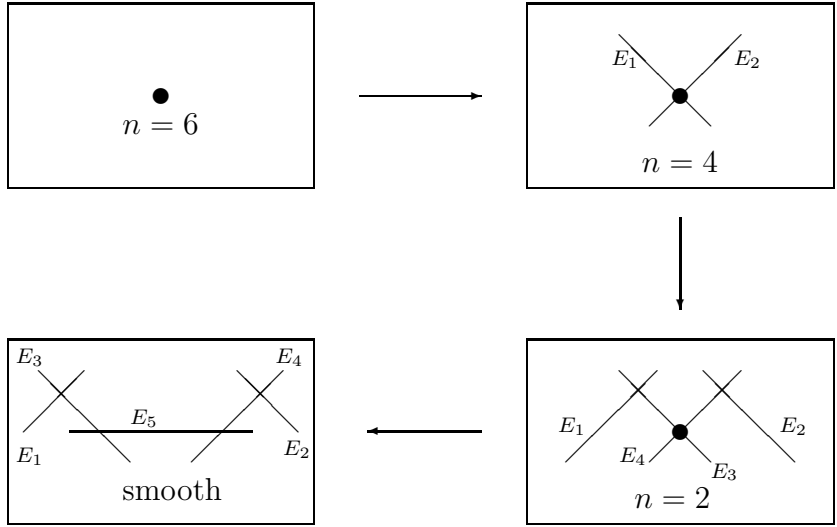
\begin{figure}
$$\setlength{\unitlength}{0.008in}%
\begin{picture}(540,380)(40,440)
\thinlines
\put( 40,700){\framebox(200,120){}}
\put(480,760){\circle*{10}}
\put(380,700){\framebox(200,120){}}
\put(480,540){\circle*{10}}
\put(380,480){\framebox(200,120){}}
\put(460,740){\line( 1, 1){ 60}}
\put(500,740){\line(-1, 1){ 60}}
\put(460,520){\line( 1, 1){ 60}}
\put(500,520){\line(-1, 1){ 60}}
\put(405,520){\line( 1, 1){ 60}}
\put(555,520){\line(-1, 1){ 60}}
\put( 40,480){\framebox(200,120){}}
\put(120,520){\line(-1, 1){ 60}}
\put( 50,540){\line( 1, 1){ 40}}
\put(160,520){\line( 1, 1){ 60}}
\put(230,540){\line(-1, 1){ 40}}
\put( 80,540){\line( 1, 0){120}}
\put(140,760){\circle*{10}}
\put(270,760){\vector( 1, 0){ 80}}
\put(120,545){\makebox(0,0)[lb]{\scriptsize$E_5$}}
\put(480,680){\vector( 0,-1){ 60}}
\put(350,540){\vector(-1, 0){ 75}}
\put(115,735){\makebox(0,0)[lb]{$n=6$}}
\put(455,710){\makebox(0,0)[lb]{$n=4$}}
\put(450,490){\makebox(0,0)[lb]{$n=2$}}
\put(115,495){\makebox(0,0)[lb]{smooth}}
\put(515,780){\makebox(0,0)[lb]{\scriptsize$E_2$}}
\put(435,780){\makebox(0,0)[lb]{\scriptsize$E_1$}}
\put(400,540){\makebox(0,0)[lb]{\scriptsize$E_1$}}
\put(545,540){\makebox(0,0)[lb]{\scriptsize$E_2$}}
\put(500,510){\makebox(0,0)[lb]{\scriptsize$E_3$}}
\put(440,520){\makebox(0,0)[lb]{\scriptsize$E_4$}}
\put( 45,520){\makebox(0,0)[lb]{\scriptsize$E_1$}}
\put(220,525){\makebox(0,0)[lb]{\scriptsize$E_2$}}
\put( 45,585){\makebox(0,0)[lb]{\scriptsize$E_3$}}
\put(215,585){\makebox(0,0)[lb]{\scriptsize$E_4$}}
\end{picture}$$
  \caption{Blowing up $\C^2/\Z_6$.}
  \label{fig:Z6}
\end{figure}

What we have shown then is that starting with the space $\C^2/\Z_n$, $n>2$,
the blow-up replaces the singularity at the origin by two $\P^1$'s
which intersect at a point but, in the case $n>3$, this point of
intersection is locally of the form $\C^2/\Z_{n-2}$. To resolve the
space completely, the procedure is clear. We simply repeat the process
until we are done. Note that when we blow up the point of intersection
of two $\P^1$'s intersecting transversely, the fact that the $\P^1$'s
approach the point of intersection at a different ``angle'' means that
after the blow-up they pass through different points of the
exceptional divisor and thus become disjoint. We show the process of
blowing-up a $\C^2/\Z_6$ singularity in figure \ref{fig:Z6}. In this
process we produce a chain of 5 $\P^1$'s, $E_1,\ldots,E_5$, when
completely resolving the singularity. We show the $\P^1$'s as lines.

Clearly we see that resolving the $\C^2/\Z_n$ singularity produces
the $A_{n-1}$ intersection matrix for the $(-2)$-curves. Thus we have
deduced the form of the $A$-series. We now ponder the $D$- and
$E$-series. 

Consider the general form of the quotient $\C^2/G$. We are interested
in the cases which occur locally in a K3 surface in which $K=0$. This requires
that $G$ leaves the holomorphic 2-form $dz_1\wedge dz_2$
invariant. This implies that $G$ must be a discrete subgroup of
$\SU(2)$. One may also obtain this result by noting that the holonomy
of the orbifold near the quotient singularity can be viewed as being
isomorphic to $G$.
The subgroups of $\SU(2)$ are best understood from the
well-known exact sequence
\begin{equation}
1\to\Z_2\to \SU(2)\to \SO(3)\to 1.
\end{equation}
Thus any subgroup of $\SU(2)$ can be projected into a subgroup of
$\SO(3)$ and considered as a symmetry of a 3-dimensional solid. The
cyclic groups $\Z_n$ may be thought of as, for example, the symmetries
of cones over regular polygons. The other possibilities are the
dihedral groups which are the symmetries of a prism over a regular
polygon, and the symmetries of the tetrahedron, the octahedron (or
cube), and the icosahedron (or dodecahedron). Each of these latter groups are
nonabelian and correspond to a subgroup of $\SU(2)$ with twice as many
elements as the subgroup of $\SO(3)$. They are thus called the
binary dihedral, binary tetrahedral, binary octahedral, and binary
icosahedral groups 
respectively. In each case the quotient $\C^2/G$ can be embedded as a
hypersurface in $\C^3$. This work was completed by Du Val (see
\cite{DuVal:sing} and references therein),
after whom the singularities are sometimes named. The case of an
icosahedron was done by Felix Klein last century \cite{Klein:sing} and so
they are also often referred to as ``Kleinian singularities''.

Once we have a hypersurface in $\C^3$ we may blow-up as before until
we have a smooth manifold.
The intersection matrices of
the resulting $(-2)$-curves can then be shown to be (minus) the Cartan
matrix of $D_n$, $E_6$, $E_7$, or $E_8$ respectively. This process is
laborious and is best approached using slightly more technology than
we have introduced here. We refer the reader to \cite{BPV:} or
\cite{Mir:fibr} for more details. The results are summarized in table
\ref{tab:ADE}(see \cite{Slod:sing} for some of the details). We have
thus shown that any degeneration of a K3 
surface that may be achieved by blowing down $(-2)$-curves leads to an
orbifold singularity.

One might also mention that in the case of the $A_n$ blow-ups,
explicit metrics are known which are asymptotically flat
\cite{EH:metric,GH:multi,Hit:poly}. Unfortunately, since the blow-up
inside a K3 
surface is not actually flat asymptotically, such metrics represent
only an approximation to the situation we desire. As we said earlier
however, lack of an explicit metric will not represent much of a problem.

This miraculous correspondence between the $A$-$D$-$E$ classification
of discrete subgroups of $\SU(2)$ and Dynkin diagrams for simply-laced
simple Lie groups must count as one of the most curious interrelations in
mathematics. We refer to \cite{Dur:15ways} or \cite{Dim:sing} for the
flavour of this subject. We will see later in section \ref{ss:enh}
that string theory will provide another striking connection.

\begin{table}\scriptsize
\def\ve{\varepsilon}
\centerline{\vbox{\tabskip=0pt\offinterlineskip
\def\tablerule{\noalign{\hrule}}
\def\gap{\omit&height2pt&&&&&&&&\cr}
\halign{\strut#&\vrule#\tabskip=1em plus2em&
\hfil#\hfil&\vrule#&
\hfil#\hfil&\vrule#&
\hfil#\hfil&\vrule#&
\hfil#\hfil&\vrule#
\tabskip=0pt\cr
\tablerule\gap
&&Group&&Generators&&Hypersurface&&Resolution&\cr
\tablerule\gap\tablerule\gap
&&Cyclic&&$\left(\begin{array}{cc}\alpha&0\\0&\alpha^{-1}\end{array}\right),
  \;\alpha=e^{\frac{2\pi i}{n}}$&&$x^2+y^2+z^n=0$&&$A_{n-1}$&\cr
\gap\tablerule\gap
&&$\hbox{Binary}\atop\hbox{Dihedral}$&&
  $\left(\begin{array}{cc}\beta&0\\0&\beta^{-1}\end{array}\right),\;
  \left(\begin{array}{cc}0&1\\-1&0\end{array}\right),\;
  \beta=e^{\frac{\pi i}{n}}$&&$x^2+y^2z+z^{n+1}=0$&&$D_{n+2}$&\cr
\gap\tablerule\gap
&&$\hbox{Binary}\atop\hbox{Tetrahedral}$&&
  $D_4,\;\frac1{\sqrt2}\left(\begin{array}{cc}\ve^7&\ve^7\\
  \ve^5&\ve\end{array}\right),\;
  \ve=e^{\frac{2\pi i}8}$
  &&$x^2+y^3+z^4=0$&&$E_6$&\cr
\gap\tablerule\gap
&&$\hbox{Binary}\atop\hbox{Octahedral}$&&
  $E_6,\;\left(\begin{array}{cc}\ve&0\\
  0&\ve^7\end{array}\right),\;
  \ve=e^{\frac{2\pi i}8}$
  &&$x^2+y^3+yz^3=0$&&$E_7$&\cr
\gap\tablerule\gap
&&$\hbox{Binary}\atop\hbox{Icosahedral}$&&
  $-\left(\begin{array}{cc}\eta^3&0\\0&\eta^2\end{array}\right),\;
  \frac1{\eta^2-\eta^3}\left(\begin{array}{cc}\eta+\eta^4&1\\
  1&-\eta-\eta^4\end{array}\right),\;
  \eta=e^{\frac{2\pi i}5}$
 	%$Sl(2,\Z_5)$
  &&$x^2+y^3+z^5=0$&&$E_8$&\cr
\gap\tablerule}}}
  \caption{$A$-$D$-$E$ Quotient Singularities.}
  \label{tab:ADE}
\end{table}

We have considered the resolution process from the point of view of
blowing-up by changing the K\"ahler form. In terms of the moduli space
of Einstein metrics on a K3 surface this is viewed as a rotation of
the 3-plane $\Sigma$ so that there are no longer any roots in the
orthogonal complement. Since $\Sigma$ is spanned by the K\"ahler form
and $\Omega$, which measures the complex structure, we may equally
view this process in terms of changing the complex structure, rather
than the K\"ahler form. In this language, the quotient singularity is
{\em deformed\/} rather than blown-up. The process of resolving is now
seen, not as giving a non-zero size to a shrunken rational curve, but
rather changing the complex structure so that the rational curve no
longer exists. This deformation process is actually very easy to
understand in terms of the singularity as a hypersurface in
$\C^3$. Consider the $A_{n-1}$ singularity in table \ref{tab:ADE}. A
deformation of this to
\begin{equation}
  x^2+y^2+z^n+a_{n-2}z^{n-2}+a_{n-3}z^{n-3}+\ldots+a_0=0
\end{equation}
will produce a smooth hypersurface for generic values of the
$a_i$'s. (Note that the $a_{n-1}z^{n-1}$ term can be transformed away by a
reparametrization.)

It is worth emphasizing that in general, when considering any
algebraic variety, blow-ups and deformations are quite different
things. We will discuss later how the difference between blowing up a
singularity and deforming it away can lead to topology changing
processes in complex dimension three, for example. It is the peculiar way in
which the complex structure moduli and K\"ahler form get mixed up in
the moduli space of Einstein metrics on a K3 surface that makes them
amount to much the same thing in this context. The relationship
between blowing up and deformations will be deepened shortly when we
discuss mirror symmetry.

%%%%%%%%%%%%%%%%%%%%%%%%%%%%%%%%%%%%%%%%%%%%%%%%%%%%%%%%%%%%%%%%%%%

\section{The World-Sheet Perspective}   \label{s:ws}

In this section we are going to embark on an analysis of string theory
on K3 surfaces from what might be considered a rather old-fashioned
point of view. That is, we are going to look at physics on the
world-sheet. One point of view that was common more than a couple of
years ago was 
that string theory could solve difficult problems by ``pulling back''
physics in the target space, which has a large number of dimensions
and is hence difficult, to the world-sheet, which is two-dimensional
and hence simple. Thus an understanding of two-dimensional physics on
the world-sheet would suffice for understanding the universe. 

More recently it has been realized that the world-sheet approach is
probably inadequate as it misses aspects of the string theory which
are nonperturbative in the string coupling expansion. Thus, attention
has switched somewhat away from the world-sheet and back to the target
space. One cannot forget the world-sheet however and, as we will see
later in section \ref{s:4d}, in some examples it would appear that the
target space point 
of view appears on an equal footing with the world-sheet point of
view. We must therefore first extract from the world-sheet as much as
we can.

\subsection{The Nonlinear Sigma Model}  \label{ss:nlsm}

``Old'' string theory is defined as a two-dimensional theory given by
maps from a Riemann surface, $\Sigma$, into a target manifold $X$:
\begin{equation}
  x:\Sigma\to X.
\end{equation}
In the conformal gauge, the action is given by (see, for example,
\cite{GSW:book} for the 
basic ideas and \cite{me:N2lect} for conventions on normalizations)
\begin{equation}
  S = \frac i{8\pi\alpha^\prime}\int_\Sigma\left(g_{ij}-B_{ij}
        \right)\partial
        x^i\bar\partial x^j\,d^2z -
	2\pi\int\Phi R^{(2)}\,d^2z+\ldots,      \label{eq:sm}
\end{equation}
where we have ignored any terms which contain fermions.
The terms are identified as follows. $g_{ij}$ is a Riemannian metric
on $X$ and $B_{ij}$ are the components of a real 2-form, $B$, on $X$. $\Phi$
is the ``dilaton'' and is a real number (which might depend on $x$)
and $R^{(2)}$ is the scalar curvature of $\Sigma$. This
two-dimensional theory is known as the ``\nlsm''.

In order to obtain a valid string theory, we require that the resulting
two-dimensional theory is conformally invariant with a specific value of the
``central charge''. (See \cite{Gins:lect} for basic notions in
conformal field theory.) Conformal invariance puts constraints on the
various parameters above \cite{Fri:sm,Cal:sm}. In general the result
is in terms of a perturbation theory in the quantity
$\alpha^\prime/R^2$, where $R$ is some characteristic ``radius''
(coming from the metric) of $X$, assuming $X$ to be compact.

The simplest way of demanding conformal invariance to leading order is
to set the dilaton, $\Phi$, to be a constant, and let $B$ be closed and
$g_{ij}$ be Ricci-flat. There are other solutions, such as the one
proposed in \cite{CHS:5b} and these do play a r\^ole in string duality
as solitons (see, for example, \cite{HS:sol}). It is probably safe to
say however that the solution we will analyze, with the
constant dilaton, is by far the best understood. 

In many ways one may regard this conformal invariance calculation to
be the string ``derivation'' of general relativity. To leading order
in $\alpha^\prime/R^2$ we obtain Einstein's field equation and then
perturbation theory ``corrects'' this to higher
orders. Nonperturbative effects, i.e., ``world-sheet instantons'',
should also modify notions in general relativity. Anyway, we see that
in this simple case, a vacuum solution for string theory is the same
as that for general relativity --- namely a Ricci-flat manifold.

There is a simple and beautiful relationship between supersymmetry in
this \nlsm\ and the K\"ahler structure of the target space manifold
$X$. We have neglected to include any fermions in the
action (\ref{eq:sm}) but they are of the form $\psi^i$ and transform,
as far as the target space is concerned, as sections of the cotangent
bundle. A supersymmetry will be roughly of the form
\begin{equation}
  \delta_\epsilon X^i=\bar\epsilon\, l^i_j\psi^j.
\end{equation}
The object of interest here is $l^i_j$. With one supersymmetry ($N=1$) on the
world-sheet one may simply reparameterize to make it equal to
$\delta^i_j$. When the $N>1$ however we have more structure. It was
shown \cite{AGZ:Ka} that for $N=2$ the second $l^i_j$ acts as an
almost complex structure and gives $X$ the structure of a K\"ahler
manifold. In the case $N=4$, we have 3 almost complex structures, as
in section \ref{ss:Em}, and this leads to a hyperk\"ahler
manifold. The converse also applies.

Note that this relationship between world-sheet supersymmetry and the
complex differential geometry of the target space required no
reference to conformal invariance. In the case that we have conformal
invariance one may also divide the analysis into holomorphic and
anti-holomorphic parts and study them separately. In this case we have
separate supersymmetries in the left-moving (holomorphic) and
right-moving (anti-holomorphic) sectors. 

The case of interest to us, of course, is when $X$ is a smooth K3
manifold. From what we have said, this will lead to an $N=(4,4)$
superconformal field theory, at least to leading order in
$\alpha^\prime/R^2$. Actually for $N=4$ \nlsm s, the perturbation
theory becomes much simpler and it can be shown \cite{AGG:N=4} that
there are no further corrections to the Ricci-flat metric after the
leading term. Additionally, one may present arguments that this is even
true nonperturbatively \cite{BS:N=4}. Thus our Ricci-flat metric on
the K3 surface is an exact solution. One must contrast this to the
$N=2$ case where there are both perturbative corrections and
nonperturbative effects in general.

\subsection{The Teichm\"uller space} \label{ss:Tsig}

The goal of this section is to find the moduli space of conformally
invariant \nlsm s with K3 target space. This may be considered as an
intermediate stage to that of the last section, where we considered
classical geometry, and that of the following sections where we
consider supposedly fully-fledged string theory. This will prove to be
a very important step however.

Firstly note that there are three sets of parameters in (\ref{eq:sm})
which may be varied to span the moduli space required. In each case we
need to know which deformations will be effective in the sense that
they really change the underlying conformal field theory. Here we will
have to make some assumptions since a complete analysis of these
conformal field theories has yet to be completed.

First consider the metric $g_{ij}$. We have seen that this must be
Ricci-flat to obtain conformal invariance. We will assume that any generic
deformation of this Ricci-flat metric to another inequivalent
Ricci-flat metric will lead to an inequivalent conformal field
theory. Since the dimension of the moduli space of Einstein metrics on
a K3 surface given in (\ref{eq:ME}) is 58, we see that the metric
accounts for 58 parameters.

Next we have the 2-form, $B$. This appears in the action in the form
\begin{equation}
  \int_{x(\Sigma)}B.
\end{equation}
Thus, since the image of the world-sheet in $X$ under the map $x$ is a
closed 2-cycle, any exact part of $B$ is irrelevant. All we
see of $B$ is its cohomology class. As $b_2$ of a K3 surface is 22,
this suggests we have 22 parameters from the $B$-field.

Lastly we have the dilaton, $\Phi$. This plays a very peculiar r\^ole
in our conformal field theory. Since $\Phi$ is a constant over $X$, by
assumption, we may pull it outside the integral leaving a contribution
of $2\Phi(g-1)$ to the action, where $g$ is the genus of $\Sigma$. In
this section we really only care about the conformal field theory for
a fixed $\Sigma$ and so this quantity remains constant. To be more
complete we should sum over the genera of $\Sigma$. Taking the limit
of $\alpha^\prime\to\infty$, the world-sheet image in the target space
will degenerate to a Feynman diagram and then $g$ will count the
number of loops. Thus we have an effective target space coupling of
\begin{equation}
\lambda=e^\Phi. 
\end{equation}
Anyway, since we want to ignore this summation of $\Sigma$ for the
time being we will ignore the dilaton. See, for example, \cite{Zw:dil}
for a further discussion of world-sheet properties of the dilaton from
a string field theory point of view.

All said then we have a moduli space of $58+22=80$ real dimensions. To
proceed further we need to know some aspects about the holonomy of the
moduli space. The local form of the moduli space was first presented
in \cite{Sei:K3} but we follow more closely here the method of
\cite{Cec:nonp}.

\def\so{\operatorname{\Goth{so}}}
\def\su{\operatorname{\Goth{su}}}
\def\gu{\operatorname{\Goth{u}}}
\def\sp{\operatorname{\Goth{sp}}}

Let us return to the moduli space of Einstein metrics on a K3 surface.
Part of the holonomy algebra of the symmetric space factor in (\ref{eq:ME}) is
$\so(3)\cong \su(2)$. This rotation in $\R^3$ comes from the choice of complex
structures given a quaternionic structure as discussed in section
\ref{ss:hol}. Thus, this part of the holonomy can be understood as
arising from the symmetry produced by the $S^2$ of complex structures.

This $\su(2)$ symmetry must therefore be present in the \nlsm. In the
case where we have a conformal field theory however, we may divide the
analysis into separate left- and right-moving parts. Thus each sector
must have an independent $\su(2)$ symmetry. Indeed, it is known from
conformal field theory that an $N=4$ superconformal field theory
contains an affine $\su(2)$ algebra and so an $N=(4,4)$ superconformal
field theory has an $\su(2)\oplus\su(2)\cong\so(4)$ symmetry.

This symmetry acts on the tangent directions to a point in the moduli
space (i.e., the ``marginal operators'') and so will be a subgroup of
the holonomy. Thus the $\so(3)$ appearing in the holonomy algebra of
the moduli space of Einstein metrics is promoted to $\so(4)$ for our
moduli space of conformal field theories.

Now we are almost done. We need to find a space whose holonomy
contains $\SO(4)$ as a factor and has dimension 80. One could suggest
spaces such as $A\times B$, where $A$ is a Riemannian manifold of
dimension 4 and $B$ is a Riemannian manifold of dimension 76. Such a
factorization is incompatible with what we know about the conformal
field theory, however. Analyzing the marginal operators in terms of
superfields shows that each one is acted upon non-trivially by at
least part of the $\SO(4)$. Given the work of Berger and Simons
therefore leaves us with only one possibility.\footnote{Actually we
should rule out the compact symmetric space possibility. This is done
by our completeness assumption, as we know we may make a K3 surface
arbitrarily large.}
\begin{theorem}
  Given the assumptions about the effectiveness of deformations on the
underlying conformal field theory, any smooth neighbourhood of the
moduli space of conformally invariant \nlsm s with a K3 target space
is isomorphic to an open subset of
\begin{equation}
  \cT_\sigma = \frac{\GO(4,20)}{\GO(4)\times \GO(20)}.  \label{eq:Tsig}
\end{equation}
  \label{th:ts-K3}
\end{theorem}

We now want to know about the global form of the moduli space. Here
we are forced to make assumptions about how reasonable our conformal
field theories can be. We will assume that from theorem \ref{th:s-K3}
it follows that the moduli space of conformal field theories is given
by $\cM_\sigma\cong G_\sigma\backslash\cT_\sigma$, where $G_\sigma$ is
some discrete 
group. That is, $\cT_\sigma$ is the Teichm\"uller space. All we are
doing here is assuming that our moduli space is ``complete'' in the
sense that there are no pathological limit points in it possibly
bounding some bizarre new region. While this assumption seems extremely
reasonable I am not aware of any rigorous proof that this is the case.

\subsection{The geometric symmetries}  \label{ss:clsym}

All that remains then is the determination of the modular group,
$G_\sigma$. To begin with we should relate (\ref{eq:Tsig}) to the
Teichm\"uller spaces we are familiar with from section
\ref{s:clas}. This will allow us the incorporate the modular groups we
have already encountered. This is a review of the work that first
appeared in \cite{AM:K3p,AM:K3m}.

The space $\cT_\sigma$ is the Grassmannian of space-like 4-planes in
$\R^{4,20}$. We saw that the moduli space of Einstein metrics on a K3
surface is given by the Grassmannian of space-like 3-planes in
$\R^{3,19}$ and this must be a subspace of $\cT_\sigma$ since the
Einstein metric appears in the action (\ref{eq:sm}). This gives us a
clear way to proceed.

Let us introduce the even self-dual lattice
$\Gamma_{4,20}\subset\R^{4,20}$. It would be nice if we could show that
this played the same r\^ole as $\Gamma_{3,19}$ played in the moduli
space of Einstein metrics. That is, we would like to show $G_\sigma\cong
\GO(\Gamma_{4,20})$. We will see this is indeed true.

First we want a natural way of choosing
$\Gamma_{3,19}\subset\Gamma_{4,20}$. To do this fix a primitive
element $w\in\Gamma_{4,20}$ such that $w.w=0$. Now consider the space,
$w^\perp\subset\R^{4,20}$, of all vectors $x$ such that
$x.w=0$. Clearly $w$ is itself contained in this space. Now project
onto the codimension one subspace $w^\perp/w$ by modding out by the
$w$ direction. We now embed $w^\perp/w$ back into $\R^{4,20}$ such that
\begin{equation}
\frac{w^\perp}{w}\cap\Gamma_{4,20}\cong\Gamma_{3,19}.
\end{equation}
It is important to be aware of the fact that all statements about
$\Gamma_{3,19}$ are dictated by a choice of $w$. The embedding
$w^\perp/w\subset\R^{4,20}$ can be regarded as a choice of a second lattice
vector, $w^*$, such that $w^*$ is orthogonal to $w^\perp/w$, $w^*.w^*=0$
and $w^*.w=1$. As we shall see, the choice of $w^*$ is not as
significant as the choice of $w$.

Now perform the same operation on the space-like 4-plane. We will
denote this plane $\Pi\subset\R^{4,20}$. First define
$\Sigma^\prime=\Pi\cap w^\perp$ and then project this 3-plane into the
space $w^\perp/w$ and embed back into $\R^{4,20}$ to give
$\Sigma$. This $\Sigma$ may now be 
identified with that of section \ref{ss:Em} to give the Einstein
metric on the K3 surface of some fixed volume.

Fixing $\Sigma$ we may look at how we may vary $\Pi$ to fill out the
other deformations. Let $\Pi$ be given by the span of $\Sigma^\prime$
and $B^\prime$, where $B^\prime$ is a vector, orthogonal to $\Sigma'$,
normalized by 
$B^\prime.w=1$. We may project $B^\prime$ into $w^\perp/w$ to give 
$B\in\R^{3,19}$. Note that $\R^{3,19}$ is the space $H^2(X,\R)$ and so
$B$ is a 2-form as desired.

Lastly we require the volume of the K3 surface. Let us decompose
$B^\prime$ as
\begin{equation}
  B^\prime = \alpha w + w^* + B,
		\label{eq:BB}
\end{equation}
We claim that $\alpha$ is related to the volume of the K3
surface. To see this we need to analyze the explicit form of the
moduli space further. 

We have effectively decomposed the Teichm\"uller space of conformal
field theories as
\begin{equation}
  \frac{\GO(4,20)}{\GO(4)\times\GO(20)} \cong
  \frac{\GO(3,19)}{\GO(3)\times\GO(19)}\times\R^{22}\times\R_+,
		\label{eq:Bdec}
\end{equation}
where the three factors on the right are identified as the
Teichm\"uller spaces for the metric on the K3 surface (given by
$\Sigma$), the $B$-field, and the volume respectively. Each of the
spaces in the equation (\ref{eq:Bdec}) has a natural metric, given by
the invariant metric for the group action in the case of the symmetric
spaces. This can be shown to coincide with the natural metric from
conformal field theory --- the Zamolodchikov metric --- given the
holonomy arguments above. The isomorphism (\ref{eq:Bdec}) will respect
this metric if ``warping'' factors are introduced as explained in
\cite{Borel:decomp}. It is these warping factors which determine the
identification of the volume of the K3 surface as
\begin{equation}
  \Vol(S) = \int_S J\wedge J = 2\alpha+B^2.  \label{eq:volK3a}
\end{equation}
This will be explained further in \cite{AM:K3m}.%
\footnote{This volume erroneously appeared as $\alpha$ in earlier
versions of these notes. I thank E.~Diaconescu for pointing out
this error. One may also determine this expression by considering how
T-dualities in the form of $\Sl(2,\Z)$ act on elliptic K3 surfaces
with a section. This is beyond the scope of these notes but is hinted
at in later sections.}

The part of $G_\sigma$ we can understand directly is the part which
fixes $w$ but acts on $w^\perp/w$. This will affect the metric
on the K3 surface of fixed volume and the $B$-field. 
We know that the modular group coming from global diffeomorphisms of
the K3 surface is $\GO^+(\Gamma_{3,19})$,
which should be viewed as $\GO^+(H^2(X,\Z))$. The
discrete symmetries for the $B$-field meanwhile can be written as
$B\cong B+e$, where $e\in H^2(X,\Z)$. To see this note that shifting
$B$ by a integer element will shift the action (\ref{eq:sm}) by
a multiple of $2\pi i$ and hence will not effect the path integral. 
Thanks to our normalization of $B^\prime$, a shift of $B$ by an
element of $\Gamma_{3,19}$ amounts to a rotation of $\Pi$ which is
equivalent to an element of $\GO(\Gamma_{4,20})$. 
Note that this can also be interpreted as a redefinition of
$w^*$. That is, the freedom of choice of $w^*$ is irrelevant once we
take into account the $B$-field shifts.

We may also
consider taking the complex conjugate of the \nlsm\ action. This has the
effect of reversing the sign of $B$ while providing the extra element
required to elevate $\GO^+(\Gamma_{3,19})$ to $\GO(\Gamma_{3,19})$.

The result is that the subgroup of $G_\sigma$ that we see directly from the
\nlsm\ action is a subgroup of $\GO(\Gamma_{4,20})$ consisting of
rotations {\em and\/} translations of
$\Gamma_{3,19}\subset\Gamma_{4,20}$. This may be viewed as the {\em
space group\/} of $\Gamma_{3,19}$, or equivalently, the semi-direct
product
\begin{equation}
  G_\sigma\supset \GO(\Gamma_{3,19})\ltimes\Gamma_{3,19}.
	\label{eq:Gclas}
\end{equation}
This is as much as we can determine from $w^\perp/w$.

\subsection{Mirror symmetry}   \label{ss:mir}

To proceed any further in our analysis of $G_\sigma$ we need to know
about elements which do not correspond to a manifest symmetry of the
\nlsm\ action (\ref{eq:sm}). This knowledge will be provided by {\em
mirror symmetry}. Mirror symmetry is a much-studied phenomenon in \CY\
threefolds (see, for example, \cite{Mrr:1} for a review). The basic idea
in the subject of threefolds is that the notion of deformation of
complex structure is exchanged with deformation of complexified
K\"ahler form. The character of mirror symmetry in K3 surfaces is
somewhat different since, as we have seen, the notion of what
constitutes a deformation of complex structure and what constitutes a
deformation of the K\"ahler form can be somewhat ambiguous. Also, we
have yet to mention the possibility of complexifying the K\"ahler form,
as that too is a somewhat ambiguous process.

Indeed mirror symmetry itself is somewhat meaningless when viewed in
terms of the intrinsic geometry of a K3 surface and does not begin to
make much sense until viewed in terms of {\em algebraic\/} K3 surfaces. The
results for algebraic K3 surfaces were first explored by Martinec
\cite{Mart:CCC}, whose analysis actually predates the discovery (and
naming) of mirror symmetry in the \CY\ threefold context. See also
\cite{Dol:K3m,Bor:K3} for a discussion of some of the issues we cover below.

Recall that the Picard lattice of section \ref{ss:alg} was defined as
the lattice of integral 2-cycles in $H^2(X,\Z)$ which were of type
$(1,1)$. The transcendental lattice, $\Lambda$, was defined as the
orthogonal complement of the Picard lattice in $H^2(X,\Z)$. The
signature of $\Lambda$ is $(2,20-\rho)$, where $\rho$ is the Picard
number of $X$.

It is clear that $\Gamma_{4,20}\cong\Gamma_{3,19}\oplus U$, where $U$
is the hyperbolic plane of (\ref{eq:U}). We will extend the Picard
lattice to the ``quantum Picard lattice'', $\Upsilon$, by defining
\begin{equation}
  \Upsilon = \Pic(X)\oplus U,   \label{eq:Ups}
\end{equation}
as the orthogonal complement of $\Lambda$ within $\Gamma_{4,20}$.

Given a K3 surface with an Einstein metric specified by $\Sigma$ and a
given algebraic structure, we know that the complex structure 2-plane,
$\Omega$, is given by $\Sigma\cap(\Lambda\otimes_\Z\R)$.\footnote{The
notation $\Lambda
\otimes_\Z\R$ denotes the real vector space generated by the
generators of $\Lambda$.} The K\"ahler form direction, $J$ is then
given by the orthogonal complement of
$\Omega\subset\Sigma$. Accordingly, $J$ lies in $\Pic(X)\otimes_\Z\R$.

We want to extend this to the \nlsm\ picture. Keep $\Omega$ defined as
above, or equivalently, 
\begin{equation}
  \Omega=\Pi\cap(\Lambda\otimes_\Z\R),
\end{equation}
and introduce a
new space-like 2-plane,
$\mho$, as the orthogonal complement of $\Omega\subset\Pi$. Clearly
we have 
\begin{equation}
  \mho=\Pi\cap(\Upsilon\otimes_\Z\R).
\end{equation}

Our notion of a mirror map will be to exchange
\begin{equation}
\mu:(\Lambda,\Omega)\leftrightarrow(\Upsilon,\mho).	\label{eq:mu}
\end{equation}
Note that as $\mho$ encodes the K\"ahler form and the value of the
$B$-field we have the notion of exchange of complex structure data
with that of K\"ahler form $+$ $B$-field as befits a mirror map.

The moduli space of \nlsm s on an algebraic K3 surface will be the
subspace of (\ref{eq:Tsig}) which respects the division of $\Pi$ into
$\Omega$ and $\mho$. This is given by
\begin{equation}
  \cT_{\sigma,{\rm alg}}\cong\frac{\GO(2,20-\rho)}{\GO(2)\times \GO(20-\rho)}
  \times\frac{\GO(2,\rho)}{\GO(2)\times \GO(\rho)},
	\label{eq:Tasig}
\end{equation}
where the first factor is the moduli space of complex structures and
the second factor is moduli space of the K\"ahler form $+$ $B$-field. 

Given an algebraic K3 surface $X$ with quantum Picard lattice
$\Upsilon(X)$ we have a mirror K3 surface, $Y$, with quantum Picard
lattice $\Upsilon(Y)$ such that $\Upsilon(Y)$ is the orthogonal
complement of $\Upsilon(X)\subset\Gamma_{4,20}$. Translating this back
into classical ideas, $\Pic(Y)\oplus U$ will be the orthogonal
complement of $\Pic(X)\subset\Gamma_{3,19}$. If $X$ is such that the
orthogonal complement of $\Pic(X)\subset\Gamma_{3,19}$ has no $U$
sublattice then $X$ has no classical mirror.

Such mirror pairs of K3 surfaces were first noticed some time ago by
Arnold (see for example, \cite{hs:}). The \nlsm\ moduli space gives a nice
setting for this pairing to appear.

Now we also have the mirror construction of Greene and Plesser
\cite{GP:orb} which takes an algebraic variety $X$ as a hypersurface
in a weighted projective space and produces a ``mirror'' $Y$ as an
quotient of $X$ by some discrete group such that $X$ and $Y$ as target
spaces produce completely identical conformal field theories. 

An interesting case is when $X$ is the hypersurface
\begin{equation}
  f = x_0^2+x_1^3+x_2^7+x_3^{42}=0
\end{equation}
in $\P_{\{21,14,6,1\}}^3$. $X$ is a K3 surface which happens to be an
orbifold due to the quotient singularities in the weighted projective
space. Blowing this up provides a K3 surface with Picard lattice
isomorphic to the even self-dual lattice $\Gamma_{1,9}$. Thus
in this case $\Lambda\cong\Upsilon\cong\Gamma_{2,10}$. What is
interesting is that in this case, the Greene-Plesser construction
says that the group by which $X$ should be divided to obtain the
mirror is trivial. This mirror map does not act trivially on the
marginal operators however but one may show actually exchanges the
factors in (\ref{eq:Tasig}). This is little delicate and we refer the
reader to \cite{AM:K3m} for details. Anyway, what this shows is that
the Greene-Plesser mirror map, which is an honest symmetry in the
sense that it produces an identical conformal field theory, really is
given by the exchange, $\mu$, in (\ref{eq:mu}).

This implicitly provides us with another element of $G_\sigma$, namely
an element of $\GO(\Gamma_{4,20})$ which exchanges two orthogonal
$\Gamma_{2,10}$ sublattices. It is now possible to show that this new
element, together with the subgroup given in (\ref{eq:Gclas}), is
enough to generate $\GO(\Gamma_{4,20})$. Thus $G_\sigma$ at least
contains $\GO(\Gamma_{4,20})$.

To prove the required result that $G_\sigma\cong\GO(\Gamma_{4,20})$ we
use the result of \cite{Allan:} which states that any further
generator will destroy the Hausdorff nature of the moduli space
contrary to our assumption.

We thus have
\begin{theorem}
  Given theorem \ref{th:ts-K3} and the assumption that the moduli space
is Hausdorff, we have that the moduli space of conformally invariant \nlsm
s on a K3 surface is
\begin{equation}
  \cM_\sigma = \GO(\Gamma_{4,20})\backslash\GO(4,20)/
		(\GO(4)\times\GO(20))
	\label{eq:s-K3}
\end{equation}
  \label{th:s-K3}
\end{theorem}

Lastly we should discuss the meaning of $\Gamma_{4,20}$. The lattice
$\Gamma_{3,19}$ was associated to the lattice $H^2(X,\Z)$. 
Extend the notion of our inner product (\ref{eq:inner}) to that of any
$p$-cycle (or equivalently $p$-form) by saying that the product of two
cycles is zero unless their dimensions add up to 4. Now one can see
that the lattice of total cohomology, $H^*(X,\Z)$, is isomorphic to
$\Gamma_{4,20}$. This gives a tempting interpretation. We will see
later that this really is the right one.

This tells us how to view $w$. A point in the moduli space
$\cM_\sigma$ determines a conformal field theory uniquely but we are
required to make a choice of $w$ before we can determine the geometry
of the K3 surface in terms of metrics and $B$-fields. This amounts to
deciding which direction in $\Gamma_{4,20}$ will be $H^0(X,\Z)$. This
choice is arbitrary and different choices will lead to potentially
very different looking K3 surfaces which give the same conformal field
theory. This may be viewed as a kind of T-duality.

\subsection{Conformal field theory on a torus}  \label{ss:tori}

Although our main interest in these lectures are K3 surfaces, it turns
out that the idea of compactification of strings on a torus will be intimately
related. We are thus required to be familiar with the situation when
the \sm\ has a torus as a target space.

The problem of the moduli space in this case was solved in
\cite{N:torus,HNW:torus}. This subject is also covered by H.~Ooguri's
lectures in this volume and we refer the reader there for further
information (see also \cite{Giv:rep}).
Here we will quickly review the result in a language
appropriate for the context in which we wish to use it.

To allow for the heterotic string case, we are going to allow for the
possibility that the left-moving sector of the \sm\ may live on a
different target space to that of the right-moving sector. We thus
consider the left sector to have a torus of $n_L$ real dimensions as
its target space and let the right-moving sector live on a
$n_R$-torus. Of course, for the simple picture of a string propagating on a
torus, we require $n_L=n_R$.

The moduli space of conformal field theories is then given by
\begin{equation}
  \GO(\Gamma_{n_L,n_R})\backslash\GO(n_L,n_R)/(\GO(n_L)\times\GO(n_R)).
	\label{eq:Nar}
\end{equation}
The even self-dual condition, required for world-sheet modular
invariance \cite{N:torus}, on the lattice $\Gamma_{n_L,n_R}$
enforces $n_L-n_R\in8\Z$. Note that whether we impose any
supersymmetry requirements or not makes little difference. The only
Ricci-flat metric on a torus is the flat one, which guarantees
conformal invariance to all orders. The trivial holonomy then means
that a torus may be regarded as K\"ahler, hyperk\"ahler or whatever so
long as the dimensions are right.

Let us assume without loss of generality that $n_L\leq n_R$.\footnote{
This means that the heterotic string will be chosen to be a
superstring in the left-moving sector and a bosonic string in the
right-moving sector. Unfortunately this differs from the usual
convention. This is imposed on us, however, when we consider
duality to a type IIA string and the natural orientation of a K3 surface.}
Our aim here is to interpret the moduli space (\ref{eq:Nar}) in terms
of the Grassmannian of space-like $n_L$-planes in $\R^{n_L,n_R}$. Let us
use $\Pi$ to denote a space-like $n_L$-plane. Any vector in
$\R^{n_L,n_R}$ may be written as a sum of two vectors, $p_L+p_R$, where
\begin{equation}
  \eqalign{p_L&\in\Pi\cr p_R&\in\Pi^\perp,\cr}
\end{equation}
and $\Pi^\perp$ is the orthogonal complement of $\Pi$. Thus
$p_L.p_L\geq0$ and $p_R.p_R\leq0$.

The winding and momenta modes of the string on the torus are now given by
points in the lattice $\Gamma_{n_L,n_R}$ according to Narain's construction
\cite{N:torus}. The left and right conformal weights of such states
are then given by $\ff12p_L.p_L$ and $-\ff12p_R.p_R$, thus allowing the mass to
be determined in the usual way.

It is easy to see how the moduli space (\ref{eq:Nar}) arises from this
point of view. The position of $\Pi$ determines $p_L$ and $p_R$ for
each mode. The modular group $\GO(\Gamma_{n_L,n_R})$ merely rearranges
the winding and momenta modes.

Let us try to make contact with the \sm\ description of the string.
We choose a null (i.e., light-like) $n_L$-plane, $W$, in
$\Gamma_{n_L,n_R}\otimes_\Z\R\cong 
\R^{n_L,n_R}$ which is ``aligned'' along the lattice
$\Gamma_{n_L,n_R}$. By this we mean that $W$ is spanned by a subset of
the generators of $\Gamma_{n_L,n_R}$. We then define $W^*$ as the null
$n_L$-plane dual to $W$, and the time-like space $V\cong\R^{n_R-n_L}$
so that 
\begin{equation}
\Gamma_{n_L,n_R}\otimes_\Z\R\cong W\oplus W^*\oplus V.
\end{equation}
This decomposition is also aligned with the lattice so that
$\Gamma_{n_L,n_R}\cap W$, $\Gamma_{n_L,n_R}\cap W^*$, and
$\Gamma_{n_L,n_R}\cap V$ generate $\Gamma_{n_L,n_R}$.

We may write $\Pi$ as given by
\begin{equation}
  \Pi = \{x+\psi(x)+A(x);\;\forall x\in W\},
	\label{eq:GBA}
\end{equation}
where
\begin{equation}
\eqalign{
  \psi:W&\to W^*\cr
  A:W&\to V\cr}
\end{equation}
Viewing $\psi$ as
\begin{equation}
  \psi:W\times W\to\R,   \label{eq:psiWW}
\end{equation}
we may divide $\psi$ into symmetric, $G$, and anti-symmetric parts,
$B$, such that $\psi=B+G$. 

Physically $W$ represents, under the
metric $G$, the target 
space in which the string lives. To be precise, the target space
$n_L$-torus is   
\begin{equation}
  \frac{W}{\Gamma_{n_L,n_R}\cap W}.
\end{equation}
$W^*$
is then the dual space in which the momenta live. Clearly $B$ is the
$B$-field. Lastly $V$ represents the gauge group generated by the
extra right-moving directions. To be precise, the group is
given by
\begin{equation}
  U(1)^{n_R-n_L} \cong\frac{V}{\Gamma_{n_L,n_R}\cap V}.
\end{equation}
We will discuss in section \ref{ss:enh} how this group can be enhanced
to a nonabelian group for particular $\Pi$'s. The quantity $A$ then
represents $\gu(1)$-valued 1-forms. We will follow convention and
refer to these degrees of freedom within $A$ as ``Wilson Lines''.

One may write $G$, $B$ and $A$ in terms of matrices to make contact
with the expressions in \cite{HNW:torus,Giv:rep}. This is a somewhat
cumbersome process but the reader should check it if they are unsure
of this construction. Here we illustrate the procedure in the
simplest case, namely that of a circle as target space. Here we have
$n_L=n_R=1$ and $\Gamma_{1,1}\cong U$ as in (\ref{eq:U}). An element
of $\GO(1,1)$ preserving the form $U$ may be written in one of the forms
\begin{equation}
  \left(\begin{array}{cc}t&0\\0&t^{-1}\end{array}\right),\;
  \left(\begin{array}{cc}-t&0\\0&-t^{-1}\end{array}\right),\;
  \left(\begin{array}{cc}0&t\\t^{-1}&0\end{array}\right),\;
  \left(\begin{array}{cc}0&-t\\-t^{-1}&0\end{array}\right),
	\label{eq:O11}
\end{equation}
where $t$ is real and positive.

We divide this space by $\GO(1)\times\GO(1)$ from the right and by
$\GO(\Gamma_{1,1})$ from the left. Both these groups are isomorphic to
$\Z_2\times\Z_2$ and are given by (\ref{eq:O11}) with $t=1$. The
result is that the moduli space is given by the real positive line mod
$\Z_2$ represented by
\begin{equation}
  \left(\begin{array}{cc}t&0\\0&t^{-1}\end{array}\right)\cong
  \left(\begin{array}{cc}t^{-1}&0\\0&t\end{array}\right),
	\label{eq:t1/t}
\end{equation}
where $t$ is real and positive.

\iffigs
\begin{figure}
  \centerline{\epsfxsize=7cm\epsfbox{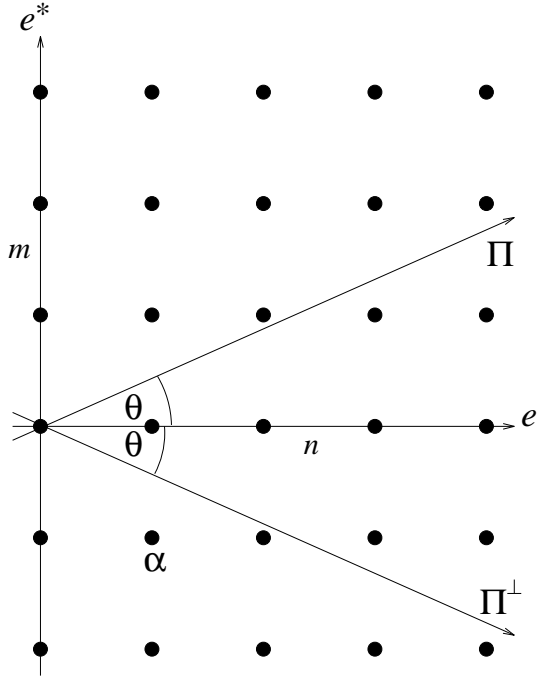}}
  \caption{The moduli space picture for a string on a circle.}
  \label{fig:O11}
\end{figure}
\fi

Let us chose a basis, $\{e,e^*\}$, for $\Gamma_{1,1}$ so that
$e.e=e^*.e^*=0$ and $e.e^*=1$. Now let the space-like direction, $\Pi$,
be given, as in figure \ref{fig:O11}, by
\begin{equation}
\Pi = \{xe + xe^*\tan\theta;\;x\in\R\},
\end{equation}
where $t=\tan\theta$. (Note that $\Pi$ and $\Pi^\perp$ may not look
particularly orthogonal in figure \ref{fig:O11} but don't forget we
are using the hyperbolic metric given by $U$!)
Denoting a state in $\Gamma_{1,1}$ by $ne+me^*$, for $m,n\in\Z$, one
can show
\begin{equation}
\eqalign{p_L.p_L &= \ff12n^2\tan\theta + nm +
\frac{m^2}{2\tan\theta}\cr
p_R.p_R &= -\ff12n^2\tan\theta + nm
-\frac{m^2}{2\tan\theta}.\cr}
\end{equation}
Since $G$ is given by $\tan\theta$, the radius of the circle is
proportional to $\tan\theta$. The identification in (\ref{eq:t1/t}) then
gives the familiar ``$R\leftrightarrow1/R$'' T-duality relation.

%%%%%%%%%%%%%%%%%%%%%%%%%%%%%%%%%%%%%%%%%%%%%%%%%%%%%%%%%%%%%%%%%%%

\section{Type II String Theory}		\label{s:II}

Now we have the required knowledge from classical geometry and ``old''
string theory on the world-sheet to tackle full string theory on a K3
surface. We begin with the most supersymmetry to make life
simple. This is the type IIA or IIB superstring which has 2
supersymmetries when viewed as a ten-dimensional field theory. The
heterotic string on a K3 surface constitutes a much more difficult
problem and we won't be ready to tackle it until section \ref{s:het}.

\subsection{Target space supergravity and compactification}
		\label{ss:supergrav} 

We begin by switching our attention from the quantum field theory that
lives in the world-sheet to the effective quantum field theory that
lives in the target space in the limit that $\alpha^\prime/R^2\to0$.
Of particular interest will be theories with $N$ supersymmetries in
$d$ dimensions (one of which is time).

A spinor is an irreducible representation of the algebra $\so(1,d-1)$ and
has dimension $2^{\left[\frac{d+1}2\right]-1}$, where the bracket
means round down to the nearest integer. This is not the only
information about the representation we require however. A spinor may
be real ($\R$), complex ($\C$), or quaternionic\footnote{Sometimes the
terminology ``pseudo-real'' is used.} ($\H$) depending on $d$. The
rule is
\begin{center}
\begin{tabular}{c@{~~if~~}l}
  $\R$&$d=1,2,3\pmod8$\\
  $\C$&$d=0\pmod4$\\
  $\H$&$d=5,6,7\pmod8$.
\end{tabular}
\end{center}
A complex representation has twice as many degrees of freedom as a real
representation. A quaternionic representation has the same number of
degrees of freedom as a complex representation (as, even though one may
define the representation naturally over the quaternions, one must
subject it to constraints \cite{FulHar:rep}). In terms of the language
often used in the physics literature: in even numbers of dimensions we
use Weyl spinors; real spinors are ``Majorana'' and quaternionic
spinors are ``symplectic Majorana''.\footnote{Note that two symplectic
Majorana spinors make up one quaternionic spinor.}

We may now list the maximal number of supersymmetries in each
dimension subject to the constraint that no particle with spin $>2$
appears. It was determined in \cite{CJ:SO(8)} that $N=1$ in $d=11$ was
maximal in this regard. To get the other values we simply maintain the
number of 
degrees of freedom of the spinors. For example, in reducing to ten
dimensions we go from 32 real degrees of freedom per spinor to 16 real
degrees of freedom per spinor. We thus need two spinors in ten
dimensions. The result is shown in table
\ref{tab:maxN}.

\begin{table}
\arraycolsep=20pt
$$\begin{array}{|c|c|l|}
  \hline
  d&N&{\rm Rep}\\
  \hline
  11&1&\R^{32}\\
  10&2&\R^{16}\\
   9&2&\R^{16}\\
   8&2&\C^8\\
   7&2&\H^8\\
   6&4&\H^4\\
   5&4&\H^4\\
   4&8&\C^2\\
   3&16&\R^2\\
  \hline
\end{array}$$
  \caption{Maximum numbers of supersymmetries.}
  \label{tab:maxN}
\end{table}

When $d\in4\Z+2$, the left and right spinors in the field theory are
independent and the supersymmetries may be separated. As we did
in section \ref{ss:nlsm}, we will denote a theory with $N_L$
left supersymmetries and $N_R$ right supersymmetries as having
$N=(N_L,N_R)$ supersymmetry. For purposes of counting total
supersymmetries, $N=N_L+N_R$.

Now we want to see what happens when we compactify such a theory down
to a lower number of dimensions. That is, we replace the space
$\R^{1,d_0-1}$ by $\R^{1,d_1-1}\times X$, for some compact manifold $X$ of
dimension $d_0-d_1$. To see what happens to the supersymmetries we
need to consider how a spinor of $\so(1,d_0-1)$ decomposes under the
maximal subalgebra
$\so(1,d_0-1)\supset\so(1,d_1-1)\oplus\so(d_0-d_1)$. The holonomy
of $X$ will act upon representations of $\so(d_0-d_1)$. Any
representation of $\so(1,d_1-1)\oplus\so(d_0-d_1)$ which is invariant
under this action will lead to representations of $\so(1,d_1-1)$ in
our new compactified target space. This tells us how to count
supersymmetries.

As the first example, consider toroidal compactification. In this case
the torus is flat and so the holonomy is trivial. Thus every
representation survives. Compactifying $N=1$ supergravity in eleven
dimensions down to $d$ dimensions results in reproducing the maximal
supersymmetries in table \ref{tab:maxN}. Note that for $d=10$ the
supersymmetry is $N=(1,1)$ and for $d=6$ the supersymmetry is
$N=(2,2)$. One simple way of deducing this immediately is from the
general result \cite{W:rKK} that compactification of eleven dimensional
supergravity on any manifold results in a non-chiral
theory.\footnote{Fortunately recent progress in M-theory appears to
tell us that compactification on spaces which are not manifolds can
circumvent this statement (see, for example, \cite{HW:E8M}).}

Naturally our next example will be compactification on a K3
surface. In this case the holonomy is $\SU(2)$. Let us consider the
case of an $N=1$ theory in ten dimensions compactified down to six
dimensions on a smooth K3 surface. The required decomposition is
\begin{equation}
  \so(1,9) \supset \so(1,5)\oplus\so(4)\cong 
		\so(1,5)\oplus\su(2)\oplus\su(2). \label{eq:10t6}
\end{equation}
The spinor decomposes accordingly as ${\bf 16}\to({\bf 4},{\bf 2},
{\bf 1}) \oplus ({\bf 4},{\bf 1},{\bf 2})$. We may take the last
$\su(2)$ in (\ref{eq:10t6}) as the holonomy. Thus the $({\bf 4},{\bf 2},
{\bf 1})$ part is preserved. It may look like we have 2 spinors in 6
dimensions at this point but remember that our spinor in ten
dimensions was real and spinors in six dimensions are
quaternionic. Thus the degrees of freedom give only a single spinor in
six dimensions. That is, $N=1$ supergravity in ten dimensions
compactified on a smooth K3 surface will give an $N=1$ theory in six
dimensions. The general rule is that compactification on a smooth K3 surface
will preserve half as many supersymmetries as compactifying on a
torus.

In the case that the number the supersymmetries or the number of
dimensions is large, the form of the moduli space of possible
supergravity theories becomes quite constrained. Holonomy is again the
agent responsible for this. Let us write the notion of extended
supersymmetry very roughly in the form
\begin{equation}
  \{\overline{Q}^i,Q^j\} = \delta^{ij}P,  \label{eq:extS}
\end{equation}
where $Q$ are the supersymmetry generators and $P$ is translation (we
may ignore any central charge for purposes of our argument). Now such
a relationship must clearly be preserved as we go around a loop in the
moduli space. However, the supersymmetries may transform among
themselves as we do this. This gives us a restriction on the holonomy
of the bundle of supersymmetries over the moduli space. Comparing 
(\ref{eq:extS}) to the analysis of invariant forms in section
\ref{ss:hol} tells us immediately what this restriction is. In the
case that the spinor is of type $\R$, $\C$, or $\H$, the holonomy
algebra will be (or contain) $\so(N)$, $\gu(N)$, or $\sp(N)$ respectively. Note
that when we have chiral spinors in $4\Z+2$ dimensions we may factor
the holonomy into separate left and right parts since these sectors
will not mix. 

Now the tangent directions in the moduli space are given by massless
scalar fields which lie in supermultiplets. These multiplets have a definite
transformation property under the holonomy group above. 
For a thorough account of this process we refer to \cite{Strath:N}.
This relates
the holonomy of the bundle of supersymmetries to the holonomy of the
tangent bundle of the moduli space. This knowledge can tell us a great
deal about the form of the moduli space.

As an example consider $N=4$ supergravity in five dimensions from
table \ref{tab:maxN}. We see immediately that the holonomy algebra of the
moduli space is $\sp(4)$. An analysis of the representation theory of
the supergravity multiplet shows that the massless scalars transforms
in a ${\bf 42}$ of 
$\sp(4)$. The only possibility from Berger and Simons result, and
Cartan's classification of noncompact symmetric spaces, is that the moduli space
is locally $E_{6(6)}/\Sp(4)_\sim$, where the tilde subscript denotes a
quotient by the central $\Z_2$. The moduli spaces for all the entries
in table \ref{tab:maxN} are given in \cite{Jul:dis}.

\subsection{The IIA string}	\label{ss:IIA}

The type IIA superstring in ten dimensions yields, in the low-energy
limit, a theory of ten-dimensional supergravity with $N=(1,1)$
supersymmetry. If we compactify this theory on a K3 surface then each
of these supersymmetries gives a supersymmetry in six dimensions and
so the result is an $N=(1,1)$ theory in six dimensions.

The local holonomy algebra from above must therefore be $\sp(1)\oplus
\sp(1)\cong\su(2)\oplus\su(2)\cong\so(4)$. There are two types of
supermultiplet in six dimensions which contain moduli fields:
\begin{enumerate}
\item The supergravity multiplet contains the dilaton which is a real
scalar.
\item Matter multiplets each contain 4 real massless scalars which
transform as a {\bf 4} of $\so(4)$.
\end{enumerate}
Thus the moduli space must factorize (at least locally) into a product
of a real line for the dilaton times the space parametrized by the
moduli coming from the matter multiplets. Thus, given the assumption
concerning completeness, the Teichm\"uller space
is of the form
\begin{equation}
\frac{\GO(4,m)}{\GO(4)\times\GO(m)}\times\R,
	\label{eq:T-IIA}
\end{equation}
where $m$ is the number of matter multiplets.

As well as the metric, $B$-field and dilaton, the type II string also
contains ``Ramond-Ramond'' states (see, for
example, \cite{GSW:book}). For the type IIA string these may be
regarded as a 1-form and a 3-form.\footnote{In a sense all odd forms
may be included \cite{Pol:D-brane}.} A $p$-form field can produce
massless fields upon compactification by integrating it over
nontrivial $p$-cycles in the compact manifold. As a K3 surface has no
odd cycles, no moduli come from the R-R sector.
Thus, comparing (\ref{eq:T-IIA}) to (\ref{eq:s-K3}) we see that
$m=20$.

The moduli space of conformal field theories may be considered as
living at the boundary of the moduli space of string theories in the
limit that $\Phi\to-\infty$, i.e., weak string-coupling. Thus
$\GO(\Gamma_{4,20})$ acts on this boundary.
Given the
factorization of the holonomy between the dilaton and the matter
fields (as they transform in different representations of $\so(4)$)
the action of $\GO(\Gamma_{4,20})$ also acts away from the boundary in
a trivial way on the dilaton. It is believed that moduli space is given simply
by
\begin{equation}
  \cM_{\rm IIA}\cong\cM_{\sigma}\times\R.
\end{equation}
That is, there are no identifications which incorporate the dilaton.
Certainly any duality which mixed the factors
would not respect the holonomy. Thus the only possibility remaining
would be an action of the form $\Phi\to-\Phi$ which would be a
strong-weak coupling duality (acting trivially on all other
moduli). Instead of such an S-duality, a far more 
curious type of duality was suggested in \cite{HT:unity,W:dyn} as we
now discuss.

The form of the moduli space of toroidal conformal field theories
(\ref{eq:Nar}) in section \ref{ss:tori} 
bares an uncanny resemblance to that of the moduli space of K3
conformal field theories, $\cM_{\sigma}$. Indeed they are identical if
$n_L=4$ and $n_R=20$. This is precisely the moduli space of conformal
field theories associated to a heterotic string compactified on a
4-torus.
Recall that the heterotic string consists of a superstring in the
left-moving sector and a bosonic string in the right-moving sector. As
such there are 16 extra dimensions in the right-moving sector which
are compactified on a 16-torus and
contribute towards the gauge group. In ten dimensions the heterotic
string is an $N=(1,0)$ theory and thus yields an
$N=(1,1)$ theory in six dimensions when compactified on a 4-torus.

The suggestion, which at first appears outrageous, is that the type IIA
string compactified on a K3 surface is the same thing, physically, as
the heterotic string compactified on a 4-torus. Although the moduli
space of conformal field theories is identical for these models, the
world-sheet formulation is so different that any notion that the two
conformal field theories could be shown to be equivalent is doomed
from the start. The claim however is not that the conformal field
theories are equivalent, but that the full string theories are
equivalent. The heterotic string has a dilaton, just like the type IIA
string, and so the moduli space of heterotic string theories is also
$\cM_{\sigma}\times\R$. The only way we can map these two moduli spaces
into each other without identifying the conformal field theories is
thus to map $\Phi$ from one theory to $-\Phi$ from the other.

This then, is the alternative to S-duality for the type IIA
string.
\begin{prop}
The type IIA string compactified on a K3 surface is equivalent to the
heterotic string compactified on a 4-torus. The moduli spaces are mapped
to each other in the obvious way except that the strongly-coupled type
IIA string maps to the weakly-coupled heterotic string and vica versa.
 \label{prop:1}
\end{prop} 

It would be nice at this point if we could prove proposition
\ref{prop:1}.  Unfortunately it appears that string theory is simply
not sufficiently defined to allow this. We should check first that
this proposition is consistent with what we do know about string
theory.

The fact that the moduli spaces of the type IIA string on a K3 surface
and the heterotic string on a 4-torus are identical is a good
start. Next we may check that the effective six-dimensional field
theory given by each is the same. This analysis was done in
\cite{W:dyn}. The result is affirmative but there is some new
information. To achieve complete agreement the flat six-dimensional
spaces given by the two string theories are not identical but instead
the metrics are scaled with respect to each other. That is, let $g_6$
denote the six-dimensional space-time metric and $\Phi$ the dilaton,
then
\def\shb#1{\hbox{\tiny #1}}
\begin{equation}
\eqalign{\Phi_{\shb{Het}} &= -\Phi_{\shb{IIA}}\cr
  g_{6,\shb{Het}} &= e^{2\Phi_{\shb{Het}}}g_{6,\shb{IIA}}.\cr}
			\label{eq:HIIA}
\end{equation}

Other checks may be performed too. Since the strings are related by a
strong-weak coupling relations, the fundamental particles in one
theory should map to solitons of the other theory. This has been
analyzed to a great extent but we refer to J.~Harvey's lectures for an
account of this.

To date nothing has been discovered in known string theory to
disprove proposition \ref{prop:1}. Assuming there are no such
inconsistencies one may wish to boldly assert that proposition
\ref{prop:1} is true {\em by definition}. That is, whatever string
theory may turn out to be, we will demand that it satisfies
proposition \ref{prop:1}. This is the point of view we will take from
now on. 

\subsection{Enhanced gauge symmetries} \label{ss:enh}

One of the interesting questions we are now fully equipped to address
is that of the gauge symmetry group of the six-dimensional theory
resulting from either a compactification of a type IIA string on a K3
surface, or a heterotic string on a 4-torus.

Firstly we may consider this question from the point of view of
conformal field theory. The type IIA string produces gauge fields from
the R-R sector. The 1-form gives a $\GU(1)$. The 3-form may be
compactified down to a 1-form over 2-cycles in $b_2(S)=22$
ways. Lastly, writing the 3-form as $C_3$, we have a dual field
$C^*_3$ given by
\begin{equation}
  dC^*_3 = *dC_3,
\end{equation}
where $C^*_3$ is a one form and also gives a $\GU(1)$. Thus, all told, we
have a gauge group of $\GU(1)^{24}$. Note that it is not possible to
obtain a nonabelian gauge theory, as far as the conformal field theory
is concerned, because the R-R fields are so reluctant to couple to any
other fields (see, for example, \cite{DKV:4d}).

Now consider the heterotic string picture. As explained in section
\ref{ss:tori}, generically we obtain a $\GU(1)^{16}$ contribution to the
gauge group from the extra 16 right-moving degrees of freedom for the
heterotic string. We also obtain 4 ``Kaluza-Klein'' $\GU(1)$ factors
from the metric from the 4 isometries of the torus. Lastly, the
$B$-field is a 2-form and so contributes $b_1(T^4)=4$ more
$\GU(1)$'s. All told we have $\GU(1)^{24}$ again as befits proposition
\ref{prop:1}.

Things become more interesting however when we realize that the
heterotic string can exhibit larger gauge groups at particular points
in the moduli space. What happens is that some winding/momentum modes
of $\Gamma_{4,20}$ may happen to give physical massless vectors for
special values of the moduli. Such states will be charged with respect
to the generic $\GU(1)^{24}$ and so a nonabelian group results.

The heterotic string is subject to a GSO projection to yield a
supersymmetric field theory and this effectively prevents any new
left-moving physical states from becoming massless. This may be
thought of as a similar statement to the assertion that the type IIA
string could never yield extra massless vectors. The right-moving
sector of the heterotic string is not subject to such constraints
however, and we may use our knowledge from section \ref{ss:tori} to
determine exactly when this gauge group enhancement occurs.

The vertex operator for one of our new massless states will be as follows. For
the left-moving part we want simply $\partial X^\mu$ to give a vector
index. For the right-moving part we require another operator with
conformal weight one. This results in a requirement that we desire a
state with $p_L=0$ and $p_R.p_R=-2$. Thus we require an element
$\alpha\in\Gamma_{4,20}$ which is orthogonal to $\Pi$ and is such that
$\alpha.\alpha=-2$.

The charge of such a state, $\alpha$, with respect to the $\GU(1)^{24}$
gauge group is simply given by the coordinates of $\alpha$. This
follows from the conformal field theory of free bosons and we refer
to \cite{Gins:lect} for details. Comparing this to the standard way of
building Lie algebras, we see that $\alpha$ looks like a root of a Lie
algebra, whose Cartan subalgebra is $\gu(1)^{\oplus24}$ (except that
the Killing form is negative definite rather than positive definite).

In the simplest case, there will be only a single pair $\pm\alpha$
which satisfy the required condition. This then will add two
generators to the gauge group charged with respect to one of the 24
$\GU(1)$'s. Thus the gauge group will be enhanced to $\SU(2)\times
\GU(1)^{23}$. Note that this enhancement to an $\SU(2)$ gauge group can
also be seen in the simple case of a string on a circle as was
pictured in figured \ref{fig:O11}. If $\theta=45^\circ$ then the
lattice element marked $\alpha$ in the figure (together with
$-\alpha$) will generate $\SU(2)$. 

The general rule then should be that the set of all vectors given by
\begin{equation}
  \cA = \{\alpha\in\Gamma_{4,20}\cap\Pi^\perp;\;\alpha.\alpha=-2\},
	\label{eq:cA}
\end{equation}
will form the {\em roots\/} of the nonabelian part of the gauge
group. Note that the rank of the gauge group always remains 24. Note
also that the roots all have the same length, i.e., the gauge group is
always {\em simply-laced\/} and falls into the $A$-$D$-$E$ classification.

This allows us to build large gauge groups. One thing one might do,
for example, is to split $\Gamma_{4,20}\cong\Gamma_{4,4}\oplus
\Gamma_8\oplus\Gamma_8$, where $\Gamma_8$ is the Cartan matrix of
$E_8$ (with a negative-definite signature). We may then consider
the case where $\Pi\subset\Gamma_{4,4}\otimes_\Z\R$. This would mean
that any element in $\Gamma_8\oplus\Gamma_8$ is orthogonal to $\Pi$
and thus the gauge group is at least $E_8\times E_8$. Looking at 
(\ref{eq:GBA}) we see that this is equivalent to putting $A=0$. Thus
we reproduce the simple result that the $E_8\times E_8$ heterotic
string compactified on a torus has gauge group containing $E_8\times
E_8$ if no Wilson lines are switched on.

Proposition \ref{prop:1} now tells us something interesting. Despite the
fact that the conformal field theory approach to the type IIA string insisted
that it could never have any gauge group other than $\GU(1)^{24}$, the
dual picture in the heterotic string dictates otherwise. There must be
some points in the moduli space of a type IIA string compactified on a
K3 surface where the conformal field theory misses part of the
story and we really do get an enhanced gauge group.

Since we know exactly where the enhanced gauge groups appear in the
moduli space of the heterotic string and we know exactly how to map
this to the moduli space of K3 surfaces, we should be able to see
exactly when the conformal field theory goes awry.

We will determine just which K3 surfaces give rise to this behaviour
for the type IIA string. To do this we are required to choose $w$ as in
section \ref{ss:clsym} so that we can find a geometric description. Let us
first assume that we may choose $w$ so that
\begin{equation}
  \alpha\in\frac{w^\perp}w,\quad\forall\alpha\in\cA.
	\label{eq:walpha}
\end{equation}
Let $w^*$ be the same vector as was introduced in equation
(\ref{eq:BB}). Any vector in $\Pi$ can be written as a sum $x+bw^*+cw$,
where $b,c\in\R$ and $x\in\Sigma$. Thus, the statement that $\alpha$
is orthogonal to $\Pi$ implies that $\alpha$ is orthogonal to
$\Sigma$. Now we use the results of section \ref{ss:orb} which tell us
that this implies that the K3 surface is an orbifold. To be precise,
the set $\cA$ corresponds to the root diagram of the $A$-$D$-$E$ 
singularity given in table \ref{tab:ADE}.

What we have just shown is a remarkable fact (and was first shown by
Witten in \cite{W:dyn}). 
\begin{prop}
If we have a K3 surface with an orbifold
singularity then a type IIA string compactified on this surface can
exhibit a nonabelian gauge group such that the $A$-$D$-$E$
classification of orbifold singularities coincides perfectly with the
$A$-$D$-$E$ classification of simply-laced Lie groups.
\end{prop}
This proposition rests on proposition \ref{prop:1} and the assumptions
that went into building the moduli spaces. As an example, we see that 
an $\SU(n)$ gauge group corresponds to a singularity locally of the
form $\C^2/\Z_n$.

Note that the orbifold singularity is not a sufficient condition for
an enhanced gauge symmetry. For $\Pi$ to be
orthogonal to $\alpha$ we also require that $B^\prime$, and hence $B$,
is also perpendicular to $\alpha$. One way of stating this is to say
that {\em the component of $B$ along the direction dual to $\alpha$ is
zero}. Note that the volume of the K3 surface does not matter in this
context. 

It is probably worth emphasizing here that this statement about the
$B$-field can be important \cite{me:enhg}. Orbifolds are well-known as
``good'' target spaces for string theory in that they lead to finite
conformal field theories. It might first appear then that we are
saying that the conformal field theory picture is breaking down at a
point in the moduli space where an enhanced gauge group appears but when
the conformal field theory appears to be perfectly reasonable. This is not
actually the case. The enhanced gauge symmetry appears when $B=0$
along the relevant direction. Conformal field theory orbifolds however
tend to give the value $B=\ff12$. Thus, the point in the moduli space
corresponding to the happy conformal field theory orbifold and the
point where the enhanced gauge group appears are not the same. There
is good reason to believe the conformal field theory at the point
where the enhanced gauge group appears is not well-behaved
\cite{W:dyn2}.

What happens if we relax our condition on $w$ given in
(\ref{eq:walpha})? Now the situation is not so clear. $\Sigma$ need not
be perpendicular to any $\alpha$ and so the K3 surface may be
smooth. If this is the case, the volume of the K3 surface
cannot be arbitrarily large. This is because in the large radius
limit, $\Pi$ is roughly the span of $\Sigma$ and $w$, but $\alpha$
is not perpendicular to $w$. In fact, the volume of the
K3 surface must be of order one in units of
$(\alpha^\prime)^2$. Thus we see, assuming the $B$-field is tuned to
the right value, that an enhanced gauge group arises
when the K3 surface has orbifold singularities and is any size, or
if the K3 surface is very small and is given just the right (possibly smooth)
shape.

Note that to get a very large gauge group we probably require the K3
surface be singular {\em and\/} it to be very small. This is
necessary, for example, if we want a gauge group $E_8\times E_8\times
\SU(2)^4\times\GU(1)^4$. Note that this latter group is ``maximal'' in
the sense that the nonabelian part is of rank 20, which is the maximal
rank sublattice that can be orthogonal to $\Pi$.

We have proven the appearance of nonabelian gauge groups by assuming
proposition \ref{prop:1}. Instead one might like to attempt some
direct justification. This is probably best seen by using Strominger's
notion of ``wrapping $p$-branes'' \cite{Str:con}. The general idea is
that the R-R solitons are associated with cycles in the target space and
the mass of these states is given by the area (or volume) of these
cycles. Thus, as the K3 surface acquires a quotient singularity an
$S^2$ shrinks down and thus a soliton becomes massless. We will not
pursue the details of this construction as solitons lie somewhat
outside our intended focus of these lectures. We also refer the reader to
\cite{BSV:D-man,DM:qiv} for further discussion.

\subsection{The IIB string}	\label{ss:IIB}

The type IIB superstring in ten dimensions yields, in the low-energy
limit, a theory of ten-dimensional supergravity with $N=(2,0)$
supersymmetry. If we compactify this theory on a K3 surface then each
of these supersymmetries gives a supersymmetry in six dimensions and
so the result is an $N=(2,0)$ theory in six dimensions.

The local holonomy algebra from above must therefore be $\sp(2)
\cong\so(5)$. There is only one type of supermultiplet in six
dimensions which contains massless scalars and that is the matter
supermultiplet. Each such multiplet contains five scalars transforming
as a {\bf 5} 
of $\so(5)$. Thus, given the assumption
concerning completeness, the Teichm\"uller space
is of the form \cite{Rom:6}
\begin{equation}
\frac{\GO(5,m)}{\GO(5)\times\GO(m)},
	\label{eq:T-IIB}
\end{equation}
where $m$ is the number of matter multiplets.

In addition to the metric, $B$-field and dilaton, moduli may also
arise from the R-R fields. The type IIB string in ten dimensions has a
0-form, a 2-form and a self-dual 4-form. The 0-form gives $b_0(S)=1$
modulus. The 2-form gives $b_2(S)=22$ moduli. The 4-form gives 
$b_4(S)=1$ modulus. One might also try to take the dual of the 4-form
to give another modulus. This would be over-counting the degrees of
freedom however, as the 4-form is self-dual. Thus, the number of
moduli are given by
\begin{center}
\tabcolsep=16pt
\begin{tabular}{r|r}
  Metric&58\\
  $B$-field&22\\
  Dilaton&1\\
  0-form&1\\
  2-form&22\\
  4-form&1\\
  \hline
  Total&105
\end{tabular}
\end{center}
To get the dimension of the Teichm\"uller space correct we require $m=21$.

To find the discrete group acting on the Teichm\"uller space we go
through a procedure remarkably similar to that of section \ref{ss:mir}
where we found the moduli space of conformal field theories. Firstly
we may consider the Teichm\"uller space as the Grassmannian of
space-like 5-planes in $\R^{5,21}$. Denote the 5-plane by $\Theta$.
Then we choose a primitive null
element, $u$, of $\Gamma_{5,21}$. Now we have
$u^\perp/u\cong\R^{4,20}$. We may define $\Pi^\prime=\Theta\cap
u^\perp$. Define the vector $R^\prime$ by demanding that
$\Pi^\prime$ and $R^\prime$ span $\Theta$, $R^\prime$ be orthogonal to
$\Pi^\prime$ and $u.R^\prime=1$. Then project $\Pi^\prime$ and
$R^\prime$ into $u^\perp/u$ to obtain the space-like 4-plane, $\Pi$,
and the vector, $R$, respectively.

Clearly we now interpret $\Pi$ in terms of the underlying conformal
field theory on the K3 surface. $R$ represents the degrees of freedom
coming from the R-R sector. The six-dimensional dilaton may be deduced from
$R^\prime-R$ just as the volume of the K3 surface was deduced from
$B^\prime-B$ in section \ref{ss:mir}.

It is important to note that this construction provides considerable
evidence for our identification of $\R^{4,20}$ with $H^*(S,\Z)$ in
section \ref{ss:mir}. This is because $R$, which represents the R-R
degrees of freedom, is a vector in $\R^{4,20}$ --- the same space in
which $\Pi$ lives. We also saw how these moduli arise from
$H^0\oplus H^2\oplus H^4\cong H^*$.

Generating the discrete modular group, $G_{\shb{IIB}}$ (which, as the
reader will have guessed, will turn out to be 
$\GO(\Gamma_{5,21})$) is slightly different to the way we built 
$\GO(\Gamma_{4,20})$ in section \ref{ss:mir}, but we show here that we
can reduce it to the same problem. First note that we have
$\GO(\Gamma_{4,20})\subset G_{\shb{IIB}}$. That is, any symmetry of
the conformal field theory will be a symmetry of the string theory
(just as any symmetry of the classical geometry is a symmetry of the
conformal field theory).

The next ingredient we use will be that of S-duality of the type IIB
string in ten dimensions \cite{HT:unity,W:dyn}. This asserts that
there is an $\Sl(2,\Z)$ symmetry acting on the ten-dimensional dilaton
and the axion (i.e., the R-R 0-form). This group is generated by a 
strong-weak coupling
interchange of the form $\Phi_{10,\shb{IIB}}\to-\Phi_{10,\shb{IIB}}$,
and a translation of the axion by one. While one might 
assert this S-duality statement as a distinct conjecture, it is
certainly intimately related to other duality statements. One simple
way of ``deriving'' it is to see that it is almost an inevitable
consequence of M-theory \cite{me:dual,Sch:d=9} (see also J.~Schwarz's
lectures). We will also see in section \ref{ss:N=4} that it follows
from proposition \ref{prop:1} and mirror symmetry.

To embed $\Sl(2,\Z)$ into $\Gamma_{5,21}$ we note that $\Sl(2,\Z)$ is
the group of automorphisms of $\Z^2$, which means that it is the group
of isometries of a null lattice of rank two. We may split
$\Gamma_{5,21} \cong \Gamma_{3,19}\oplus\Gamma_{2,2}$ and then let
the $\Sl(2,\Z)$ act on a null 2-plane in the $\Gamma_{2,2}$ part. If
we identify $\Gamma_{3,19}$ with $H^2(S,\Z)$ then we see that
the group of classical symmetries of the K3 surface (i.e.,
$\GO(\Gamma_{3,19})$) commutes with S-duality. This is exactly what we
desire from the effective target space theory.

Now the shift of the axion is a shift in the $H^0(S,\Z)$
direction. Since $\GO(\Gamma_{4,20})$ acts transitively on primitive
null vectors, we immediately see that any shift by an element of
$\Gamma_{4,20}$ is a symmetry of the string theory. This is an
analogue of the ``integral $B$-field shift''. That is, {\em shifting
any R-R modulus by an element of $H^*(S,\Z)$ is a symmetry of the
string theory}.

The other generator of $\Sl(2,\Z)$ may be taken as one which exchanges
the two null vectors generating the null 2-plane in
$\Gamma_{2,2}$. That is, we swap the axion direction, $H^0(S,\Z)$,
with $u$, a null vector outside $\R^{4,20}$. This is the exact analogue of
mirror symmetry (which exchanged an element of $H^2(S)$ with
$H^0(S)$). We have reduced the problem to one completely analogous 
to finding the modular group for conformal field theories on a K3
surface. Thus we deduce 
that we can generate all of $\GO(\Gamma_{5,21})$ (as asserted first in 
\cite{W:dyn}). Assuming the moduli space is Hausdorff we have
\begin{prop}
  The moduli space of type IIB string theories compactified
on a K3 surface is
\begin{equation}
  \cM_{\shb{IIB}} = \GO(\Gamma_{5,21})\backslash\GO(5,21)/
		(\GO(5)\times\GO(21)).
	\label{eq:s-IIB}
\end{equation}
  \label{prop:IIB}
\end{prop}
This proposition depends on the S-duality conjecture for the type IIB
string in ten dimensions, theorem \ref{th:s-K3}, and the completeness
and Hausdorff constraints.

\iffigs
\begin{figure}
  \centerline{\epsfxsize=7cm\epsfbox{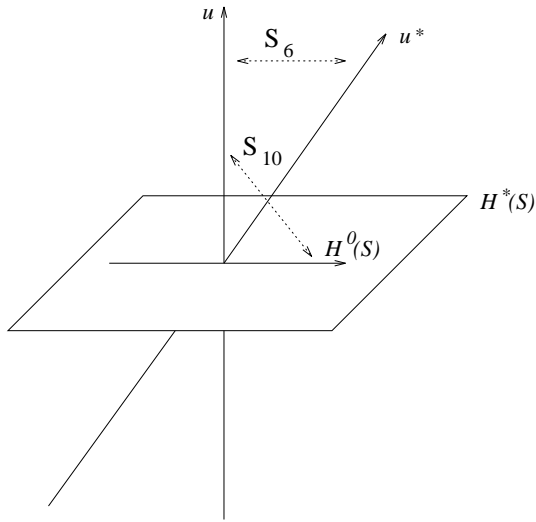}}
  \caption{The S-dualities of the type IIB string.}
  \label{fig:IIB}
\end{figure}
\fi

We should mention that there is a strong-weak coupling duality in the
resulting six-dimensional theory. Let $u^*$ be a null vector in
$\Gamma_{5,21}$ dual to $u$ such that $u$ and $u^*$ span the
$\Gamma_{1,1}$ sublattice orthogonal to
$H^*(S,\Z)\cong\Gamma_{4,20}$. One of the elements in
$\GO(\Gamma_{5,21})$ exchanges $u$ and $u^*$ and has the effect of
reversing the sign of the six-dimensional dilaton. Note that this is not
at all the same as the element of $\GO(\Gamma_{5,21})$ which
changed the sign of the {\em ten\/}-dimensional dilaton as the latter
exchanged $u$ with $H^0(S,\Z)$ as shown in figure \ref{fig:IIB}. In
general an S-duality in a given 
number of dimensions will not give rise to an S-duality in a lower
number of dimensions upon compactification. In the type IIB string 
however, we see S-duality in both six and ten dimensions.

The behaviour of the type IIA and type IIB string can be
contrasted. The strongly-coupled type IIA string on a K3 surface is
dual to a different string theory (the heterotic string) which is
weakly-coupled. The type IIB strongly-coupled string on a K3 surface
is dual to a weakly-coupled version of itself. 

Finally in this section let us mention some strange properties of the
type IIB string on a K3 surface. We know that when the type IIA string
is compactified on an orbifold with $B$=0 then an enhanced gauge
symmetry appears. This indicated some divergence within the underlying
conformal field theory. It should be true therefore that a type IIB
string compactified on the same space must have some interesting
nonperturbative physics since it is associated with the same divergent
conformal field theory. It was explained in \cite{W:dyn2} that these
new theories are associated with massless string-like solitons which
appear at these points in the moduli space.

%%%%%%%%%%%%%%%%%%%%%%%%%%%%%%%%%%%%%%%%%%%%%%%%%%%%%%%%%%%%%%%%%%%

\section{Four-Dimensional Theories}   \label{s:4d}

Now let us explore what happens when we compactify string theories
down to four dimensions. This process need not involve a K3 surface in
general but we will find that in all the easy cases K3 surfaces will
be present in abundance!

\subsection{$N=4$ theories}   \label{ss:N=4}

Supersymmetries are not chiral in four dimensions and so, in contrast
to the six-dimensional case, there is only one kind of $N=4$ theory.

The local holonomy algebra must be $\gu(4)
\cong\gu(1)\oplus\su(4)\cong\gu(1)\oplus\so(6)$. There are two types of
$N=4$ supermultiplet in four dimensions which contain moduli fields:
\begin{enumerate}
\item The supergravity multiplet contains the dilaton-axion field
which is a complex object under the $\gu(1)$ holonomy.
\item Matter multiplets each contain 6 real massless scalars which
transform as a {\bf 6} of $\so(6)$.
\end{enumerate}
Thus the moduli space must factorize into a product
of a complex plane, for the dilaton-axion, times the space parametrized by the
moduli coming from the matter multiplets. Thus, given the assumption
concerning completeness, the Teichm\"uller space
may be written in the form
\begin{equation}
\frac{\GO(6,m)}{\GO(6)\times\GO(m)}\times\frac{\Sl(2)}{\GU(1)},
	\label{eq:T-4}
\end{equation}
where $m$ is the number of matter multiplets. 

A very simple way to arrive at this theory is to compactify a
heterotic string on a 6-torus. The first factor of (\ref{eq:T-4}) is
then clearly the moduli space of the conformal field theories on the
torus with 6 left-moving dimensions and 22 right-moving
dimensions --- that is, $m=22$. The $\Sl(2)/\GU(1)$ term then comes
from the dilaton-axion system: the dilaton being the string dilaton as
usual and the axion from dualizing the $B$-field to obtain a scalar.

The type IIA string may be compactified on K3${}\times T^2$ to obtain an
$N$=4 theory too. The conformal field theory on a K3 has 80
real deformations and for $T^2$ it has 4 deformations. The 1-form R-R
field gives $b_1(\mbox{K3}\times T^2)=2$ moduli and the 1-form R-R
field gives $b_3(\mbox{K3}\times T^2)=44$ moduli. The 3-form may also
be compactified down to a 2-form in $b_1(\mbox{K3}\times T^2)=2$ ways
and then dualized to give 2 more scalars. Adding the dilaton-axion we
have $80+4+2+44+2+2=134$. This implies $m=22$ again. Actually
proposition \ref{prop:1} tells us that this must be the same
theory as the heterotic string on a 6-torus.

Let us examine the moduli space of conformal field theories on a
2-torus. As far as the Teichm\"uller space is concerned we have
\begin{equation}
  \frac{\GO(2,2)}{\GO(2)\times\GO(2)}\cong
      	\frac{\Sl(2)}{\GU(1)}\times\frac{\Sl(2)}{\GU(1)},
\end{equation}
up to $\Z_2$ identifications. One of the $\Sl(2)/\GU(1)$ factors may
be regarded as the complex structure of the torus and the other
$\Sl(2)/\GU(1)$ factor represents the K\"ahler form and $B$-field on
the torus. We refer to \cite{Giv:rep} for a review of this. What does
the second factor in (\ref{eq:T-4}) represent? There are many ways of
approaching this problem. Here we use a trick following \cite{Duff:S}
that will come in use later on.

Begin with a six-dimensional field theory given by the heterotic
string compactified on a 4-torus.
The effective field theory in six dimensions will be
roughly of the form
\begin{equation}
  S = \int d^6X\,\sqrt{g_6}e^{-2\Phi_{6,\shb{Het}}}(R+\ldots)
\end{equation}
Now compactify over a 2-torus of area $A_{\shb{Het}}$ (as measured by
the heterotic string).
\begin{equation}
\eqalign{
  S &= \int d^4X\,\sqrt{g_4}\,e^{-2\Phi_{6,\shb{Het}}}A_{\shb{Het}}
		(R+\ldots)\cr
  &= \int d^4X\,\sqrt{g_4}\,e^{-2\Phi_{4,\shb{Het}}}
		(R+\ldots),\cr
  &= \int d^4X\,\sqrt{g_4}\,A_{\shb{IIA}}
		(R+\ldots),\cr}   \label{eq:DD}
\end{equation}
where $A_{\shb{IIA}}$ is the area of the 2-torus when we consider building
the same theory by compactifying the type IIA string on K3${}\times
T^2$. We have made use of (\ref{eq:HIIA}) to derive this. Looking at
the second line of (\ref{eq:DD}) we see that the area of the $T^2$ is
actually playing the r\^ole of the coupling constant. Thus, in going
from the heterotic string description of the situation to the type IIA
description of the same situation, the dilaton of the heterotic string
has been replaced by the area of the $T^2$ of the type IIA
string. Thus, {\em the $\Sl(2)/\GU(1)$ factor in \mbox{(\ref{eq:T-4})}
must represent the 
K\"ahler form and $B$-field of the 2-torus in the type IIA picture}.

We know from conformal field theory that $\Sl(2,\Z)$ acts on the
$\Sl(2)/\GU(1)$ part of the Teichm\"uller space giving the K\"ahler
form and $B$-field of the 2-torus. Combining this knowledge with what
we found from the heterotic string, we see that $\GO(\Gamma_{6,22})
\times\Sl(2,\Z)$ acts as the modular group for our $N=4$ theory.
\begin{prop}
The type IIA string compactified on K3${}\times T^2$ is equivalent to
the heterotic string compactified on a 6-torus and they form the
moduli space
\begin{equation}
  \cM_{N=4} = \left(\GO(\Gamma_{6,22})\backslash\GO(6,22)/
		(\GO(6)\times\GO(22))\right)\times\left(\Sl(2,\Z)\backslash
		\Sl(2)/\GU(1)\right).
	\label{eq:K3xT2}
\end{equation}
  \label{prop:K3xT2}
\end{prop}
This rests on the same assumptions as proposition \ref{prop:IIB}.%
\footnote{Depending on one's tastes, in the case of the moduli space
of the heterotic string on a 
6-torus one may wish to consider this statement as more fundamental
than proposition \ref{prop:1} as it may be analyzed directly in terms
of solitons \cite{Sen:DMb}.}

The $\Sl(2,\Z)$ factor of the modular group acts as an S-duality in the
effective four-dimensional theory. Thus we have {\em derived}, from
proposition \ref{prop:1} the existence of Montonen-Olive S-duality 
\cite{MO:S,Os:N=4,Sen:4d} for $N=4$ theories in four
dimensions. (Again we are going to neglect to discuss solitons --- see
\cite{Sen:DMb} for such analysis.)

Lastly we may consider the type IIB string compactified on K3${}\times
T^2$. Again there are a multitude of ways of arriving at the desired
result. One of the easiest ways is to take following proposition from
\cite{DLP:IIAB,DHS:IIAB}:
\begin{prop}
  The type IIA superstring and type IIB superstring compactified down
to nine dimensions on a circle are equivalent except that the radii of
the circles are inversely related.
\end{prop}
One also needs to shift the dilaton of one theory relative to the
other to achieve the same target space effective field theory for the
two string theories.

Thus, since the $T^2$ of K3${}\times T^2$ contains a circle, the type
IIB theory compactified on K3${}\times T^2$ can also be bundled into
proposition \ref{prop:K3xT2}. Note that an \RoR\ transformation on one
of the circles in the $T^2$ is a mirror map in the sense that the
notions of deformation of complex structure and complexified K\"ahler
form are interchanged. Thus, the $\Sl(2,\Z)\backslash\Sl(2)/\GU(1)$
factor in the moduli 
space in (\ref{eq:K3xT2}) represents the complex structure moduli
space of the 2-torus in the case of the type IIB string.

Thus the $\Sl(2,\Z)\backslash\Sl(2)/\GU(1)$ factor in the moduli space
in (\ref{eq:K3xT2}) can play 3 r\^oles \cite{AM:Ud,Duff:tri}:
\begin{enumerate}
\item The dilaton-axion variable in the case of the heterotic string.
\item The area and $B$-field of the 2-torus in the case of the type
IIA string.
\item The complex structure of the 2-torus in the case of the type
IIB string.
\end{enumerate}

Comparing the heterotic string to the type IIA string in this setup
may be regarded as fairly profound. The dilaton of the heterotic
string, i.e., the coupling of the space-time field theory, is mapped to
an area in the type IIA theory, i.e., the coupling of the world-sheet
field theory. Thus, in a sense {\em we are mapping the space-time
field theory associated to one string theory to the world-sheet field
theory associated to another}. One might take this as evidence
that neither the target space point of view nor the world-sheet point
of view of string theory may be regarded as more fundamental than the
other since they may be exchanged. 

The $\Sl(2,\Z)$ S-duality of the type IIB string in ten dimensions is
now sitting in the group $\GO(\Gamma_{6,22})$. In fact, the above
analysis can be used to ``prove'' the existence of this S-duality
group. This can be viewed as an analogue of the deduction of this same
S-duality group from M-theory as was done in \cite{me:dual,Sch:d=9}.

Note that we can play the same game as in section \ref{ss:enh} to find
the enhanced gauge groups. In this case we have a space-like 6-plane
in $\Gamma_{6,22}\otimes_\Z\R$ and we look for roots perpendicular to
this plane. The gauge group is always of rank 28.

\subsection{More $N=4$ theories}   \label{ss:mN=4}

Does the moduli space (\ref{eq:K3xT2}) represent all possible $N=4$
theories in four dimensions? It seems unlikely as one expects to be
able to build theories with a gauge group of rank $<28$. Consider
compactification of the type II string. All we demand to obtain the
desired theory in four dimensions is that the manifold on which we
compactify have 
$\SU(2)$ holonomy. The only complex surface with $\SU(2)$ holonomy is
a K3 surface. We have more possibilities in complex dimension three
however. 
Thus we expect that the moduli spaces we discussed in section
\ref{s:II} to give the complete story for $N=2$ theories in six
dimensions but we are not done yet for $N=4$ theories in four dimensions.

Note that this is a similar statement to the one that Seiberg gave using
anomalies \cite{Sei:K3}. When using a conformal field theory to
compactify a ten-dimensional theory to six dimensions one may consider
the case of a type IIB string compactified to a chiral $N=(2,0)$
six-dimensional theory and analyze the anomalies. The Hodge numbers of a
K3 surface are found to be necessary for a consistent theory.

We should add that one can find further theories in six dimensions if
one is willing to drop the requirement that the compactification has
some conformal field theory description (and switching to something
like M-theory instead). See \cite{SS:pairs} for an
example. 

Let us consider how to build a complex threefold with holonomy
$\SU(2)$. First we note the existence of a covariantly-constant
holomorphic 2-form and thus $h^{2,0}=1$. The Dolbeault index
\cite{EGH:dg} may then be used to establish $h^{1,0}=1$. Thus our
manifold cannot be simply-connected. Now we may use the
Cheeger--Gromoll theorem \cite{CG:th} which tells us that the universal
cover of the manifold is isometric to $M\times\R^n$ for some compact
simply-connected manifold, $M$. It is clear that we require
that the universal cover to be K3${}\times\R^2$. In other words, {\em
any complex threefold with holonomy $\SU(2)$ is isometric to
K3${}\times T^2$ or some quotient thereof}.

To build more $N=4$ theories we will consider compactifying type II
strings on a quotient of K3${}\times T^2$. This quotient must of course
preserve the global $\SU(2)$ holonomy and thus any element of the
quotienting group must preserve the holomorphic 2-form on the K3
surface. Any such action has fixed points on the K3 surface. Thus, to
avoid getting a quotient singularity, any such action on the K3
surface must be accompanied by a translation on the $T^2$. That is,
the quotienting group must have translations in $\R^2$ as a faithful
representation. An immediate consequence is that the quotienting group
must be abelian.

\begin{table}[b]
\centerline{\begin{tabular}{|c||c|c|c|c|c|c|c|c|c|c|c|c|}
\hline
$G$&$\Z_2$&$\Z_2\times\Z_2$&$\Z_2\times\Z_4$&$\Z_2\times\Z_6$
  &$\Z_3$&$\Z_3\times\Z_3$
  &$\Z_4$&$\Z_4\times\Z_4$&$\Z_5$&$\Z_6$&$\Z_7$&$\Z_8$\\
\hline
$M$&8&12&16&18&12&16&14&18&16&16&18&18\\
\hline
\end{tabular}}
\caption{Nikulin's K3 quotienting groups.}
\label{tab:K3aut}
\end{table}

The classification of such groups, $G$, has been done by Nikulin
\cite{Nik:K3aut} and we list the results in table \ref{tab:K3aut}. $M$ is
the rank of the maximal sublattice of $H^2({\rm K3},\Z)$ that
transforms nontrivially under $G$. 

This action of $G$ on
$\Gamma_{6,22}\cong\Gamma_{4,20}\oplus\Gamma_{2,2}$ is now
determined. Firstly, the action on K3 is a geometric symmetry and so
must preserve $w$ and $w^*$. The K3 part of the action is then
determined by the action on $H^2({\rm
K3},\Z)\cong\Gamma_{3,19}\subset\Gamma_{4,20}$
For the explicit form of the action
on the lattice $H^2({\rm K3},\Z)$ we refer the reader to
\cite{CL:KT/G}. 
Lastly we need the action on $\Gamma_{2,2}$. This encodes the action
of $G$ on $T^2$ which we require to be a translation. We also want
this action to be geometric and therefore left-right symmetric. This
forces the shift to be a {\em null\/} direction. This is sufficient to
determine the shift up to isomorphism.

Now that we know the action of $G$ on $\Gamma_{6,22}$ we may copy the
description of the quotienting procedure over into the heterotic
string picture. The result is that we are now describing an {\em
asymmetric orbifold\/} of a heterotic string on $T^6$. Such objects
were first analyzed in \cite{NSV:asym}. An asymmetric orbifold is an
string-theory orbifold in which the left-movers and right-movers of
the conformal field theory are not treated identically. Because of
this the geometric description of the quotienting process in terms of
target space geometry is obscure. Also the chiral nature of
quotienting can produce anomalies. One manifestation of this can be
lack of modular invariance of the resulting conformal field theory.

Let us consider an asymmetric orbifold of a toroidal theory built on a
lattice $\Lambda$. Consider an element of the quotienting group, $g\in
G$, and represent it as a rotation of the lattice followed by a shift,
$\delta\in\Lambda\otimes_\Z\R$. Let the eigenvalues of the rotation be
of the form $\exp(2\pi i.r_j)$. A necessary condition for modular
invariance is that \cite{NSV:asym,FV:anom}
\begin{equation}
  |g|\left(\ff14\sum_j r_j(1-r_j) + \ff12\delta.\delta\right)\in\Z;
	\quad\forall g\in G,   \label{eq:anom}
\end{equation}
where $|g|$ is the order of $g$. One may check the groups in table
\ref{tab:K3aut} and show that this condition is indeed satisfied in
every case.

Did we have the right to 
expect that (\ref{eq:anom}) should be satisfied for all the groups in
table \ref{tab:K3aut}? To this author the result seems a little
mysterious. Indeed, it is the case that if one considers quotients
which destroy the $N=4$ supersymmetry then one need not be so lucky
\cite{FHSV:N=2,VW:pairs}. For now though, since we are concerned with
$N=4$ theories at this point, we may content ourselves with the
knowledge that the anomalies appear to be looking after themselves and
press on.

Now let us determine the moduli space. Firstly any deformation of the
original theory of the type II string on K3${}\times T^2$ or the
heterotic string on $T^6$ which is invariant under $G$ will be a
deformation of the resulting quotient. Secondly we need to worry in
string theory that we may introduce some ``twisted marginal
operators'' --- that is, massless modes associated with fixed
points. Since there are no fixed points of $G$ (at least in the type II
picture) we may ignore the latter possibility. To obtain the first
type of deformation we may simply restrict attention to the invariant
sublattice $\Lambda_G\subset\Gamma_{6,22}$ under (the rotation
part of) $G$. Note that 
all of the space-like directions of $\Gamma_{6,22}$ are not rotated by
$G$ and so the resulting invariant sublattice, $\Lambda_G$, will have signature
$(6,m)$, where $m=22-M$ from table \ref{tab:K3aut}.

What we have done is to build a moduli space of the form
\begin{equation}
  \GO(\Lambda_G)\backslash\GO(6,m)/(\GO(6)\times\GO(m)),
	\label{eq:6m}
\end{equation}
of an $N=4$ theory in four dimensions which we viewed either as a type
IIA string on a freely-acting quotient of K3${}\times T^2$ or an
asymmetric orbifold of a heterotic string on $T^6$.

\iffigs
\begin{figure}
  \centerline{\epsfxsize=12cm\epsfbox{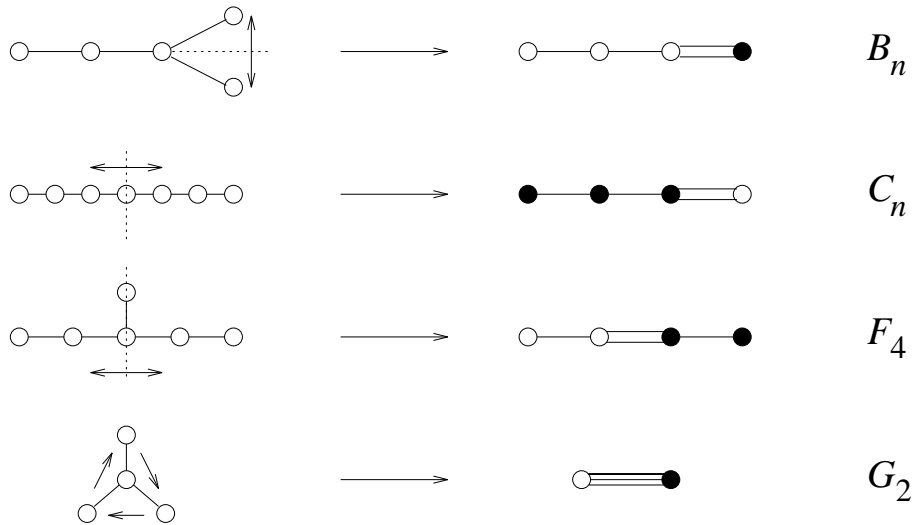}}
  \caption{Quotients of simply-laced groups.}
  \label{fig:oaut}
\end{figure}
\fi

Are there any further possibilities beyond those listed in table
\ref{tab:K3aut}? We restricted ourselves to classical symmetries of
the K3 surface. There should certainly be more symmetries from the
stringy geometry of the K3 surface when it is Planck-sized. One may
also look at M-theory to provide more possibilities. We refer the
reader to \cite{CL:KT/G,CL:mons} for further discussion.

Although (\ref{eq:6m}) looks suspiciously like a Narain moduli space
for a heterotic string on a torus, it is important to notice that
$\Lambda_G$ is, in general, not self-dual. Any attempt to describe
this theory in a straight forward way as a toroidal compactification
is doomed as it would imply that the theory is not modular invariant.

On a similar point we have to be careful when considering which
enhanced gauge groups can appear. One can view the moduli space as a
Grassmannian of space-like 6-planes in $\R^{6,m}$ but it is no longer
the case that only elements, $v\in\Lambda_G$, perpendicular to this
6-plane with $v.v=-2$ will give massless vector fields. The simplest
way to approach the question of enhanced gauge groups is as follows
(see \cite{CHL:bigN,CP:ao} for the original analysis in terms of
heterotic strings). Consider the original theory before we divide by
$G$. In this case, we know what the roots of the enhanced gauge group
are, given the space-like 6-plane. As $G$ has a nontrivial action on the
lattice $\Gamma_{6,22}$, it may also act on the roots of the gauge
group. Since our desired theory is the invariant part of the original
theory under the quotient by $G$, the resulting gauge group will be
the invariant part of the original gauge group under the action of the
discrete group $G$. 

The problem we have therefore is as follows. Given a simply-laced
gauge group and an action of a discrete group, $G$, on the roots of
this gauge group, find the subgroup of the gauge group which is invariant 
under this 
action. This will be the enhanced gauge group of the desired quotient
theory. Fortunately this is a well-known problem in Lie group theory
(see, for example, exercise 22.24 in \cite{FulHar:rep}). The outer
automorphism of the 
group, given by an action on the roots can be written as a symmetry of
the Dynkin diagram in the obvious way.\footnote{Except for the case of
$\su(2n+1)$ in which case the outer automorphism yields $\so(2n+1)$ as the
invariant subalgebra.}
The results are shown in figure
\ref{fig:oaut} and show that non-simply-laced Lie groups can
result. In particular one may show that {\em any\/} Lie group (of
sufficiently small rank) can appear as an enhanced gauge symmetry.

\iffigs
\begin{figure}
  \centerline{\epsfxsize=10cm\epsfbox{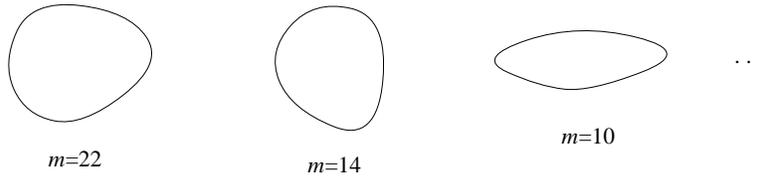}}
  \caption{The moduli space of $N=4$ theories.}
  \label{fig:N=4}
\end{figure}
\fi

Lastly let us note an important point about the $N=4$ moduli
spaces. This is that they are {\em disconnected\/} from each other as
shown in figure \ref{fig:N=4}. If
components of the moduli space with different values of $m$ were to
touch each other then, at such a point of contact, we would have a
theory with very special properties. As one approached such a theory
from within the interior of one of the regions, extra states would
become massless to furnish the deformations into the other
region. This does not happen according to the conformal field
description of either the heterotic string or the type II
strings. Thus it would appear unreasonable to expect it to occur in the
full string theory. This is to be contrasted with the behaviour of
$N=2$ theories in four dimensions, as we discuss in section \ref{ss:ext}.

\subsection{Generalities for $N=2$ theories}     \label{ss:N=2}

Now that we have understood the main features of the moduli space of
$N=4$ theories in four dimensions we are ready to embark on a study of
the much richer field of $N=2$ theories. Much of the recent interest
in duality was sparked by Seiberg and Witten's work on $N=2$
Yang-Mills field theory \cite{SW:I,SW:II}. Here we are hoping to
analyze full string theory in the same context. Thus we expect the
subject to be at least as rich as Seiberg-Witten theory. In the short
period that $N=2$ theories have been studied in the context of string
duality, the subject is already vast and it will be difficult to do
justice to it here. As in the rest of these lectures, we will attempt
to confine our attention to matters related to the moduli space of
theories.

How can we obtain an $N=2$ theory in four dimensions from string
theory? Two answers immediately appear given the usual holonomy
argument. Firstly one may take a heterotic string theory in ten
dimensions and compactify it on a complex threefold with $\SU(2)$
holonomy. We have already discussed such manifolds in section
\ref{ss:mN=4} and found that they are of the form K3${}\times T^2$, or
some free quotient thereof. Secondly one might take a type II string
and compactify it on a complex threefold of $\SU(3)$ holonomy, i.e., a
\CY\ manifold. Given the story for $N=4$ theories above it is tempting
to conjecture that there may be dual pairs of such theories. That is,
we wish that a heterotic string when compactified in a specific way on
K3${}\times T^2$ be physically equivalent to a type II string
compactified on a \CY\ threefold. This story began with the papers
of \cite{KV:N=2,FHSV:N=2} and, as we shall see, the full picture is
still to be uncovered. 

There is one immediately apparent curiosity which is associated with
such a conjecture. This is that there are a very large number of
topological classes of \CY\ threefolds. The exact number is not known
since a classification remains elusive. Indeed, one cannot rule out
the possibility that the number is infinite. Contrasted to this are the
few manifolds of K3${}\times T^2$ and its quotients. At first it might
appear that only a tiny fraction of the type II compactifications can
have heterotic partners. This argument is flawed, however, as the
heterotic string requires more data to specify its class than just the
topology of the space on which it is compactified.

The heterotic string in ten dimensions has a gauge group which is either
$E_8\times E_8$ or $\spnh$. Let us consider the $E_8\times
E_8$ string for purposes of discussion. This ``primordial'' gauge
group must be compactified in
addition to the extra dimensions. The generally accepted way to do
this is to take a vector bundle $E\to X$ over the compactification
manifold, $X$, with a structure group contained in $E_8\times
E_8$. The embedding of this structure group into the heterotic
string's 
gauge group then gives a recipe for compactifying the heterotic string
on $X$ including the gauge degrees of freedom.

This is exactly what we were doing in section \ref{ss:tori}. In the
case of a heterotic string we
considered a rank 16 principal $\GU(1)^{16}$-bundle over a torus. The structure
group was embedded as the Cartan subgroup of $E_8\times E_8$ and the
connection on the bundle was specified by the 
parameters of the matrix $A$. The equations of motion demand that $A$
be a constant and so the bundle is flat. The Narain moduli space then
gives the full moduli space of such flat vector bundles on a torus.

In the case of compactification over a more general manifold, one way
of solving the equations of motion \cite{W:issues} is to demand that
the vector 
bundle be {\em holomorphic\/} and that the curvature satisfies
\begin{equation}
  g^{i\bar\jmath}F_{i\bar\jmath}=0.
\end{equation}
One also requires $c_1(E) \in H^2(X,2\Z)$ and that
\begin{equation}
  c_2(E) = c_2(T_X),
\end{equation}
for anomaly cancelation. Clearly the torus fits into this picture if
we replace the principal $\GU(1)^{16}$-bundle by the associated sum of
holomorphic line bundles.

The analysis of this bundle for the case of a heterotic string on a K3
surface is going to be much harder than the toroidal case because now
the bundle {\em cannot\/} be flat as it must satisfy $c_2(E)=24$.%
\footnote{As is common, we assume integration over the base K3 surface
in this notation.} We can see hope then that the large number of choices of
possible \CY\ manifolds for compactification of the type II string
might be matched by the large number of choices of suitable bundles
over K3${}\times T^2$ for the heterotic string compactification.

It will take a fairly long argument before we are able to give an
explicit example of such a pair so we will discuss the situation in
general first. Let us start in our usual way by thinking about the
holonomy of the moduli space. For an $N=2$ theory in four dimensions
the holonomy algebra is $\gu(2)\cong\gu(1)\oplus\sp(1)$. There are two
types of $N=2$ supermultiplet which contain massless scalars:
\begin{enumerate}
\item The vector multiplets each contain 2 real fields
which form a complex object under the $\gu(1)$ holonomy.
\item The hypermultiplets each contain 4 real massless scalars which
transform as a quaternionic object under the $\sp(1)$ holonomy.
\end{enumerate}
Thus, at least away from points where the manifold structure may break
down, we expect the moduli space to be in the form of a product
\begin{equation}
  \cM_{N=2} = \cM_V\times\cM_H,
\end{equation}
where $\cM_V$ is a K\"ahler manifold spanned by moduli in vector
supermultiplets and $\cM_H$ is a quaternionic K\"ahler manifold spanned
by moduli in hypermultiplets. It can be shown \cite{BagW:N=2} that $\cM_H$ is
not hyperk\"ahler.

The effective four-dimensional theory always contains a dilaton-axion
system. This will govern the string-coupling. As it plays such an
important r\^ole we will look first at whether the dilaton-axion lives
in a hypermultiplet or a vector multiplet. One way to do this is
simply to count the dimensions of the moduli space and use the fact
that $\cM_V$ has an even number of real dimensions and $\cM_H$ has a
multiple of four dimensions. For a more direct way of justifying which
kind of supermultiplet the dilaton-axion lives in see \cite{VW:pairs}.

Consider first the type IIA superstring on a \CY\ manifold $X$. First
consider the moduli space of underlying conformal field theories (see,
for example, \cite{me:N2lect} for a full discussion). We have
$h^{1,1}(X)$ complex dimensions of moduli space coming from the
deformations of K\"ahler form and $B$-field and we have $h^{2,1}(X)$
complex dimensions coming from deformations of complex structure. The
R-R sector moduli come from a 3-form giving $b_3(X)=2(h^{2,1}(X)+1)$
real deformations. Finally we have 2 real deformations given by the
dilaton-axion. Given that $h^{1,1}(X)$ and $h^{2,1}(X)$ can be even or
odd, the only way to arrange these deformations in a way consistent
with the dimensionality of the moduli space is to arrange:
\begin{itemize}
   \item There are $h^{1,1}(X)$ vector supermultiplets.
   \item There are $h^{2,1}(X)+1$ hypermultiplets, one of which contains
the dilaton-axion.
\end{itemize}

Next consider the type IIB superstring compactified on a \CY\
manifold, $Y$. The moduli space of conformal field theories is as for
the type IIA string with $h^{1,1}(Y)$ complex deformations of complexified
K\"ahler form and $h^{2,1}$ complex deformations of complex structure.
The R-R moduli consist of one from the 0-form, $b_2(Y)=h^{1,1}(Y)$
from the 2-form, $b_4(Y)=h^{1,1}(Y)$ from the self-dual 4-form and one
more from dualizing the 2-form. Lastly we have two more moduli from
the dilaton-axion. Now we are forced to arrange as follows.
\begin{itemize}
   \item There are $h^{2,1}(Y)$ vector supermultiplets.
   \item There are $h^{1,1}(Y)+1$ hypermultiplets, one of which contains
the dilaton-axion.
\end{itemize}

Note that the type IIA picture and the type IIB picture are related by
an exchange
\begin{equation}
  \eqalign{h^{1,1}(X)&\leftrightarrow h^{2,1}(Y)\cr
  h^{2,1}(X)&\leftrightarrow h^{1,1}(Y).\cr}
\end{equation}
If we have a pair of \CY\ varieties, $X$ and $Y$, such that type IIA
string theory compactified on $X$ is equivalent to type IIB string
theory on $Y$ then we may use this as definition of the statement that
``$X$ and $Y$ are a mirror pair''. See \cite{Mor:MII} for further
discussion of this point.

Lastly we consider the heterotic string compactified on a product of a
K3 surface and a 2-torus. First let us assume that there are no nasty
obstructions in the moduli space and we can count the number of
deformations of the K3 surface, the bundle over the K3 surface, the
2-torus, and the bundle over the 2-torus and simply sum the
result. There are 80 deformations of the K3 surface as far as
conformal field theory is concerned and the resulting moduli space is
a quaternionic K\"ahler manifold. It was shown by Mukai \cite{Muk:symp}
that the moduli space of holomorphic vector bundles over the K3 will
be hyperk\"ahler. Thus it appears very reasonable to expect that the
complete moduli space coming from the K3 surface will be a
quaternionic K\"ahler manifold. Certainly its dimension is a multiple
of four assuming unobstructedness. The moduli space coming from the
torus will be locally of the 
form $\GO(2,m)/(\GO(2)\times\GO(m))$ and has an even number of
dimensions. Thus we assemble the supermultiplets as follows
\begin{itemize}
  \item The vector multiplets come from the 2-torus, together with
its bundle, and the dilaton-axion.
  \item The hypermultiplets are associated to the K3 surface, together with
its bundle.
\end{itemize}
Note that we are assuming we can give simple geometric interpretations
to all the moduli. We will see later that there may be other vector or
hypermultiplet moduli we have not included in the above lists.

\subsection{K3 fibrations}   \label{ss:K3f}

Suppose we are able to find a pair of theories, one a type II string
on a \CY\ manifold and the other a heterotic string theory
compactified over some bundle on K3${}\times T^2$. The first thing one
would do would be to line up the moduli spaces of the two theories so
that the parameters of one theory could be understood in terms of the
other. We begin by analyzing what would happen in the case of the
vector multiplet moduli space. The is the first step required before
we can actually propose a dual pair of such theories. We now follow an
argument first presented in \cite{AL:ubiq}.

For definiteness we choose a type IIA string rather than a type
IIB. The reason for this is that we will ultimately be able to tie our
analysis to proposition \ref{prop:1}, which was also phrased in terms
of the type IIA string.

The holonomy argument leads to a factorization of the moduli space,
which in turn has another consequence given the fact that all the matter
fields are related to the moduli by supersymmetry. This is that the
couplings between fields in the vector multiplets can only depend on
the moduli from the vector multiplets and the couplings between fields
in the hypermultiplets can only be affected by the moduli from the
hypermultiplets. This may also be deduced directly from the
supergravity Lagrangian \cite{deW:prod}.

\def\barcF{\overline{\cF}}
Let us consider just the moduli space, $\cM_V$, coming from the
scalars in the vector multiplets. We know this is a complex K\"ahler manifold.
Further analysis of the supergravity Lagrangian puts more
constraints on the geometry of the moduli space
\cite{dWvP:spec}. A manifold satisfying these extra conditions is
called ``special K\"ahler''.\footnote{The quaternionic K\"ahler
manifold parametrized by the hypermultiplets is also subject to extra
constraints \cite{CFG:II}.} The main importance of special K\"ahler
geometry is the fact that all the information we require about the
theory is encoded in a single holomorphic function $\cF$ on the moduli
space. If 
we use specific complex coordinates, the ``special coordinates'', on the
moduli space, the metric is of the form
\begin{equation}
\eqalign{K&=-\log\left(2(\cF+\barcF)-(q^i-\bar q^{\bar\imath})
  \left(\frac{\partial\cF}{\partial q^i}-\frac{\partial\barcF}
  {\partial \bar q^{\bar\imath}}\right)\right)\cr
  g_{i\bar\jmath}&=\frac{\partial K}{\partial q^i\partial \bar
  q^{\bar\jmath}}.\cr}
\end{equation}

When viewed from the point of view of the heterotic string, we expect
the dilaton-axion to be contained in this moduli space. Let us suppose
for the time being that all the other moduli can be understood from the
world-sheet perspective of the heterotic string. This should mean that
we have a moduli space of conformal field theories spanning all but
one of the complex directions in the moduli space with the extra
dimension being given by the dilaton-axion system. In the limit that
the string coupling becomes very small, i.e., the dilaton approaches
$-\infty$, we expect that the moduli space as described by the
conformal field theory becomes exact. Thus, in the limit of small
dilaton, the moduli space should factorize into a product of the
moduli space of conformal field theories, and the extra bit spanned by
the dilaton-axion. In \cite{FvP:Ka} precisely this problem was
analyzed. It was discovered that the only way a special K\"ahler
manifold could factorize was if it became locally a product of the form
\begin{equation}
  \frac{\GO(2,m)}{\GO(2)\times\GO(m)}\times\frac{\Sl(2)}{\GU(1)}.
	\label{eq:sKsplit}
\end{equation}
This of course is excellent news. The first term in (\ref{eq:sKsplit})
looks suspiciously like the Narain moduli space for a 2-torus and the
second term looks like a dilaton-axion. This is exactly what we
wanted. Note that the form (\ref{eq:sKsplit}) is only expected in the
limit that the dilaton approaches $-\infty$. Away from this limit we
expect the two factors to begin to interfere with each other.

Now we want to carry this information over to the type IIA
compactification on the \CY\ manifold, $X$. All of the vector
multiplet moduli are expected to be associated to deformations of the
K\"ahler form and $B$-field on $X$. Let us fix some notation. In
contrast to the K3 surface, degrees of freedom of the K\"ahler form
and the $B$-field for the \CY\ threefold, $X$, can be nicely
paired-up. Introduce a basis of divisors, or 4-cycles, $\{D_k\}$,
spanning $H_4(X,\Z)$, where $k=0,\ldots, h^{1,1}(X)-1$. Dual to the dual
of these divisors we have a basis of 2-forms, $\{e_k\}$, generating
$H^2(X,\Z)$.\footnote{For simplicity let us assume all cohomology is
torsion-free.}
Expand out the K\"ahler form and $B$-field as 
\begin{equation}
  B+iJ = \sum_{k=0}^{h^{1,1}-1}(B_k+iJ_k)e_k,
\end{equation}
for real numbers $B_k,J_k$. We will take $e_0$ to correspond to the
generator associated to the direction in moduli space given by the
heterotic dilaton-axion.

Our information for the heterotic side is in terms of the local form of
the moduli space. Thanks to special K\"ahler geometry we can translate
this into information concerning couplings between certain fields. The
fields we are interested in are the superpartners of the moduli of the
vector superfields --- i.e., the gauginos, the vector bosons and the
moduli themselves. One may consider couplings in the effective action
of the form of ``Yukawa couplings'', i.e., $\kappa_{ijk}=\langle
a_i\psi_j\psi_k\rangle$, or other terms equivalent by
supersymmetry. It was shown in \cite{SW:coup} that, to leading order
in the \nlsm, the coupling between three fields is given by
\begin{equation}
  \kappa_{ijk} = \#(D_i\cap D_j\cap D_k),  \label{eq:trint}
\end{equation}
where the $D$'s are the divisors in $X$ associated to the fields. The
reader is also referred to \cite{DG:exact} for an account of this.%
\footnote{Note that \cite{DG:exact} explicitly refers to a heterotic
string compactified on a \CY\ manifold whereas we are considering a
type IIA string. Most of the calculations are unaffected however.}
It is also known from special K\"ahler geometry that
\begin{equation}
  \kappa_{ijk} = \frac{\partial\cF}{\partial q_i\partial q_j\partial q_k}.
\end{equation}

We now have some approximate knowledge about both the heterotic string
and the type IIA string. In the case of the heterotic string we know
that, in the small dilaton limit, the moduli space factorizes in the
form (\ref{eq:sKsplit}) and in the case of the type IIA string, we
know that, in the small $\alpha'/R^2$ limit, the couplings are of the
form (\ref{eq:trint}). We can make a useful statement about a
heterotic-type II dual pair if both of these approximations happen to
be simultaneously true. Note that, for the heterotic string, the
dilaton lies in a vector multiplet and that, in the type IIA string,
the size parameters lie in vector multiplets. Thus we wish to assert
that the moduli spaces of the theories are aligned in the right way so
that as the dilaton in the heterotic string approaches $-\infty$, some
size in the \CY\ space on which the type IIA string is compactified is
becoming very large.

To picture this presumed aligning of the moduli spaces it is best to
picture exactly what makes corrections to the approximations we are
considering. In the case of the heterotic string, corrections arise
from instantons in the Seiberg-Witten theory \cite{SW:I}. The action
of such an instanton becomes very large, and hence the contribution to
any physical quantity becomes very small, when the dilaton becomes
close to $-\infty$. In the case of the \nlsm, corrections come from
world-sheet instantons \cite{DSWW:}. A world-sheet instanton takes the
form of a holomorphic map from the world-sheet into the target
space. We will be interested only in tree-level effects and so as far
as we are considered {\em world-sheet instantons are rational curves}.
The action for such an instanton is simply the area of the
curve. Thus the effect of an instanton gets weaker as the K\"ahler
form is varied so as to make the rational curve bigger.

The picture one should have mind therefore is that as the heterotic
dilaton is decreased down to $-\infty$, some rational curve (or some
set or family of rational curves) is getting bigger. The important
thing is that no curve should shrink down during this process.

Let us assume we are now at the edge of our moduli space where,
in the type IIA interpretation, all of the rational curves are very
large compared to $\alpha'$. Thus, by assumption, the heterotic
string's dilaton is close to $-\infty$. We can take the moduli space
given by (\ref{eq:sKsplit}) and deduce the form of $\cF$. We may then
translate this into a statement about $\kappa_{ijk}$ and thus about
the topology of $X$ from (\ref{eq:trint}). The result is that \cite{FvP:Ka}
\begin{equation}
  \eqalign{
  \#(D_0\cap D_0\cap D_0)&=0\cr
  \#(D_0\cap D_0\cap D_i)&=0,\quad i=1,\ldots,h^{1,1}-1\cr
  \#(D_0\cap D_i\cap D_j)&=\eta_{ij},\quad i,j=1\ldots,h^{1,1}-1,\cr}
		\label{eq:fib1}
\end{equation}
where $\eta_{ij}$ is a matrix of nonzero determinant and signature 
$(+,-,-,\ldots,-)$.

Suppose that $X$ is such that there is some smooth complex surface
embedded in $X$ whose class, as a divisor, is $D_0$. From
(\ref{eq:fib1}) we see that $D_0\cap D_0=0$ and so the normal bundle
for this surface is trivial. It then follows from the adjunction
formula, and the fact that $X$ is a \CY\ space, that the tangent
bundle for this surface has trivial $c_1$. Thus, the surface
representing $D_0$ must either be a K3 surface or an algebraic 4-torus (also
known as an ``abelian surface''). The fact
that the normal bundle is trivial also suggests that the K3 or abelian
surface can be ``moved'' parallel to itself to sweep out the entire
space $X$. That is, $X$ is a fibration where the generic fibre is
either a K3 surface or an abelian surface.

We can make the above more rigorous by appealing to a theorem by
Oguiso \cite{Og:K3f} which states that
\begin{theorem}
Let $X$ be a minimal \CY\ threefold. Let $D$ be a nef divisor on
$X$. If the numerical $D$-dimension of $D$ equals one
then there is a fibration $\Phi:X\to W$, where $W$ is $\P^1$ and the
generic fibre is either a K3 surface or an abelian surface.
\end{theorem}
The numerical $D$-dimension of a divisor is the maximal number of times it may
be intersected with itself to produce something nontrivial. We see
above that $D_0$ has $D$-dimension equal to one. The statement that
$D$ is ``nef'' is the assertion that for any algebraic curve, $C\subset
X$, we have that
\begin{equation}
  \#(D\cap C)\geq0, \quad\forall C\subset X.
     \label{eq:nef}
\end{equation}
We can come very close to proving that $D_0$ is nef following our
assumption about the way that moduli spaces are aligned. The special
coordinates on a our moduli space of type IIA string theories may be
written in the form
\begin{equation}
    q_k = e^{2\pi i(B_k+iJ_k)}.
\end{equation}
We are in an area of moduli space where all $q_k\ll1$.
The contribution of a curve, $C$, to the instanton sum will then scale
roughly as
\begin{equation}
  \prod_{k=1}^{h^{1,1}} q_k^{\#(D_k\cap C)}.   \label{eq:mon}
\end{equation}
Clearly, if $C$ is not nef then negative powers of $q_k$ appear and
the instanton sum will fail to converge, in contradiction to
expectation.

\begin{difficult} 
The above argument that $C$ is nef is not actually complete. When one
computes the instanton sum, one also has to compute the coefficient in
front of the monomial of the form (\ref{eq:mon}). This can be done
using the methods discussed in \cite{W:matrix,AM:rat}. In the simple
case of an isolated curve, the coefficient is simply one (although
extra contributions arise from multiple covers). $C$ may not always be
isolated however. One may be able to deform $C$ into a whole family of
rational curves. In this case the coefficient might be zero. One
cannot then rule out that $\#(D\cap C)<0$ as the field theory is
simply unaffected by $C$.

An example of a case where rational curves don't count in this way is
that of $N=4$ theories. Compactifying the type IIA string on
K3${}\times T^2$, there are no instanton corrections but the K3
surface may contain rational curves. Each rational curve will clearly
be in the form of a family $C\times T^2$ inside K3${}\times T^2$. An
algebraic surface of the form $\P^1\times T^2$ is known as an
``elliptic scroll''.\footnote{An ``elliptic curve'' is the algebraic
geometer's name for an algebraic curve of genus 1.} Since the
calculation in \cite{W:matrix,AM:rat} 
is essentially local we may generalize this result to any $N=2$ \CY\
compactification. That is we claim that {\em rational curves in an
elliptic scroll do not contribute to the instanton sum}. Note that
rational curves in a K3 surface are unstable in the sense that a
generic deformation of complex structure of the K3 surface will kill
them. The work of Wilson \cite{Wil:Kc} showed that essentially the same
is true for elliptic scrolls in \CY\ manifolds. Thus, if we choose a
generic complex structure on the \CY\ manifold then there will be no
curves lying in an elliptic scroll. 

If we assume that the only curves contributing zero to the instanton
sum lie in an elliptic scroll then we may complete our proof that
$D_0$ is nef by choosing a generic complex structure. We do not know
if this assumption is valid but it seems reasonable.
\end{difficult}

Note that the fibration $\Phi:X\to W$ is allowed to ``degenerate'' at a
finite number of points in $W$. Indeed, if this were not the case then
$X$ could not be a \CY\ variety. Note that the degenerate fibre need
not be some degenerate limit of a K3 surface. It could be a perfectly
smooth manifold (which is neither a K3 or abelian surface).

Finally we would like to know if the generic fibre is a K3 surface or
an abelian surface. We can determine this by finding $c_2$, and hence
the Euler characteristic, of the
fibre. This may be determined by using the {\em holomorphic anomaly\/} of
\cite{BCOV:ell}. The result is that the Euler characteristic of the
generic fibre is 24 
and so the fibration is of the K3 type. We refer the reader to
\cite{AL:ubiq} for details of this calculation.

We thus arrive at the following:
\begin{prop}
Given a dual pair of theories, one of which is a heterotic string
compactified on K3${}\times T^2$ (or some free quotient thereof) and
the other is a type IIA string compactified on a \CY\ manifold $X$,
then if there is a region of moduli space in which both the respective
perturbation theories converge, then $X$ must be of the form of a
K3-fibration over $\P^1$.  \label{prop:K3f}
\end{prop}
This proposition depends upon our statement about zero contributions
from rational curves only arising from elliptic scrolls. The fact that
the base of the fibration is a $\P^1$ may be deduced from the fact
that $H^1$ of the base injects\footnote{This follows from the Leray
spectral sequence for a fibration.} into the total cohomology of $X$ and
that $X$ has $b_1=0$.

Not only do we know that $X$ is a K3 fibration, but we also know that
the divisor $D_0$ is the generic K3 fibre. This tells us immediately
which deformation of $X$ corresponds to a deformation of the heterotic
dilaton --- it is the component of the K\"ahler form that
affects the area of the curve dual to $D_0$.

Let us first assume that the K3-fibration of $X$ has a global
section. This means that as well as the fibration map:
\begin{equation}
 \Phi:X\to W,
\end{equation}
we also have a holomorphic embedding 
\begin{equation}
 \gamma:W\to X.
\end{equation}
The image of this embedding is a rational curve in $X$. Clearly this
curve is dual to $D_0$. Thus the value of the dilaton in the heterotic
string is given by the area of this rational curve. The instanton sums
converge as the dilaton approaches $-\infty$ and as this rational curve
gets very large. Thus the weakly-coupled limit of the heterotic string
is given by the limit of $X$ in which the base $\P^1$ swells up to
infinite size. 

Note that if the K3 fibration does not have a global section, but
rather a multi-valued section, we may play a similar game. In this case
the multi-section will not define an embedding of the base $\P^1$ into
$X$ but rather an embedding of a multiple cover of $X$. Thus, the
algebraic curve within $X$ whose area gives the heterotic dilaton will
be of genus greater than zero.

One should note that what we have discussed here for the $N=2$ dual
pairs is actually very
similar to the $N=4$ case discussed above. In that case we had a type
IIA string compactified on K3${}\times T^2$ or a quotient
thereof. This space may be viewed as a smooth K3-fibration over $T^2$
(that is, an elliptic curve). This fibration is trivial in the case
that the target space is K3${}\times T^2$ but becomes nontrivial when
we take a free quotient. We saw in section \ref{ss:N=4} that the
dilaton for the heterotic string is given by the area of the $T^2$ ---
i.e., the size of the base of the fibration. 

The general picture we see is the following. If we have a
heterotic-type IIA dual pair in four dimensions, we may expect that
the type IIA string is compactified on
\begin{itemize}
  \item a K3-fibration over a 2-torus in the case of $N=4$
supersymmetry, or,
  \item a K3-fibration over a $\P^1$ in the case of $N=2$
supersymmetry.
\end{itemize}
In either case, the area of the base of the fibration gives the value of the
heterotic dilaton.

The link to proposition \ref{prop:1} is clear in the $N=4$ case. For
the $N=2$ case we should note that the K3 surface on which the
heterotic string is compactified might be viewable as an elliptic
fibration itself --- that is, a fibration over $\P^1$ with generic
fibre given by a $T^2$. Thus the K3${}\times T^2$ (or quotient
thereof) upon which the heterotic string is compactified may be viewed
as an ``abelian fibration'' over $\P^1$ --- i.e., a fibration over
$\P^1$ with generic fibre given by a $T^4$. When viewed in this way
the relationship between the $N=2$ theories in four dimensions given
by the heterotic string and the type IIA string may be viewed as a {\em
fibre-wise\/} application of the duality of proposition \ref{prop:1}.

Such ``fibre-wise duality'', which was first suggested in
\cite{VW:pairs}, is a potentially very powerful tool. It has been
extended to fibre-wise mirror symmetry in \cite{OV:Db} and has recently
been applied to the problem of mirror symmetry itself in
\cite{SYZ:mir}. Both of these developments deserve to be covered here
in some detail since they both have direct relevance to K3
surfaces, but we do not have time to do so. We refer the reader to
\cite{GW:SYZ,Mor:SYZ} for more details of the latter in the context of
K3 surfaces.

Let us note that the general appearance of
K3-fibrations in the area of heterotic-type II duality was first noted
in \cite{KLM:K3f} where some naturalness arguments based on the work
of \cite{LY:K3f} were presented. It is interesting to note that in
this context it was really the type IIB string that being studied
rather than the type IIA. This raises the question as to whether the
mirror of a K3 fibration is another K3 fibration, which again raises
the possibility of some fibre-wise duality argument.

\subsection{More enhanced gauge symmetries}  \label{ss:N2enh}

So far we have identified one of the moduli lying in a vector
multiplet. This is the dilaton-axion in the heterotic string, or the
component of the complexified K\"ahler form associated to the base of
the K3-fibration on which the type IIA string is compactified. Now we
wish to turn our attention to some of the other vector multiplet
moduli. 

Let us first look at things from the type IIA perspective. We know that
our \CY\ space, $X$, is a K3-fibration and we wish to analyze elements
of $H^2(X)$, or equivalently, $H_4(X)$. In general, given a fibration
$X\to W$, the cohomology of 
$X$ may be determined from the cohomology of the base together with
the cohomology of the fibre. The mechanism by which this happens is
called a ``spectral sequence''. We do not wish to get involved with the
technicalities of spectral sequences here and refer the reader to 
\cite{McL:ss} for the general idea. 

\iffigs
\begin{figure}
  \centerline{\epsfxsize=10cm\epsfbox{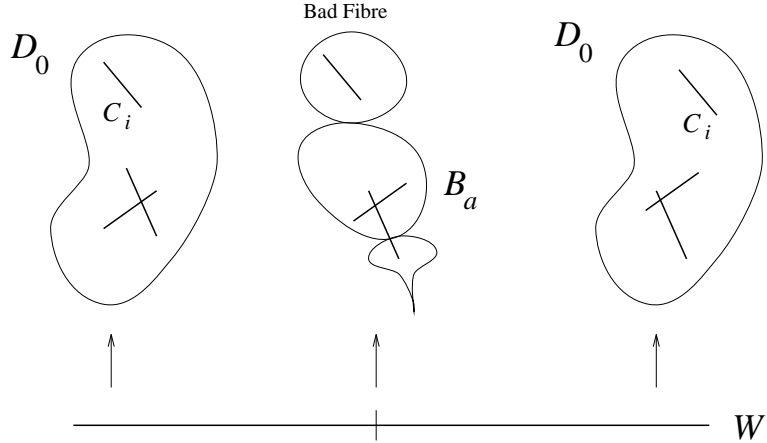}}
  \caption{$X$ as a K3 fibration over $W$.}
  \label{fig:fib}
\end{figure}
\fi

The result is that the contributions to $H_4(X)$ are as follows
(see figure \ref{fig:fib}):
\begin{enumerate}
\item The generic fibre, $D_0$, will generate an element of $H_4(X)$.
\item Take an algebraic 2-cycle in the fibre, i.e., an element of
$H_2(D_0)$ and 
use it to ``sweep out'' a 4-cycle in $X$ by transporting it around the
base, $W$. Note that this 2-cycle needs to be monodromy invariant for
this to make sense. Thus the 2-cycle might be an irreducible curve in
$D_0$ which is monodromy invariant or it may be the sum of two curves
which are interchanged under monodromy, etc.
\item When we have a bad fibre which is a reducible divisor in $X$, we
may vary the volumes of the components of this bad fibre
independently. Thus such fibres will contribute extra pieces to
$H_4(X)$.
\end{enumerate}

The second class above is clearly generated by elements of the Picard
group of the generic fibre. A monodromy-invariant element of the group
$C_i$ will sweep out a divisor $D_i$, where $C_i=D_0\cap D_i$. Thus
\begin{equation}
  \eqalign{\#(D_0\cap D_i\cap D_j)_X &= 
           \#((D_0\cap D_i)\cap(D_0\cap D_j))_{D_0}\cr
	&= \#(C_i\cap C_j)_{D_0},\cr}
\end{equation}
where the subscript denotes the space within which we are considering
the intersection theory. This agrees very nicely with our earlier
result of (\ref{eq:fib1}). We see that $\eta_{ij}$ is simply the
natural inner product of the monodromy-invariant part of the Picard
lattice of the generic fibre. As we mentioned in section \ref{ss:alg},
the Picard lattice has signature $(+,-,-,\ldots,-)$. When we take the
monodromy-invariant part we retain the positive eigenvalue as the
K\"ahler form restricted to the generic K3 fibre must be monodromy
invariant and the generic fibre has positive volume. Thus $\eta_{ij}$
has the correct signature.

Now consider the third class. We denote such an element by
$B_a$. Since this class is supported away from the generic fibre, we
have $D_0\cap B_a=0$. Now comparing to (\ref{eq:fib1}) we are in
trouble. The moduli coming from such vector supermultiplets cannot
live in the space (\ref{eq:sKsplit}) we expected from the heterotic
string. What can have gone wrong? The only place our argument was
flawed was in the {\em perturbative\/} analysis of the heterotic string. It
turns out that the classes $B_a$ cannot be understood perturbatively from
the perspective of the heterotic string. We will encounter such
objects later in section \ref{ss:ell} but in this section we will
restrict our attention to perturbative questions.

We now know exactly how to interpret the Teichm\"uller space
\begin{equation}
  \frac{\GO(2,m)}{\GO(2)\times\GO(m)}\times\frac{\Sl(2)}{\GU(1)},
	\label{eq:sKsplit2}
\end{equation}
both in terms of the heterotic string compactified on K3${}\times T^2$
and its supposed type IIA dual compactified on $X$. The second term
comes from the heterotic dilaton-axion and the complexified K\"ahler
form associated to the class $D_0\in H_4(X)$. The first term is
considered to be associated to the Narain moduli space of the
heterotic string on $T^2$, or, from what we have just said, the
complexified K\"ahler form on the (monodromy-invariant part of) the K3
fibre. We computed the stringy moduli space of the K\"ahler-form and
$B$-field on an algebraic K3 surface earlier and obtained the second
term of (\ref{eq:Tasig}). Fortunately it turns out to have just the right
form! 

Note that we have reduced our problem once again to a comparison of a
heterotic string on a torus (this time a 2-torus) with a type IIA
string on a K3 surface (this time an algebraic fibre in $X$). We
should be able to use the same old arguments we used in sections 
\ref{ss:enh} and \ref{ss:mN=4} to obtain the enhanced gauge group. 

Let us first assume there is no monodromy acting on the Picard group
of the K3 fibre for $X$.
We expect the slice of the moduli space coming from varying the
complexified K\"ahler form on the generic K3 fibre to be of the form
\begin{equation}
   \GO(\Upsilon)\backslash \GO(2,\rho)/(\GO(2)\times\GO(\rho)),
	\label{eq:N2Nr}
\end{equation}
where $\Upsilon$ is the quantum Picard lattice introduced in
(\ref{eq:Ups}). 
Identifying this with the Narain moduli space of $T^2$ that we expect to
see in the weak-coupling limit of the heterotic string, we see that
the Narain lattice for the $T^2$ is isomorphic to $\Upsilon$.%
\footnote{Note that there is no reason to expect that $\Upsilon$
should be self-dual. In many proposed dual pairs (e.g., some of those of
\cite{KV:N=2}) it is not self-dual. On the heterotic side this says
that the heterotic string on $T^2$ is not modular invariant. We shall
assume that modular invariance is satisfied once the K3 factor is
taken into account too. In general one might worry that strange
effects such as those encountered in \cite{FHSV:N=2} might cause
problems with this. We will assume here that modular invariance looks
after itself.}
That is, we have a gauge bundle of rank $\rho-2$ compactified over
$T^2$.

To obtain the enhanced gauge groups we take (\ref{eq:N2Nr}) to be the
Grassmannian of space-like 2-planes in $\R^{2,\rho}$ and look for
points where the 2-plane becomes orthogonal to roots in the lattice
$\Upsilon$. The roots will give the root diagram of the enhanced gauge
group in the usual way.

Let us clarify all this general discussion with an example taken from
\cite{KV:N=2}. Let us
take $X_0$ to be the hypersurface
\begin{equation}
  z_1^2+z_2^3+z_3^{12}+z_4^{24}+z_5^{24}=0,
		\label{eq:KV2}
\end{equation}
in the weighted projective space $\P^4_{\{12,8,2,1,1\}}$. The weighted
projective space contains various quotient singularities. We will blow
these up and take the proper transform of $X_0$ to be $X$. There is a
$\Z_2$-quotient singularity along the locus $[z_1,z_2,z_3,0,0]$. We
may blow this up, replacing each point in the locus by $\P^1$. The
projection of $X$ onto to this $\P^1$ will be the K3-fibration. That
is, roughly speaking, we view $[z_4,z_5]$ as the homogeneous
coordinates of the base $W\cong\P^1$. To find the fibre fix a point in
the base by fixing $z_4/z_5$. This projects $\P^4_{\{12,8,2,1,1\}}$
onto the subspace $\P^3_{\{12,8,2,1\}}$. Now $\P^3_{\{12,8,2,1\}}$ may
be viewed as a $\Z_2$-quotient of $\P^3_{\{6,4,1,1\}}$ by taking
$z_4\mapsto-z_4$. Such codimension quotients are equivalent to
reparametrizations in complex geometry and so the fibre may be taken
to be the hypersurface
\begin{equation}
  z_1^2+z_2^3+z_3^{12}+z_4^{12}=0,
\end{equation}
in the weighted projective space $\P^3_{\{6,4,1,1\}}$. This is a K3
surface as expected. Note that we still have a $\Z_2$-quotient
singularity in the fibre along $[z_1,z_2,0,0]$. This intersects the
generic K3 fibre at a single point. Thus for a smooth $X$ we blow-up
again to introduce a single $(-2)$-curve into each generic fibre.

Now let us work out the Picard lattice of the generic fibre. A generic
hyperplane in $\P^3_{\{6,4,1,1\}}$ may be written as $az_3+bz_4=0$ for
some $a,b\in\C$. Any two such hyperplanes will intersect at the point
$[z_1,z_2,0,0]$ but this is exactly where we are blowing up. Thus the
hyperplane doesn't intersect itself at all but will intersect the
$(-2)$-curve (i.e., the exceptional divisor) once. The Picard lattice,
for a generic value of complex structure of $X$, will be generated by
the hyperplane and the single $(-2)$-curve and has intersection matrix
\begin{equation}
  \left(\begin{array}{cc}0&\phantom{-}1\\1&-2\end{array}\right).
\end{equation}
A simple change of basis shows that this is $\Gamma_{1,1}\cong U$. Note
that neither the hyperplane nor the exceptional $(-2)$-curve are
effected by monodromy around $W$ and so we needn't worry about
monodromy invariance in this case.

The bad fibres occur when we fix a point on the base $W$ such that
$z_4^{24}+z_5^{24}=0$. Thus, at 24 points, we have fibres
\begin{equation}
  z_1^2+z_2^3+z_3^{12}=0,
\end{equation}
in $\P^3_{\{6,4,1,1\}}$. Although singular, this equation does not
factorize and so the bad fibres are irreducible. Thus there are no
contributions to $h^{1,1}$ from bad fibres. Thus, $h^{1,1}=1+2=3$
since we have the generic fibre itself together with $\rho=2$.
 
Now $\Upsilon=U\oplus U\cong\Gamma_{2,2}$ and the moduli space of
vector multiplets coming from the quantum Picard lattice will be
\begin{equation}
  \GO(\Gamma_{2,2})\backslash\GO(2,2)/(\GO(2)\times\GO(2)).
		\label{eq:O22}
\end{equation}
This is exactly the moduli space for a string on $T^2$. Thus if the
type IIA string compactified on $X$ is dual to a heterotic string on 
K3${}\times T^2$ then the $T^2$ part of the latter has none of the
gauge group from the ten-dimensional string wound around it. All must
be wound around the K3 factor.

The lattice $\Gamma_{2,2}$ contains root diagrams for
$\su(2)\oplus\su(2)$ and $\su(3)$ and so we should be able to obtain
these gauge symmetries for suitable choices of vector moduli. Clearly the
$(-2)$-curve in the generic K3 fibre may be shrunk down to a point to
obtain $\SU(2)$. To obtain more gauge symmetry one must shrink the K3
fibre itself down to a size of order $(\alpha')^2$ as discussed in
section \ref{ss:enh}.

There is an important point we should note in general about enhanced
gauge groups. It is known that in $N=2$ theories quantum effects may
break the gauge group down to a Cartan subgroup of the classical
gauge group. Exactly how this happens depends on whether any extra
hypermultiplets are becoming massive when the point of classical
enhanced gauge symmetry occurs. We don't want to discuss hypermultiplets
yet but, at least in simple cases, there are usually a small number, if
any, of such massless particles and the theory will be
``asymptotically free''. In this case the gauge group will be broken.

Thus we only really expect the nonabelian gauge group to appear in the
heterotic string for the case that the coupling tends to zero. In the
type IIA picture this corresponds to the base $\P^1$ blowing up to
infinite size. This means that we are ``decompactifying'' the type IIA
picture so that it is compactified, not on $X$, but on the
generic fibre --- a K3 surface. In this respect we really are saying
little more than we already said in section \ref{ss:enh}. We can
obtain enhanced gauge symmetries when we have a theory in six dimensions
from a type IIA string compactified on a certain K3 surface. This is
approximately true in four dimensions so long as the base $\P^1$ is
very large.

For the heterotic string, the gauge group is broken in the quantum
theory by Yang-Mills instantons. For the type IIA string we note that
world-sheet instantons wrapping around the base $\P^1$ presumably play
an analogous r\^ole. This is another example of one string's target
space field theory being another string's world-sheet field theory as
we saw in section \ref{ss:N=4}. It would be interesting to see if
these instantons can be explicitly mapped to each other.

To determine the gauge group we should say what happens when there is
monodromy in the Picard lattice of the generic K3 fibre as we move
about $W$. There really is no difference between this and the $N=4$
analogue we discussed in section \ref{ss:mN=4}. The monodromy of the
Picard lattice should be translated into an action on the heterotic
string on $T^2$ and divided out. Thus we expect an {\em asymmetric
orbifold\/} of $T^2$ for the heterotic dual. When finding the enhanced
gauge group we should take the monodromy-invariant subdiagram of the
root diagram. This may lead to non-simply-laced gauge groups again.

One should also worry about the global form of the gauge
group. That is, one may have the simply connected form of the group or
one may have to mod out by part of the center, e.g., the gauge group
might be $\SU(2)$ or $\SO(3)$. All we have said above
is only really enough to determine the {\em algebra\/} of the gauge
symmetry. We will evade this issue where possible in these lectures by
only specifying gauge algebras rather than gauge groups. There will be
times later on in these lectures when we have to confront this problem, however.

Let us mention here that compactifying the type IIB string, rather
than the type IIA string, on a K3 fibration, can lead to a very direct
link between the geometry of the \CY\ threefold and Seiberg-Witten
theory as explored in \cite{KKL:limit,KLM:hSW}. Therefore an
understanding of mirror symmetry within K3 fibrations may shed
considerable light on some of the details of string duality.

Finally let us note that further analysis may be done on the moduli
space of vector multiplets to check that string duality is working as
expected. We refer the reader to \cite{KLT:limit,AGNT:dF} for examples
and especially to \cite{HM:alg} where further direct links to geometry
were established.

\subsection {Heterotic-heterotic duality}   \label{ss:hh}

We have seen how a type IIA string compactified on a \CY\ space with a
K3 fibration may have as a dual partner a heterotic string
compactified on K3${}\times T^2$. An obvious question springs
immediately to mind. What happens if $X$ can admit more than one K3
fibration? One might expect it may be dual to more than one heterotic
string. This implies that we can find pairs of heterotic strings that
are dual to each other. We will follow this line of logic to analyze a
case introduced in \cite{DMW:hh}. This geometric approach was
discussed originally in \cite{AG:mulK3,MV:F}.

We are going to consider the example we introduced in the last section
based on the hypersurface in $\P^4_{\{12,8,2,1,1\}}$ given by
(\ref{eq:KV2}). In the last section we projected onto the last 2
coordinates of the $\P^4_{\{12,8,2,1,1\}}$ to obtain a K3
fibration over $\P^1$. Now let us project into the last 3
coordinates. Our base space will now be $\P^2_{\{2,1,1\}}$. The fibre
will be an algebraic 2-torus, that is, an elliptic curve. Thus $X$ may
be considered as an elliptic fibration as well as a K3 fibration.

The space $\P^2_{\{2,1,1\}}$ is singular. Writing the homogeneous
coordinates as $[x_0,x_1,x_2]$, there is a $\Z_2$ quotient singularity
at the point $[1,0,0]$. This may be blown up, introducing a
$(-2)$-curve. This exceptional curve provided the base of the K3
fibration in the last section. We may also use it to write the blow-up
of $\P^2_{\{2,1,1\}}$ as a fibration. From the same argument as we
used in the last section, it is straight forward to see that the fibre
will be $\P^1$. Thus our blown-up $\P^2_{\{2,1,1\}}$ is a fibration with base
space given by $\P^1$ and with fibre given by $\P^1$. Note that the
fibre never degenerates. Such complex surfaces are called ``Hirzebruch
surfaces''. These objects will turn out to be important for the
analysis of the heterotic string on a K3 surface so we discuss the
geometry here in some detail.
%sections?
\def\HS#1{{\bf F}_{#1}}

\iffigs
\begin{figure}
  \centerline{\epsfxsize=10cm\epsfbox{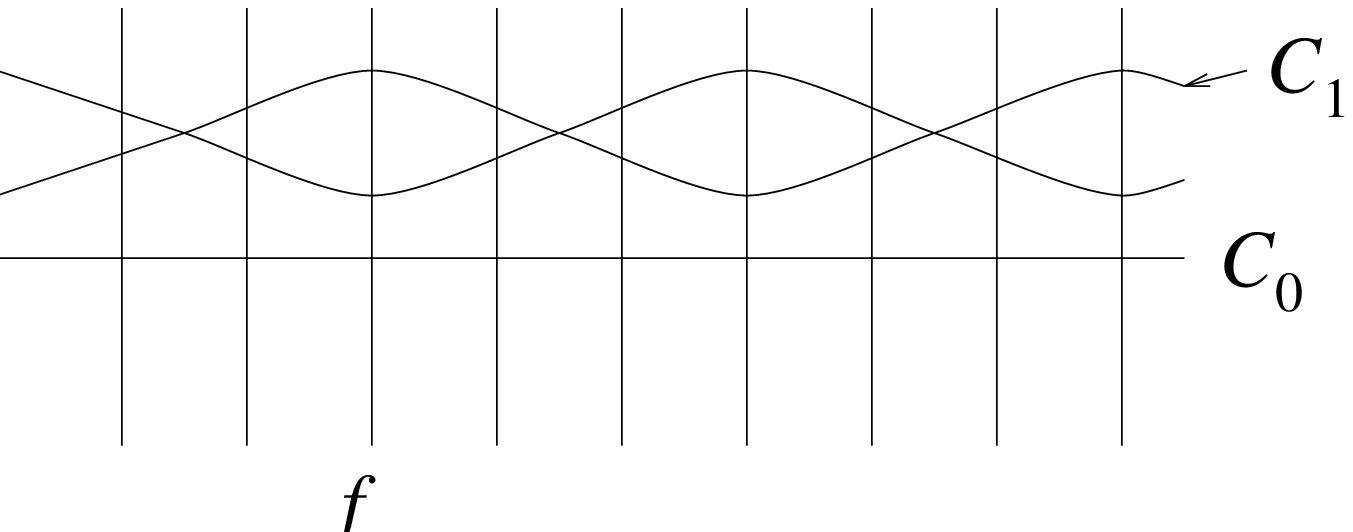}}
  \caption{The Hirzebruch surface $\HS n$.}
  \label{fig:HS}
\end{figure}
\fi

A $\P^1$ bundle over $\P^1$ may be regarded as the compactification of
a complex line bundle over $\P^1$ by adding a point to each fibre ``at
infinity''. Such line bundles are classified by an integer --- the
first Chern class of the bundle integrated over the base $\P^1$. We
use the notation $\HS n$, to denote the Hirzebruch surface built from
the line bundle with $c_1=-n$. Assume first that $n\geq 0$. Denote the
base rational curve by 
$C_0$. The line bundle with $c_1=-n$ represents the normal bundle to
$C_0$ and so the self-intersection of $C_0$ equals $-n$. Thus $C_0$ is
isolated (assuming $n>0$). Denote the
fibre by $f$. Clearly $\#(f\cap f)=0$ and $\#(f\cap C_0)=1$. We may
introduce a class $C_1=C_0+nf$. This intersects $f$ once and $C_0$ not
at all. $C_1$ is a section of the bundle, and hence a rational
curve, away from the isolated section $C_0$. The self intersection of
$C_1$ is $+n$. Note that $C_1$ can be deformed. Thus we view $\HS n$
as a $\P^1$ bundle over $\P^1$ with one isolated section, of
self-intersection $-n$, and a family of sections, disjoint from the
isolated section of self-intersection $+n$. This picture of a
Hirzebruch surface, with two sections in the class $C_1$ shown, is
drawn in figure  
\ref{fig:HS}. If $n=0$ this picture degenerates and we simply have
$\HS 0\cong\P^1\times\P^1$. 

Note that if $n<0$ we may exchange the r\^oles of $C_0$ and $C_1$ and
recover the surface $\HS{-n}$. Thus we may assume $n\geq0$.

For the case we are concerned with in this section $C_0$ is the
exceptional divisor with self-intersection $-2$ and so we have $\HS2$
as the base of $X$ as an elliptic fibration. It turns out however that
$\HS2$ is somewhat ``unstable'' in this context. Note that one may
embed $\P^2_{\{2,1,1\}}$ into $\P^3$ as follows. Denote the homogeneous
coordinates by $[x_0,x_1,x_2]$ and $[y_0,y_1,y_2,y_3]$
respectively. By putting 
\begin{equation}
  \eqalign{y_0 &= x_0\cr y_1 &= x_1^2\cr y_2 &= x_2^2\cr
   y_3 &= x_1x_2,\cr}
\end{equation}
we embed $\P^2_{\{2,1,1\}}$ as the hypersurface
\begin{equation}
  y_1y_2-y_3^2=0.
\end{equation}
This is singular, as expected, but we may deform this hypersurface to
a generic quadric. It is well-known that a generic quadric in $\P^3$
is isomorphic to $\P^1\times\P^1\cong \HS0$ (see, for example,
\cite{GH:alg}). Thus $\P^2_{\{2,1,1\}}$ may be blown-up to give $\HS2$
but deformed to give $\HS0$. A generic point in the moduli space of
$X$ is actually an elliptic fibration over $\HS0$ rather than $\HS2$.%
\footnote{To see this note that when we blow-up the fibration of $X$
over $\P^2_{\{2,1,1\}}$ to a fibration over $\HS2$ we introduce an
elliptic scroll just as in section \ref{ss:K3f}. The results of
\cite{Wil:Kc} tell us that the $(-2)$-curve will therefore vanish for
a generic complex structure.}

We have arrived at the result that $X$ is an elliptic fibration over
$\HS0$. This may be viewed as a two-stage fibration. $X$ may be
projected onto a $\P^1$ to produce a K3 fibration. The fibre of this
map may be projected onto the other $\P^1$ to write the K3 as an
elliptic fibration.

Since the base space of the elliptic fibration is $\P^1\times\P^1$, if
one of the $\P^1$'s in the base may be viewed as 
the base for $X$ as a K3 fibration, then so may the other $\P^1$. Thus
$X$ can be written as a K3 fibration in two different ways. This
suggests that there should be two heterotic strings dual to the type
IIA theory on $X$ and hence dual to each other. What is the
relationship between these two heterotic strings?

One of the heterotic strings will be compactified on $S_1\times
T^2_1$ and the other on $S_2\times T^2_2$, where $S_1$ and $S_2$ are
K3 surfaces. We examined the moduli space coming from the vector
multiplets in section \ref{ss:N2enh} and saw that the 3 moduli
described the dilaton-axion and the Narain moduli of the $T^2$ (without any
Wilson lines switched on). The dilaton is given by the area of the
base $\P^1$ in $X$. Let us determine the area of the heterotic $T^2$
in terms of the moduli of the K3 fibre within $X$.

The generic K3 fibre within $X$ has Picard number 2 and has a moduli
space of K\"ahler form and $B$-field given by (\ref{eq:O22}). Let
$\Gamma_{2,2}$ be generated by the null 
vectors $w$ and $v$ together with their duals $w^*$ and $v^*$.
For simplicity we will avoid switching on any
$B$-fields. Thus we consider a point in the moduli space
(\ref{eq:O22}) to be given by a space-like 2-plane, $\mho$, spanned by
$w^*+\alpha w$ and $v^*+\beta v$, for $\alpha,\beta>0$. Following our
construction to determine the K\"ahler form, we have that
$\mho\cap w^\perp$ is spanned by $v^*+\beta v$ which is contained in
$w^\perp/w$. Thus, as promised, $B$ is zero. From our analysis in
section \ref{ss:clsym} we see that the K\"ahler form determines the
volume to be
\begin{equation}
  J.J = 2\alpha+B^2=2\alpha.
\end{equation}
We also have that the direction of $J$ is given by $v^*+\beta v$. Thus
the K\"ahler form is
\begin{equation}
  J = \sqrt{\frac{\alpha}{\beta}}v^*+\sqrt{\alpha\beta}v.
		\label{eq:hhJ}
\end{equation}

We have seen that the generic K3 fibre is itself an elliptic fibration
over $\P^1$. Knowing the intersection numbers of the curves within
this fibration together with the positivity of the K\"ahler class
determines the class of the base $\P^1$ to be $v^*-v$ and that of the
elliptic fibre to be $v$. Thus, the area of the $\P^1$ within the K3
is given by
\begin{equation}
  J.C = \sqrt{\alpha\beta}-\sqrt{\frac{\alpha}{\beta}}.
	\label{eq:AC1}
\end{equation}

Now let us reinterpret $\mho$ in terms of the moduli of the
$T^2$. From section \ref{ss:tori} we obtain a map from
$W$ into $W^*$ given by%
\footnote{Note that $W$ is spanned by $w^*$ and $v^*$; and $W^*$ is
spanned by $w$ and $v$. There is no
simple choice of conventions which would have circumvented this
notational irritation!}
\begin{equation}
  \psi = \left(\begin{array}{cc}\alpha&0\\0&\beta\end{array}\right).
\end{equation}
This is symmetric and thus gives the metric. Therefore, the area of
the $T^2$ is given by $\sqrt{\det(\psi)}=\sqrt{\alpha\beta}$.

We may now obtain an interesting result confirming the analysis of
\cite{DMW:hh} by going to the limit where $\alpha$ and $\beta$ are
taken to be very large. In this case, our heterotic string is
compactified on K3${}\times T^2$ where the 2-torus is very large. In
the type IIA string, the generic K3 fibre contains a rational curve
which becomes very large. From the equation (\ref{eq:AC1}), the area of this
rational curve is proportional to the heterotic string's 2-torus in
this limit.

This limit as $\alpha,\beta\to\infty$ may be viewed as a
decompactification of the model to a six-dimensional theory
given by the heterotic string compactified on a K3 surface. Let us
consider the coupling of this six-dimensional theory,
$\lambda_6$. This is given by 
the four-dimensional coupling prior to decompactification and the area
of the $T^2$. The former is given by the size of the base $\P^1$ of
the K3 fibration, which we refer to as $\P^1_1$. The latter is
given by the area of the base of the K3 fibre itself written as an
elliptic fibration, which we will refer to as $\P^1_2$. We have that
\begin{equation}
  \lambda_6^2 = \lambda_4^2.\Area(T^2)\sim
	\frac{\Area(\P^1_2)}{\Area(\P^1_1)}.  \label{eq:Shh}
\end{equation}

The other heterotic string is obtained by exchanging the r\^oles of
$\P^1_1$ and $\P^1_2$. Thus we have
\begin{prop}
Let $X$ be the \CY\ manifold given by a resolution of the degree 24
hypersurface in $\P^4_{\{12,8,2,1,1\}}$.
The two heterotic string theories dual to
the type IIA string theory on $X$ decompactify to two heterotic string
theories compactified on K3 surfaces. From (\ref{eq:Shh}) these two
six-dimensional theories 
are S-dual in the sense that the coupling of one theory is inversely
proportional to the coupling of the other theory. 
\end{prop}
The main assumption underlying this proposal is that these heterotic
string theories actually exist. We will get closer to identifying
these theories in section \ref{ss:si}.

We may vary the complex structure of $X$ and this should correspond to
deformations of the K3 together with its vector bundle on the
heterotic side. Note that the K3 of one of the heterotic strings will,
in general, have quite different moduli than the K3 of the other
heterotic string since exchanging the r\^oles of $\P^1_1$ and $\P^1_2$
need not be a geometrical symmetry of $X$ --- only a topological
symmetry. Before we can explicitly give the map between these K3
surfaces and their vector bundles we need to map out the moduli space
of the moduli from the hypermultiplets. This remains to be done.

There are many examples of \CY\ manifolds which admit more than one K3
fibration. This will lead to many examples of heterotic-heterotic
duality. In most cases however the result will be a rather tortuous
mapping between four-dimensional theories and will not be as simple as
the above example.

\subsection{Extremal transitions and phase transitions}
  \label{ss:ext}

We will take our first tentative steps into the moduli space of
hypermultiplets in this section. This will deal with the simplest
aspects of the bundle over K3 --- namely when this bundle, or part of
this bundle, becomes trivial.

Our heterotic string theory is compactified on a bundle over
K3${}\times T^2$ which we view as the product of a bundle over K3 and
a bundle over $T^2$. Generically one would expect the deformation space
of this bundle structure to be smooth. Where this can break down
however is when part of the bundle becomes trivial. To deform away
from such a bundle we may wrap the trivial part around either the K3
surface or the $T^2$. Thus we obtain branches in the moduli space. A
deformation in the K3 part will be a hypermultiplet modulus while a
deformation in the $T^2$ part will be a vector modulus. Our picture
therefore for a transition across this branch will consist of moving
in the moduli space of hypermultiplets until suddenly a vector scalar
becomes massless and can be used as a marginal operator to move off into
a new branch of the moduli space, at which point some of the moduli in
the hypermultiplets in the original theory may acquire mass.

When we compactify the $E_8\times E_8$ or $\spnh$ heterotic string
on a vector bundle, the original gauge group is broken by the holonomy
of the bundle. Thus, when we move to a transition point where part of
the vector bundle becomes trivial we may well expect the holonomy
group to shrink and thus the observed gauge group is enhanced. Since
we have only $N=2$ supersymmetry this observed gauge group enhancement
may get killed by quantum effects but should be present in the
zero-coupling limit.

The picture is the geometric version of the ``Higgs'' transitions
explored in, for example, \cite{SW:II}. We wish to see how this
transition appears from the type IIA picture.

We have already seen what the type IIA picture is for moving within
the moduli space of vector scalars to a point at which an enhanced
gauge symmetry appears. This is where we vary the K\"ahler form on the
generic K3 fibre to shrink down some rational curves (or go to Planck
scale effects). Thus, every K3 fibre becomes singular. That is, we
have a curve of singularities in $X$. Now given such a singular \CY\
manifold, it may be possible to deform this space by a deformation of
complex structure to obtain a new smooth \CY\ manifold. This would
correspond to a deformation of each singular K3 fibre to obtain a
smooth K3 fibre. This process will decrease the Picard number of the
generic fibre (as we have lost the class of rational curves we shrank
down) and will change the topology of the underlying \CY\ threefold. 

In the type IIA language then, this Higgs transition consists of
deforming the K\"ahler form on $X$ to obtain a singular space and then
smoothing by a deformation to another smooth manifold. Such a
topology-changing process is called an ``extremal transition''. 

One example of an extremal transition is the ``conifold'' transition
of \cite{GH:con}. A conifold transition consists of shrinking down
isolated rational curves and then deforming away the
singularities. These were explored in the context of full string
theory in \cite{Str:con,GMS:con} (see also B.~Greene's lectures). In
our case however we are not shrinking down isolated curves but whole
curves of curves and so we are not discussing a conifold transition. 

There has been speculation \cite{CGH:con,GMS:con} that the moduli
space of all \CY\ threefolds is connected because of extremal
transitions (based on an older, much weaker, statement by Reid
\cite{Reid:never}). 
See \cite{CGGK:srch,ACJM:srch} for recent results in this direction.
Certainly no counter example is yet known to this
hypothesis. The heterotic picture of these extremal transitions is
simple to understand in the case of singularities developing within
the generic fibre. Such specific extremal transitions are certainly
not sufficient to connect the moduli space and we will require an
understanding of nonperturbative heterotic string theory to complete
the picture. One obvious example to worry about is when the \CY\
threefold on the type IIA side goes through an extremal transition
from something that is a K3 fibration to something that is not. It is
not difficult to see that such a transition must involve shrinking
down the base, $W$, of the fibration and thus going to a
strongly-coupled heterotic string. It is not surprising therefore that
we cannot understand such a transition perturbatively in the heterotic
picture.

Given a dual pair of a type IIA string compactified on a \CY\ manifold
and a heterotic string compactified on K3${}\times T^2$ we may
generate more dual pairs by following each through these phase
transitions we do understand perturbatively. Such ``chains'' of dual
pairs were first identified in 
\cite{AFIQ:chains} and many examples have been given in \cite{CaFo:web}.
In order to understand where these chains come from, and indeed the
original Kachru--Vafa examples of dual pairs in \cite{KV:N=2},
we need to confront the
issue of compactifying the heterotic string on a K3 surface, a subject
we have done our best to avoid up to this point!

%%%%%%%%%%%%%%%%%%%%%%%%%%%%%%%%%%%%%%%%%%%%%%%%%%%%%%%%%%%%%%%%%%%

\section{The Heterotic String}		\label{s:het}

This section will be concerned with the heterotic string compactified
on a K3 surface. In particular we would like to find a string theory
dual to this. An obvious answer is another heterotic string
compactified on another K3 surface, as we saw in section
\ref{ss:hh}. We will endeavor to find a type II dual. It turns out the
we will not be able to find a type II dual directly but will have to
go via the construction of section \ref{s:4d}. This process (in its
various manifestations) is often called ``F-theory''. A great deal of
the following analysis is based on work by Morrison and Vafa
\cite{Vafa:F,MV:F,MV:F2}.

\subsection{$N=1$ theories}  \label{ss:N=1}

The heterotic string compactified on a K3 surface gives a theory of
$N=1$ supergravity in six dimensions. From section \ref{ss:supergrav}
the holonomy algebra of the moduli space coming from supersymmetry
will be $\sp(1)$. There are two types of supermultiplet in six
dimensions which contain moduli fields:
\begin{enumerate}
\item The hypermultiplets each contain 4 real massless scalars which
transform as quaternionic objects under the $\sp(1)$ holonomy.
\item The ``tensor multiplets'' each contain one real massless scalar.
\end{enumerate}
Thus, at least away from points where the manifold structure may break
down, we expect the moduli space to be in the form of a product
\begin{equation}
  \cM_{N=1} = \cM_T\times\cM_H,
\end{equation}
(possibly divided by a discrete group)
where $\cM_T$ is a generic Riemannian manifold spanned by moduli in tensor
supermultiplets and $\cM_H$ is a quaternionic K\"ahler manifold spanned
by moduli in hypermultiplets.

For a review of some of the aspects of these theories we refer to
\cite{SW:6d}. The six-dimensional dilaton lives in a tensor
multiplet. An interesting feature of these theories is that it appears
to be impossible to write down an action for these theories unless
there is exactly one tensor multiplet.\footnote{This is because the
gravity supermultiplet (of which there is always exactly one) contains
a self-dual two-form and the tensor multiplets contain anti-self-dual
two-forms. It is problematic to write down Lorentz-invariant actions
for theories with net (anti-)self-dual degrees of freedom
\cite{MSch:ASD}.} Thus moduli in tensor multiplets other than the
dilaton should be regarded as fairly peculiar objects.

It is useful to compare $N=1$ theories in six dimensions with the
resulting $N=2$ theory in four dimensions obtained by compactifying
the six-dimensional theory on a 2-torus. To explain the moduli which
appear in four dimensions we need to consider another
supermultiplet of the $N=1$ theory in six dimensions. This is the
vector multiplet which contains a real vector degree of freedom but no
scalars.
Each supermultiplet in six dimensions produces moduli in four
dimensions as follows:
\begin{enumerate}
\item The 4 real scalars of a hypermultiplet in six dimensions simply
produce the 4 real scalars of a hypermultiplet in four dimensions.
\item The 1 real scalar of a tensor multiplet together with the
anti-self-dual two-form compactified on $T^2$ produce the two real
scalars of a vector multiplet in four dimensions.
\item The vector field of a vector multiplet compactified on the
two 1-cycles of $T^2$ of produces the two real
scalars of a vector multiplet in four dimensions.
\end{enumerate}
That is, hypermultiplets in six dimensions map to hypermultiplets in
four dimension but both tensor multiplets and vector multiplets in six
dimensions map to vector multiplets in four dimensions.

We should emphasize that quantum field theories in four and six
dimensions are quite different. In particular, conventional arguments
imply that six-dimensional quantum field theories should always be
infra-red free and therefore rather boring. This notion has been
revised recently in light of many of the results coming from string
duality \cite{SW:6d,Sei:6d} where it is now believed that nontrivial
theories can occur in six dimensions as a result of ``tensionless
string-like solitons'' appearing. Such theories potentially appear
when one goes 
through extremal transitions between tensor multiplet moduli and
hypermultiplet moduli.

Despite the strange properties of these exotic six-dimensional field
theories we will be able to avoid having to explain them here. This is
because all of our discussion will really happen in four dimensions
--- the six-dimensional picture is only considered as a large $T^2$
limit. It is important to realize however that some of the
things we will say, based on four-dimensional physics, may well be
rather subtle in six dimensions. An example of this will be when
certain enhanced gauge symmetries are said to appear when the
hypermultiplet moduli are tuned to a certain value. If massless tensor
moduli also happen to appear at the same time (which will happen for
$E_8$ as we shall see) then any conventional description of the
resulting six-dimensional field theory is troublesome. Declaring what
the massless spectrum of such a theory is not an entirely
well-defined question and one should move the theory slightly by
perturbing either a massless hypermultiplet modulus or a tensor
modulus before asking such a question.

\subsection{Elliptic fibrations}   \label{ss:ell}

Our method of approach will be to consider a type IIA string
compactified on a \CY\ threefold, $X$, dual to a heterotic string
compactified on K3${}\times T^2$ and then take the volume of the
2-torus to infinity thus decompactifying our theory to a heterotic
string on a K3 surface. Actually, decompactification is a rather
delicate process and perhaps we should be more pragmatic and say that
we will consider a heterotic string on K3${}\times T^2$ and try to
systematically ignore all aspects of the type IIA string on $X$
coming from the $T^2$ part of the heterotic string.

Much of the analysis we require we have already covered in section
\ref{ss:hh}. There, for a specific example, we did precisely the
decompactification we require. We need to consider how general we can
make this process. Clearly we should insist that our moduli space of vector
multiplets in the four-dimensional theory corresponding to the
heterotic string on K3${}\times T^2$ contain the space
\begin{equation}
  \GO(\Gamma_{2,2})\backslash\GO(2,2)/(\GO(2)\times\GO(2)).
		\label{eq:O22-1}
\end{equation}
That is, we have the moduli space of the string on $T^2$. If this were
not the case, we could not claim that we had really compactified on a
true product of K3${}\times T^2$. From the analysis of section
\ref{ss:K3f} we expect $X$ to be a K3 fibration. Given the appearance
of $\Gamma_{2,2}$ in (\ref{eq:O22-1}) we expect from section
\ref{ss:N2enh} that the Picard lattice of the generic fibre of $X$
must contain the lattice $\Gamma_{1,1}$.

A K3 surface with a Picard lattice containing $\Gamma_{1,1}$ is an
elliptic fibration. To see this, let $v$ and $v^*$ be a basis for 
$\Gamma_{1,1}$. The class $v-v^*$ is a primitive element of
self-intersection $-2$ and thus (see, for example, \cite{BPV:}) either
this class, or minus this class, corresponds to a rational curve
within the K3. Now either $v$ or $v^*$ is a nef curve of zero
self-intersection. This will be the class of the elliptic fibre. Note
that we have a rational curve in the K3 surface intersecting each
fibre once --- our elliptic fibration has a global section. What's more,
it is easy to see that this section is unique as there is no other
$(-2)$-curve.\footnote{Actually we have cheated here. We have assumed
that the bundle over K3 may be chosen so that it breaks the $E_8\times
E_8$ or $\spnh$ gauge group. Then the Narain moduli space for the
$T^2$ really is given by (\ref{eq:O22-1}). There can be times however,
as we shall see soon, when this is not possible and then the Narain
moduli space for the 2-torus becomes larger. This allows for more than
one section of the elliptic fibration. We claim this is not important
for the examples we discuss in these lectures, however, as we may begin
in a case where the gauge group is broken by the K3 bundle and then
proceed via extremal transitions to the case we desire, preserving the
section of the elliptic fibration if $X$.}

We thus claim that $X$ is a K3 fibration and that each K3 fibre can be
written as an elliptic fibration with a unique section. Together these
give \cite{AG:sp32}
\begin{prop}
  If a heterotic string compactified on K3${}\times T^2$ is dual to a
type IIA string on $X$ and there are no obstructions in the moduli space
to taking the size of $T^2$ to infinity and thereby ignoring the effects
associated to it, then $X$ is an elliptic fibration over a complex surface
with a section.
\end{prop}
This proposition is subject to the same conditions as proposition
\ref{prop:K3f} and to the caveat in the footnote above.

Let us denote this elliptic fibration as $p:X\to\Theta$. $\Theta$
itself is a $\P^1$ fibration over $\P^1$, $\Theta\to W$. The simplest
possibility is that $\Theta$ is the Hirzebruch surface $\HS n$. This
need not be the case however --- there may be some bad fibres over
some points in $W$. 

The subject of elliptic fibrations has been studied intensively by
algebraic geometers, thanks largely to Kodaira \cite{Kod:ell}. This
can be contrasted to the subject of K3 fibrations, about which
relatively little is known. 

Let us fix our notation. Let $W$, the base of $X$ as a K3 fibration, be
$\P^1$ with homogeneous coordinates $[t_0,t_1]$. We will also use the
affine coordinate $t=t_1/t_0$. $\Theta$ is a $\P^1$ fibration over
$W$. Over a generic point in $W$, the $\P^1$ fibre will have
homogeneous coordinates $[s_0,s_1]$ and affine coordinate
$s=s_1/s_0$. Now we want to write down the fibre of $X$ as an elliptic
fibration over $\Theta$. Any elliptic curve may be written as a cubic
hypersurface in $\P^2$. Writing this in affine coordinates we may put
the fibration in ``Weierstrass form'' for a generic point $(s,t)\in\Theta$:
\begin{equation}
  y^2 = x^3 + a(s,t)x + b(s,t),    \label{eq:Wf}
\end{equation}
where $x$ and $y$ are affine coordinates in a patch of $\P^2$ and $a$
and $b$ are arbitrary polynomials. If $\Theta$ itself has bad fibres
--- that is, it is not $\HS n$ --- then more coordinate patches need to be
introduced to give a global description of the elliptic fibration.

Note that not any elliptic fibration can be written in Weierstrass
form. Homogenizing the coordinates of $\P^2$ putting $x=x_1/x_0$ and
$y=x_2/x_0$, we see that $[0,0,1]$ always lies in (\ref{eq:Wf}), which
gives the fibration a global section. Thus only elliptic fibrations
with a section can be written in this form. Luckily this is the case
we are interested in.

The $j$-invariant of the elliptic fibre (\ref{eq:Wf}) is $4a^3/\delta$,
where $\delta$ is the discriminant of (\ref{eq:Wf}) given by
\begin{equation}
  \delta = 4a^3+27b^2.
\end{equation}
If $\delta=0$ then the elliptic fibre is singular. The hypersurface, or divisor,
$\delta(w,z)=0$ in $\Theta$ thus gives the locus of bad fibres. It is
commonly called the discriminant locus.

\iffigs
\begin{figure}[t]
  \centerline{\epsfxsize=12cm\epsfbox{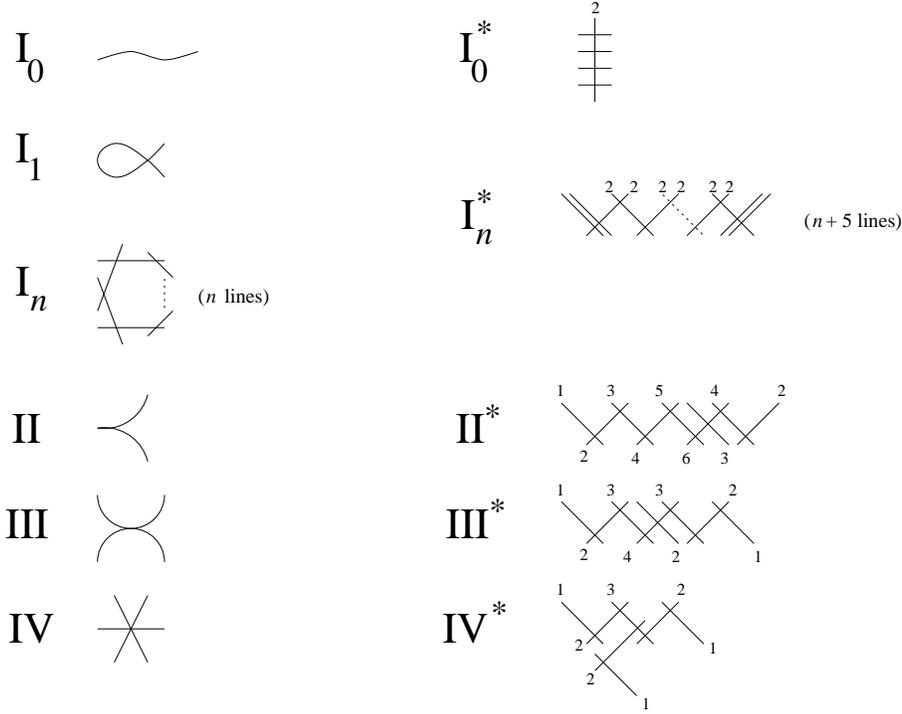}}
  \caption{Classification of elliptic fibres.}
  \label{fig:Kod}
\end{figure}
\fi

\def\pII{\vphantom{\rm II}}
There are only so many things that can happen at a generic bad fibre
of an elliptic K3 surface and these
have been classified (see, for example, \cite{BPV:}). Let us take a
small disc $D\subset\C$, embedded in $\Theta$, with coordinate $z$
which cuts a generic point 
on the discriminant locus transversely. Let $z=0$ be the location of
the discriminant locus. 
Let us analyze the restriction of the fibration of $X$ to the part
which is fibred over $D$,
We are thus considering an open set of a
fibration of a complex {\em surface}.
Assuming that the total space of the fibration
is smooth and there are no $(-1)$-curves, which
will certainly be true if we are talking about a \CY\ manifold, the
possibilities for what happens to the fibre when $z=0$ is shown in
figure \ref{fig:Kod}. The case ${\rm I}_0$ is when there is no zero of
$\delta$ and the elliptic fibre is smooth. In all other cases the
lines and curves in figure \ref{fig:Kod} represent rational
curves. Case ${\rm I}_1$ is a rational curve with a ``double point'',
i.e., it looks locally like $y^2=x^2$ at one point and case II is a
rational curve with a ``cusp'', i.e., it looks locally like $y^2=x^3$
at one point. All the other cases consist of multiple rational curves.
All of the singular fibres should be homologous to the smooth
fibre. To achieve this, some of the rational curves in the bad fibre
must be counted more than once to obtain the correct homology. The
multiplicity of the curves are shown as the small numbers in figure
\ref{fig:Kod}. If omitted, multiplicity one is assumed.\footnote{There
is also a possibility that ${\rm I}_n$ fibres can appear with multiplicity
greater than one. We ignore this as the canonical class of such a
fibration cannot be trivial.}

The possibilities listed in figure \ref{fig:Kod} can also be
classified according to the Weierstrass form. On our disc, $D$, the
polynomials, $a$ and $b$, in (\ref{eq:Wf}) will be polynomials in
$z$. Let us define the non-negative integers $(L,K,N)$ by
\begin{equation}
  \eqalign{a(z) &= z^La_0(z)\cr b(z) &= z^Kb_0(z)\cr 
	\delta(z) &= z^N\delta_0(z),\cr}
\end{equation}
where $a_0(z)$, $b_0(z)$, and $\delta_0(z)$ are all nonzero at
$z=0$. The triple $(L,K,N)$ then determines which fibre we have
according to table \ref{tab:Mir}. See \cite{Mir:fibr} for an
explanation of this. To be precise, the Weierstrass form of the
fibration in $x$, $y$ and $z$ will produce a surface singularity
which, when blown-up, will have the fibres in figure
\ref{fig:Kod}. This is closely linked to the results in table
\ref{tab:ADE}.

\begin{table}
$$\begin{array}{|c|c|c|c|c|}
\hline
L&K&N&\mbox{Fibre}&\Upsilon'\\
\hline
\geq0&\geq0&0&{\rm I}_0&\\
0&0&>0&{\rm I}_N&A_{N-1}\\
\geq 1&1&2&{\rm II}&\\
1&\geq2&3&{\rm III}&A_1\\
\geq 2&2&4&{\rm IV}&A_2\\
\geq 2&\geq 3&6&{\rm I\pII}^*_0&D_4\\
2&3&\geq 7&{\rm I\pII}^*_{N-6}&D_{N-2}\\
\geq 3&4&8&{\rm IV}^*&E_6\\
3&\geq 5&9&{\rm III}^*&E_7\\
\geq 4&5&10&{\rm II}^*&E_8\\
\hline
\end{array}$$
\caption{Weierstrass classification of fibres.}
\label{tab:Mir}
\end{table}

It is important to note that this classification procedure only
applies to a smooth point on the discriminant locus. Only in this case
can we characterize the bad fibre in terms of the family of elliptic
curves over our small disc, $D$. When the discriminant is singular
the nature of the bad fibre need not be expressible in terms of the
geometry of a complex surface --- it will be higher-dimensional in
character. In general, for \CY\ threefolds, we should expect
to encounter some singular fibres not listed above. Such exotic fibres
are important in string theory but we will try to avoid such examples
here as it makes the analysis somewhat harder.

For our elliptic fibration, $p:X\to\Theta$, a knowledge of the explicit
Weierstrass form is enough to calculate the canonical class,
$K_X$. This may be done as follows. In homogeneous coordinates, the
Weierstrass form is
\begin{equation}
  x_0x_2^2 = x_1^3 + ax_0^2x_1 + bx_0^3,    \label{eq:Wf2}
\end{equation}
giving a cubic curve in $\P^2$. Now since the elliptic fibration is not
trivial, this $\P^2$ will vary nontrivially as we move over $\Theta$.
We may describe such a $\P^2$ as the
projectivization of a sum of three line bundles over $\Theta$. 
We are free to declare that $x_0$ is a section of a trivial line
bundle. We may then find a line bundle, $\cL$, such that $x_1$ is a
section of $\cL\,{}^2$ and $x_2$ is a section of $\cL\,{}^3$ in order to
be compatible with (\ref{eq:Wf2}). It also follows 
that $a$ is a section of $\cL\,{}^4$ and $b$ is a section of $\cL\,{}^6$. 

Now let us consider the normal bundle of the section, $\sigma$, given
by $[x_0,x_1,x_2]=[0,0,1]$, embedded in $X$. We may use affine
coordinates $\xi_1=x_1/x_2$ and $\xi_2=x_0/x_2$ whose origin gives the
section $\sigma$. Note that $\xi_1$ is a section of $\cL\,{}^{-1}$ and $\xi_2$
is a section of $\cL\,{}^{-3}$. Near $\sigma$, (\ref{eq:Wf2}) becomes
\begin{equation}
  \xi_2 = \xi_1^3.
\end{equation}
Thus, $\xi_1$ is a good coordinate to describe the fibre of the normal
bundle of $\sigma$ in $X$. This implies that this normal bundle is given by
$\cL\,{}^{-1}$. 

We may now use the adjunction formula for $\sigma\subset X$ to give
\begin{equation}
K_X|_\sigma = K_\sigma + \cL\,.
\end{equation}
Actually, this is the only contribution towards the canonical class of
$X$. That is,
\begin{equation}
K_X = p^*(K_\Theta+\cL\,).   \label{eq:KX}
\end{equation}
For the case we will be interested in, we want $K_X=0$ and so
$\cL=-K_\Theta$. We will give examples of such constructions later.

Note that if there is a divisor within $\Theta$ over which $a$
vanishes to order $\geq4$ and $b$ vanishes to order $\geq6$, we may
redefine $\cL$ to ``absorb'' this divisor and lower the degrees of $a$ and
$b$ accordingly. This is why no such fibres appear in table
\ref{tab:Mir}. As this will change $K_X$, such occurrences cannot
happen in a \CY\ variety.

Let us consider a K3 surface written as an elliptic fibration with a
section. The 
Picard number of the K3 is at least two --- we have the section and a
generic fibre as algebraic curves. If we have any of the fibres
${\rm I}_n$, for $n>2$, or ${\rm I\pII}^*_n$, ${\rm III}$,
${\rm IV}$, ${\rm II}^*$, ${\rm III}^*$, or ${\rm IV}^*$ we will also
have a contribution to the Picard group from reducible fibres. Each of
these fibres contains rational curves in the form of a root lattice of
a simply-laced group. Let us denote this lattice $\Upsilon'$. The
possibilities are listed in table \ref{tab:Mir}. Thus, shrinking down
these rational curves will induce the corresponding gauge group for a
type IIA string. 

We know that for a \CY\ manifold compactified on a K3 fibration, the
moduli coming from varying the K\"ahler form on the K3 fibre map to
the $T^2$ part of the heterotic string compactified on K3${}\times
T^2$. In particular, the act of blowing-up rational curves in the K3 to resolve
singularities, and hence break potential gauge groups, is identified
with switching on Wilson lines on $T^2$. Thus, to ignore Wilson
lines, these rational curves must all be blown down and held at zero
area. That is, {\em any of the fibres ${\rm I}_n$, for $n>2$, or ${\rm
I\pII}^*_n$, ${\rm III}$, ${\rm IV}$, ${\rm II}^*$, ${\rm III}^*$, or
${\rm IV}^*$ appearing in the elliptic fibration will produce an
enhanced gauge symmetry in the theory.}

From section \ref{ss:hh} and, in particular (\ref{eq:hhJ}), the size
of an elliptic fibre within this K3 will be fixed to some constant
$\sqrt{\alpha/\beta}$ as $\alpha,\beta\to\infty$ to make the $T^2$
infinite area. Thus this size is ``frozen out'' as a degree of
freedom. To ignore the 2-torus degrees of freedom for the type
IIA string compactified on $X$ we should take the K3 fibre within $X$,
consider it as an elliptic fibration with an elliptic fibre of frozen
area and blow down any rational curves which may take the Picard
number of the K3 fibre beyond 2. In summary, any degrees of freedom
coming from sizes within the elliptic fibre structure are ignored.

Consider the base, $\Theta$, as a $\P^1$ bundle over $W$. Suppose we
have bad fibres in this case. These must
correspond to reducible fibres. Now, when we build $X$ as a K3
fibration over $W$, these reducible fibres in $\Theta$ will build
reducible fibres in $X$. This is exactly case 3 listed at the
beginning of section \ref{ss:N2enh}. That is, varying the size of
irreducible parts of these reducible fibres will give moduli in vector
supermultiplets which cannot be understood perturbatively on the
heterotic side. The important point to note here is that these degrees
of freedom will not go away when we unwind any Wilson lines around
$T^2$ and take its area to infinity. Thus, these degrees of freedom
must be associated to the six-dimensional theory of the heterotic
string compactified on a K3. From section \ref{ss:N=1} these moduli
must therefore come from tensor multiplets in the six-dimensional
theory. It is the peculiar nature of tensor moduli which prevented us
from having a perturbative understanding of these moduli in section
\ref{ss:N2enh}.

We saw in section \ref{ss:hh} how the size of $W$ and the size of the
fibre of $\Theta$ were used to produce the area of the heterotic
string's $T^2$ and the size of its dilaton. The area of the $T^2$ is
lost as a degree of freedom in our six-dimensional theory. We see then
{\em that the number of tensor multiplets in our theory will be the Picard
number of $\Theta$ minus one.} In the case that $\Theta$ is $\HS n$,
that is, there are no bad fibres, the number of tensor multiplets will
be one and this single multiplet contains the heterotic dilaton as a
modulus.

Questions concerning hypermultiplets between the four-dimensional
theory and the six-dimensional theory are unchanged. In particular, we
retain the relationship from section \ref{ss:N=2} that the number of
hypermultiplets is given by $h^{2,1}(X)+1$.

Now we have counted massless tensor multiplets and hypermultiplets, let us
count massless vector multiplets. We know that any vector multiplet in the
four-dimensional theory must have its origin in either a vector
multiplet or a tensor multiplet in six dimensions. Thus we can count
the number of six-dimensional vectors by subtracting
$h^{1,1}(\Theta)-1$ from the 
number of four-dimensional vectors.

\begin{difficult}
Many of these vectors can be seen directly in terms of the enhanced
nonabelian gauge symmetry but there is an additional contribution.
In section \ref{ss:N2enh} we saw how $H^2(X)$, where $X$ is a K3
fibration, could be built from elements from the base, from the generic
fibre and from the bad fibres in a fairly obvious way. Now we want to
consider if the same thing is true for an elliptic fibration. Analyzing the
spectral sequence one can see that
$H^2(\Theta,\Z)$ contributes, for the base, to give the
tensor multiplets and $H^0(\Theta,R^2p_*\Z)$ contributes for the
fibres. This latter piece accounts for the enhanced nonabelian
symmetry discussed above. The object of
interest will be the term $H^1(\Theta,R^1p_*\Z)$. In the case of K3
fibrations, this is trivial since $H^1({\rm K3})=0$. This ceases to be
true for an elliptic fibration however. This contribution may be
associated to the group of sections of the bundle as seen in 
\cite{CZ:gag}. Later on, in section \ref{ss:so32}, we will consider a
case where this is a finite group, but here we note that if this group
is infinite, then its rank will contribute to the dimension of
$H^2(X)$.

We see then that if the elliptic fibration has an infinite number of
sections, there will be massless vector fields beyond those accounted
for from the nonabelian gauge symmetry from bad fibres. These may
contribute extra $\gu(1)$ terms to the gauge symmetry \cite{MV:F2} and,
conceivably, more nonabelian parts. As this part of the gauge group is
rather difficult to analyze we will restrict ourselves to examples in
these lectures where there is no such contribution.
\end{difficult}

\subsection{Small instantons}   \label{ss:si}

Our goal in this section will be to find the map from at least part of
the moduli space of hypermultiplets for a type IIA string on a \CY\
manifold to the moduli space of hypermultiplets of the heterotic
string compactified on a K3{}$\times T^2$. 
As we have seen, the $T^2$
factor is irrelevant for the hypermultiplet moduli space and we are
free to consider the latter as a heterotic string on a K3 surface so long
as the \CY\ is an elliptic fibration with a section.
We will be able to go some way to determining ``which'' heterotic
string a given elliptic threefold is dual to.

The best policy when finding a map between two moduli spaces is to
start with a particularly special point in the moduli space of one
theory with hopefully unique properties which allow it to be
identified with a correspondingly special point in the other theory's
moduli space. This special point will usually be very symmetric in
some sense. The trick, invented in \cite{MV:F} and inspired by the work of
\cite{W:small-i}, is to look for very large gauge symmetries resulting
from collapsed instantons in the heterotic string.

Recall that the K3 surface, $S$, on which the heterotic string is
compactified comes equipped with a bundle, $E\to S$, with
$c_2(E)=24$. Fixing $S$, the bundle $E$ will have moduli. A useful
trick when visualizing the moduli space of bundles is to try to
flatten out as much of the connection on the bundle as possible. Of
course, the fact that $c_2(E)=24$ makes it impossible to completely
flatten out $E$ but we can concentrate the parts of the bundle with
significant curvature into small regions isolated from each other. In
this picture, an approximate point of view of at least part of the
moduli space of the bundle can be viewed as 24 ``instantons''
localized in small regions over $S$ each of which contributes one to
$c_2(E)$. 

As well as its position, each instanton will have a degree of freedom
associated to its size --- that is, the characteristic length away from
the centre of the instanton where the curvature becomes small. At
least from a classical point of view, there is nothing to stop one
shrinking this length scale down to zero.\footnote{Although there is
no reason to suppose that one may do this independently for all 24 instantons.}
Such a process naturally takes one to 
the boundary of the moduli space of instantons. 

Let us consider the $E_8\times E_8$ heterotic string.
The observed gauge group of a heterotic string theory compactified on
$S$ will be the part of the original $E_8\times E_8$ 
which is not killed by the holonomy of $E$. It is the ``centralizer''
of the embedding of the holonomy in $E_8\times E_8$ ---
i.e., all the elements of $E_8\times E_8$ which commute
with the holonomy. What is the holonomy around a collapsed instanton?
In general the global holonomy is generated by contractable loops due
to the curvature of the bundle, and from non-contractable loops in the
base. The curvature is zero everywhere when the instanton has become a
point. Also, if $S$ is smooth we may look at a 3-sphere surrounding
the instanton to look for holonomy effects. Since $\pi_1(S^3)=0$ we
cannot generate non-contractable loops. Thus, the holonomy of a
point-like instanton is trivial.\footnote{Note that this need not be
the case if $S$ has an orbifold singularity and the instanton sits at
this point. Then we can only surround the instanton by a lens space
which is not simply connected. One should note that we have
conveniently ignored those loops which happen to go through the point
where the point-like instanton sits.}

One possibility is to shrink all 24 instantons down to zero
size. When we do this, $E$ will have no holonomy whatsoever and the
resulting heterotic string theory will retain its full $E_8\times E_8$
gauge symmetry. As this is such a big gauge group it is a
good place to start analyzing our duality map.

We thus want to find a \CY\ space on which we may compactify the type
IIA string to give an $E_8\times E_8$ gauge symmetry. That is, we want an
elliptic fibration over $\HS n$ such that when viewed as a K3
fibration, the generic K3 fibre has two $E_8$ singularities. Let us
discuss what this implies about the discriminant locus within
$\Theta\cong\HS n$.

First, let us be a little sloppy with notation and not distinguish
between curves and their divisor class (roughly speaking, homology
class) in $\Theta$. Thus we use $C_0$ to denote the class of base,
i.e., the $(-n)$-curve within $\Theta$, and $f$ to denote the class of
the generic $\P^1$ fibre. Let us determine $K_\Theta$ in terms of these
classes. Consider the adjunction formula for a curve
$C\in\Theta$. Integrating this over $C$ we obtain its Euler characteristic
\begin{equation}
   \chi(C) = -C.(C+K_\Theta).
\end{equation}
Knowing that $C_0$ and $f$ are spheres is enough to determine
\begin{equation}
  K_{\HS n} = -2C_0 -(2+n)f.
\end{equation}
Let us use the letters $A$, $B$, and $\Delta$ to denote the divisors
associated to the equations $a=0$, $b=0$, and $\delta=0$ respectively.
From (\ref{eq:KX}), the classes of these divisors will be
\begin{equation}
  \eqalign{A &= 8C_0+(8+4n)f\cr
  B &= 12C_0+(12+6n)f\cr\Delta &= 24C_0+(24+12n)f,\cr}
		\label{eq:ABD}
\end{equation}
in order that $X$ be \CY. 

The locus of $E_8$ singularities in $X$ will map to curves in
$\Theta$. As these $E_8$'s are independent,
we want to make these curves disjoint sections of $\Theta$,
that is, one curve will be in the class $C_0$ (the isolated zero section of the
Hirzebruch surface as a $\P^1$ bundle over $\P^1$) and the other in the
class $C_\infty=C_0+nf$ (a section in the non-isolated class which we
view as a section ``at infinity'' of the Hirzebruch surface). From
table \ref{tab:Mir} we see that we want ${\rm 
II}^*$ fibres over these curves.

These two curves of ${\rm II}^*$ fibres will account for a large
portion of the $A$, $B$, and $\Delta$ divisors. Let us write $A'$, $B'$,
and $\Delta'$ for the remaining parts of the divisors not contained in
the curves $C_0$ and $C_\infty$. From table \ref{tab:Mir} we have
\begin{equation}
  \eqalign{
  A' &= A-4(C_0)-4(C_0+nf) = 8f\cr
  B' &= B-5(C_0)-5(C_0+nf) = 2C_0+(12+n)f\cr
  \Delta' &= \Delta-10(C_0)-10(C_0+nf) = 4C_0+(24+2n)f.\cr}
\end{equation}

This means that what is left of the discriminant, $\Delta'$, will
collide with the curve $C_0$ a total number of $C_0.\Delta'=2(12-n)$
times and with the curve $C_\infty$ a total number of $2(12+n)$
times. These 48 points of intersection are not independent. The reason
that $\Delta'$ collides with the curves of ${\rm II}^*$ fibres is
precisely because $B'$ also collides with these curves. Each time $B'$
hits these curves, the degree of the discriminant will rise by 2 and
hence $\Delta'$ hits them twice in the same place. To see exactly what
shape this intersection is one may explicitly write out the
equations. The result is that $\Delta'$ crosses itself transversely at
these points as well as hitting $C_0$ or $C_\infty$. We see then that,
generically, $\Delta'$ collides with $C_0$ at $12-n$ points and with
$C_\infty$ at $12+n$ points. Within $\Delta'$,
away from these collisions, we expect the discriminant to behave
reasonably nicely.\footnote{Although it will have cusps.}
The result is shown in the upper part of figure
\ref{fig:curly}.

\iffigs
\begin{figure}[t]
  \centerline{\epsfxsize=10cm\epsfbox{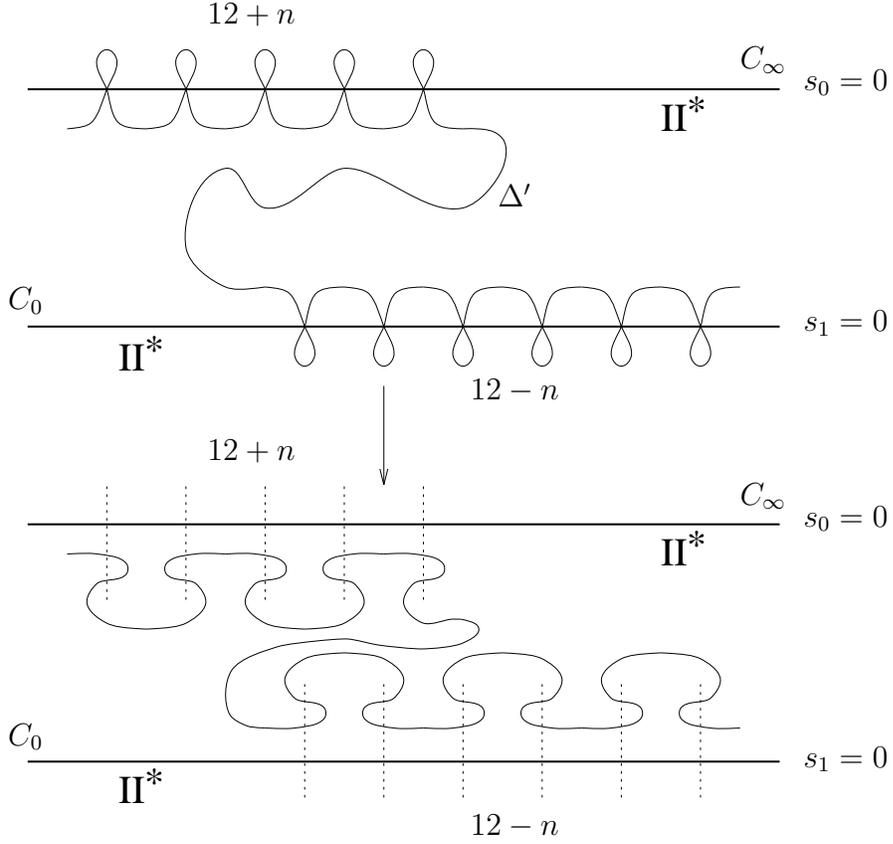}}
  \setlength{\unitlength}{1mm}
  \begin{picture}(0,0)
     \put(32,14){\makebox(0,0){$C_0$}}
     \put(130,46){\makebox(0,0){$C_\infty$}}
     \put(141,11){\makebox(0,0){$s_1=0$}}
     \put(141,43){\makebox(0,0){$s_0=0$}}
     \put(97,2){\makebox(0,0){$12-n$}}
     \put(62,52){\makebox(0,0){$12+n$}}
     \put(32,72){\makebox(0,0){$C_0$}}
     \put(130,104){\makebox(0,0){$C_\infty$}}
     \put(141,69){\makebox(0,0){$s_1=0$}}
     \put(141,101){\makebox(0,0){$s_0=0$}}
     \put(97,60){\makebox(0,0){$12-n$}}
     \put(62,110){\makebox(0,0){$12+n$}}
     \put(97,86){\makebox(0,0){$\Delta'$}}
  \end{picture}
  \caption{24 small instantons in the $E_8\times E_8$ string.}
  \label{fig:curly}
\end{figure}
\fi

We know how to deal with all of the points on the discriminant from
Kodaira's list in figure \ref{fig:Kod} except for the 24 points of
collision of $\Delta'$ with the two lines of ${\rm II}^*$ fibres. Here
we have to work harder to obtain a smooth model for $X$. 

Let us focus on one of these points where either $C_0$ or $C_\infty$
hits $\Delta'$ twice within $\Theta$. Blow up this point of
intersection, $\pi:\widetilde\Theta\to\Theta$. This will introduce a
new rational curve, $E$, in the blown-up 
surface $\widetilde\Theta$ such that $K_{\widetilde\Theta}=\pi^*K_\Theta
+E$. Pulling back $\cL$ onto $\widetilde\Theta$ will show that $a$ now
vanishes to degree 4 on $E$ and $b$ vanishes to degree 6 on $E$.
Introduce $\cL\,{}'=\pi^*\cL-E$ and write $a$ and $b$ in
terms of $\cL\,{}'$ instead of $\cL$. Now $a$ and $b$ will not vanish
at all on a generic point on $E$. Note that the effect on $K_X$ of
blowing up the base, $\Theta$, and then subtracting $E$ from $\cL$
nicely cancels out and so $X$ is still \CY.

Thus, to obtain a smooth $X$ we need to blow up all 24 points of
collision. This process is shown in the lower part of figure
\ref{fig:curly}. The dotted lines represent the 24 new $\P^1$'s
introduced into the base. Note that they are not part of the
discriminant.

From our interpretation of moduli in section \ref{ss:ell} we see that
in terms of the underlying six-dimensional field theory we have 24 new
moduli from tensor multiplets as soon as we try to enhance the gauge
group to $E_8\times E_8$ in this way. Remember that in the language of
the heterotic string this large gauge group was meant to be the result
of shrinking down 24 instantons to zero size. What we have therefore
shown here is that shrinking down 24 instantons to zero size results
in the appearance of 24 new tensor moduli.

Given that the appearance of tensor multiplets should be a
nonperturbative phenomenon in the heterotic string, it would seem
unreasonable to expect them to appear when the target space and vector
bundle is smooth. At least in the case that the underlying K3 surface
is large, it is then clear that each tensor modulus can be tied to
each shrunken instanton in this picture. That is, one point-like
instanton will result in one tensor modulus appearing.

Suppose we try to give size to some of the instantons. This should
correspond to a deformation of the complex structure of the \CY\
threefold. We know that
this should result in the disappearance of (at least) one of the tensor
multiplets and thus (at least) one of the blow-ups in $\Theta$. Thus the
deformation has to disturb one of the curves of ${\rm II}^*$ and thus
lower the size of the effective gauge group. What is important to
notice is that $12+n$ of the small instantons are thus embedded in the
$E_8$ factor associated to $C_\infty$ and $12-n$ of the instantons are
embedded in the $E_8$ associated to $C_0$. To put this another way,
when we smooth everything out to obtain a theory with no extra tensor moduli
we expect to have an $(E_8\times E_8)$-vector bundle which is a sum of
two $E_8$-bundles, 
one of which has $c_2=12+n$ and the other with $c_2=12-n$.

This pretty well specifies exactly which heterotic string our type IIA
string is dual to. For a $G$-bundle on a K3 surface, where $G$ is
semi-simple and simply-connected, the topological class of this bundle
is specified by a
map $H_4({\rm K3})\to\pi_3(G)$ (for a clear explanation of this see
\cite{W:tools}).  This may be viewed as
given by the second Chern class of each sub-bundle associated to each
factor of $G$. In our case we fix the total second Chern class and so
the only freedom remaining is specified by how $c_2$ of the
bundle is split between the factors of $G$. Thus $n$
determines the class of our $E_8\times E_8$ bundle. We have arrived at the
following:
\begin{prop}
  Let $E$ be a sum, $E_1\oplus E_2$, of two $E_8$-bundles on a smooth
K3 surface 
such that $c_2(E_1)=12+n$ and $c_2(E_2)=12-n$. Then a heterotic string
compactified on this bundle on K3 is dual to (a limit of) a type IIA
string compactified on a \CY\ threefold which is an elliptic fibration
with a section over the Hirzebruch surface $\HS n$.
		\label{prop:HS}
\end{prop}
Our main assumption here is that the type IIA string on the \CY\
manifold really is dual to a heterotic string. If it is, then we have
certainly identified the correct one subject to the provisos of
proposition \ref{prop:K3f}.

This proposition first appeared in \cite{MV:F}. Although virtually all
of the mathematics above has been copied from that paper our
presentation has been slightly different. Rather than take the limit
of decompactifying a $T^2$ in a heterotic string on K3${}\times T^2$,
the line of attack in \cite{MV:F} was effectively to decompactify the
dual type IIA string (or its mirror partner, the type IIB string) to a
twelve-dimensional theory compactified on a \CY\ manifold. It is not
clear whether this twelve-dimensional ``F-theory'' exists in the usual
sense of ten-dimensional string theory or eleven-dimensional M-theory
or whether it serves simply as useful mnemonic for the above analysis.
Another point of view of F-theory is to think of it as the type IIB string
compactified down to six dimensions on $\Theta$ (see, for example,
\cite{Sen:ort} for some nice results along these lines). The fact that
$\Theta$ is not a \CY\ space is corrected for by placing fixed D-branes
within it. Again this is essentially equivalent to the above. The key
ingredient to associate to the term ``F-theory'' is the elliptic
fibration structure. Whether one wishes to think of this in terms of a
mysterious twelve-dimensional theory or a type IIA string or a type IIB
string is up to the reader.

The above proposition establishing the link between how the 24 of the
second Chern class is divided between the two $E_8$'s, and over which
Hirzebruch surface $X$ is elliptically fibred, is in agreement with all
the relevant conjectured dual pairs of \cite{KV:N=2} and
\cite{AFIQ:chains} for example. The case of $n=12$ was established in
\cite{HM:alg}. 

The appearance of
tensor moduli for small instantons was first noted in
\cite{SW:6d}. Since we are not allowing ourselves to appeal to
M-theory or D-branes in these lectures we will not reproduce the
argument here but just note that all necessary information appears to be
contained in the type IIA approach we use here.

\subsection{Aspects of the $E_8\times E_8$ string}  \label{ss:E8}

Let us now follow the analysis of \cite{MV:F,MV:F2} and continue to
explore the duality between the type IIA string on $X$ and the
$E_8\times E_8$ heterotic string.

In the previous section we looked at the case of fixing hypermultiplet
moduli in order to break none of the $E_8\times E_8$ gauge group. We
should ask the opposite question of what the gauge group is at a
generic point in the hypermultiplet moduli space. We shall do this as
follows. If any of the $E_8\times E_8$ gauge group remains unbroken we
expect either of our curves $C_0$ or $C_\infty$ to contain part of the
discriminant locus, $\Delta$. Let us focus on $C_0$. Split off the
part of the discriminant not contained in $C_0$ by putting
\begin{equation}
  \Delta = NC_0+\Delta',
\end{equation}
where $N\geq0$ and $\Delta'$ does not contain $C_0$. Since the only
way that the intersection number of two algebraic curves in an algebraic
surface can be negative is if one of the curves contains the other, we
have
\begin{equation}
  \Delta'.C_0\geq0.
\end{equation}
Following (\ref{eq:ABD}) we have, for $n\geq0$,
\begin{equation}
  N \geq 12-\frac{24}n.   \label{eq:Nlim}
\end{equation}
Similarly we may analyze the divisors $A$ and $B$ to obtain the
respective orders, $L$ and $K$, to which $a$ and $b$ vanish on
$C_0$. This gives
\begin{equation}
  \eqalign{L &\geq 4-\frac8n\cr
  K &\geq 6-\frac{12}n.\cr}  \label{eq:LKlim}
\end{equation}
We may now use table \ref{tab:Mir} to determine the fibre over a
generic point in $C_0$. Repeating this procedure for the other
``primordial'' $E_8$ along $C_\infty$ shows that no singular fibres are
required there for $n\geq0$.

We see that in the case $n>2$, we will have singular fibres over
$C_0$ generating a curve of singularities within $X$. Thus we expect
an enhanced gauge group. Loosely speaking, the gauge group can be read
from the last column of table \ref{tab:Mir}. The only thing we have to worry
about is the monodromy of section \ref{ss:mN=4} --- it may be that
there is monodromy on the singular fibres as we move about $C_0$. If
$\Delta'.C_0=0$ then there can be no monodromy since the fibre is the
same over every point of $C_0$. This occurs for $n=2,3,4,6,8,12$. When
$n=7,9,10,11$, the fibre admits no symmetry and thus there cannot be
any monodromy. Therefore, the only time we have to worry about
monodromy is for the ${\rm IV}^*$ fibre in the case $n=5$. There is
indeed monodromy in this case \cite{AG:sp32}. (See \cite{BKV:enhg} for
an account of this in terms of ``Tate's algorithm'' or \cite{me:sppt}
for an alternative approach.) Thus, whereas one associates $E_6$
with a type  
${\rm IV}^*$ fibre, this becomes $F_4$ from figure \ref{fig:oaut} when
$n=5$. The gauge algebras for generic moduli are summarized in table
\ref{tab:ggg}.

\begin{table}
$$\begin{array}{|c|c|c|c|c|c|c|c|}
\hline
n&L&K&N&\mbox{Fibre}&\mbox{Mon.}&G&H_0\\
\hline
\leq2&0&0&0&{\rm I}_0&&&E_8\\
3&2&2&4&{\rm IV}&&\su(3)&E_6\\
4&2&3&6&{\rm I\pII}^*_0&&\so(8)&\so(8)\\
5&3&4&8&{\rm IV}^*&\Z_2&F_4&G_2\\
6&3&4&8&{\rm IV}^*&&E_6&\su(3)\\
7&3&5&9&{\rm III}^*&&E_7&\su(2)\\
8&3&5&9&{\rm III}^*&&E_7&\su(2)\\
\geq9&4&5&10&{\rm II}^*&&E_8&\\
\hline
\end{array}$$
\caption{Generic gauge symmetries, $G$.}
\label{tab:ggg}
\end{table}

This agrees nicely with the heterotic picture. For a given value
of $c_2$ of a bundle, find the largest possible structure group, $H$, of
a vector bundle. Then the desired gauge group, $G$, will be the
centralizer of $H$ within $E_8\times E_8$. One ``rough and ready''
approach to this question is as follows. Consider an $H$-bundle $E$ with
fibre in an irreducible representation, $R$, of the structure
group. The Dolbeault index theorem on the K3 surface then gives
\cite{EGH:dg}
\begin{equation}
  \eqalign{\dim H^0(E)-\dim H^1(E)+\dim H^2(E) &= \int_Std(T)\wedge ch(E)\cr
   &= 2\rank(R)-l(R)c_2(E),\cr}   \label{eq:DolE}
\end{equation}
where $l(R)$ is the index of $R$ using the conventions of
\cite{Slansky:}. If $E$ really is a strict $H$-bundle with fibre $R$,
it should have 
no nonzero global sections, otherwise the structure group would be a strict
subgroup of $H$. Thus $H^0(E)$ is trivial. Similarly, by Serre
duality, we expect the same for $H^2(E)$. Thus, the right hand side of
(\ref{eq:DolE}) must be non-positive. That is,
\begin{equation}
  c_2(E) \geq \frac{2\rank(R)}{l(R)},   \label{eq:c2lim}
\end{equation}
for any irreducible representation, $R$.
The bundle with $c_2=12+n$ may have the full $E_8$ as structure group
so one $E_8$ of the $E_8\times E_8$ will be broken generically for any
$n\geq0$. Table \ref{tab:ggg} then shows how the other $E_8$ is broken
down to $G$ by a bundle with structure group $H_0$ and
$c_2=12-n$. For example, the {\bf 3} of $\su(3)$ has $l({\bf 3})=1$ and
so $c_2$ of a generic $\su(3)$-bundle with no global sections is at
least 6. This is why it
appears on the row for $n=6$ in the table. 

Note that there need not exist a bundle that saturates the bound in
(\ref{eq:c2lim}) and so we cannot reproduce all of the rows in
table \ref{tab:ggg}. For example, in the case $n=3$ one sees that
(\ref{eq:c2lim}) does not rule 
out a bundle with the full $E_8$ structure group but we see that only an
$E_6$-bundle is expected. See \cite{DMW:hh} for a discussion of
this. It is interesting to note that proposition \ref{prop:HS}
implies that $\su(3)$ {\em must\/} appear as the gauge symmetry in the
case $n=3$ and so the $E_8$-bundle must not exist. If a smooth
$E_8$-bundle on a K3 surface with $c_2=9$ is discovered it will
violate proposition \ref{prop:HS}.

The cases $n=9,10,11$ are somewhat peculiar since we appear to be
suggesting that we have a bundle with trivial structure group and yet
$c_2>0$. Clearly this is not possible classically. The fact that
classical reasoning is breaking down somewhat can be seen from the
fact that the reasoning of section \ref{ss:si} applies to this case
and we have $12-n$ tensor multiplets. That is, we have $12-n$
point-like instantons which cannot be given nonzero size.

The fact that $n\leq12$ can be understood from both the type IIA side
and the heterotic side. In the case of our elliptic fibration over
$\HS n$, if $n>12$ then $(L,K,N)$ as determined from (\ref{eq:Nlim})
and (\ref{eq:LKlim}) is at least $(4,6,12)$. As discussed in section 
\ref{ss:ell}, this means that we may redefine $\cL$ to absorb $C_0$ to
reduce the fibre to something in the list in figure
\ref{fig:Kod}. This kills $K_X=0$ however and so we do not have a \CY\
space. On the heterotic side, $c_2<0$ would be a clear violation of
(\ref{eq:c2lim}). One might worry about ``point-like anti-instantons''
but such objects would break supersymmetry and as such do not solve
the equations of motion.

In section \ref{ss:ext} we discussed extremal transitions between \CY\
manifolds which, in the heterotic language, corresponded to unwrapping
part of the gauge bundle around the K3 surface and rewrapping it
around the $T^2$. Such transitions are not of much interest to us in
this section as we are concerned only with the K3 part of the
story. There are other possible extremal transitions which will effect
us though. One kind which is of interest are ones which will take us
from an elliptic fibration over $\HS n$ to another elliptic fibration
over $\HS{n-1}$. In the heterotic string this will correspond to
a transition from
splitting the $E_8\times E_8$ bundle into two bundles with $c_2$ equal
to $12+n$ and $12-n$, to a splitting of $12+n-1$ and $12-n+1$
respectively. In this way we may ``join up'' all the theories
considered so far into one connected moduli space. This phenomenon was
first observed in \cite{SW:6d} but we will again follow the argument
as presented in \cite{MV:F,MV:F2}.

As explained in section \ref{ss:si}, when the $E_8\times E_8$
heterotic string has a point-like instanton, we expect the dual \CY\
space for the type IIA string to admit a blow-up in the base,
$\Theta$, of the elliptic fibration. When all 24 instantons are
point-like we saw this as a collision between a curve of ${\rm II}^*$
fibres and other parts of the discriminant locus as shown in figure
\ref{fig:curly}. Let us concentrate on what happens when one of these
points is blown up. 

\iffigs
\begin{figure}[t]
  \centerline{\epsfxsize=13cm\epsfbox{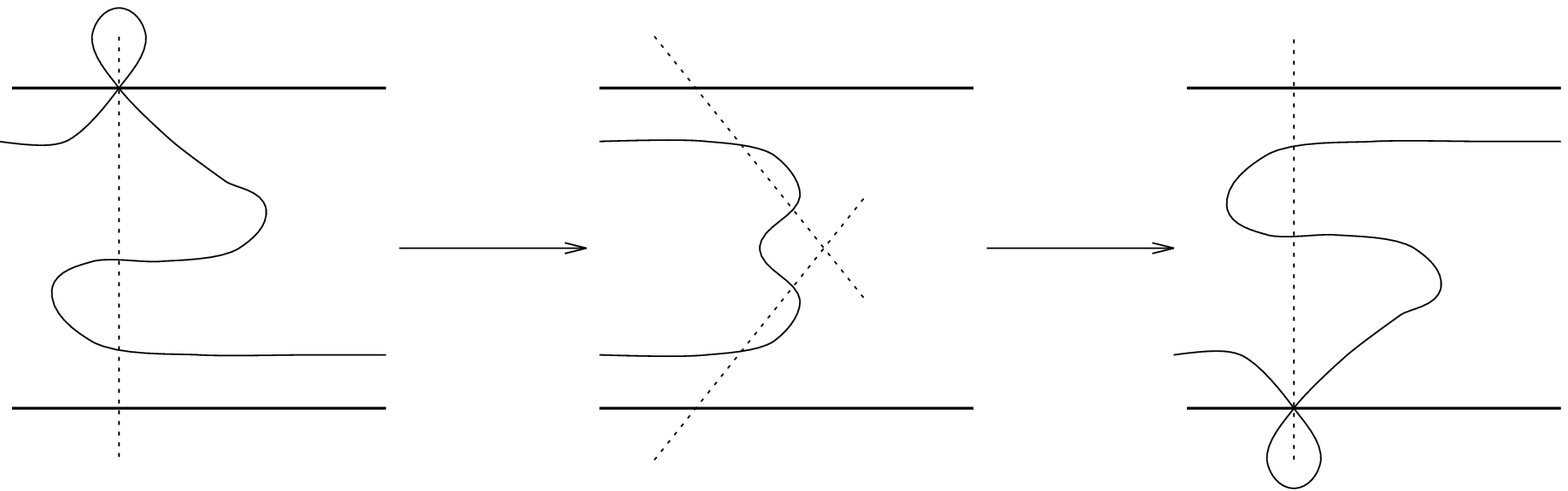}}
  \setlength{\unitlength}{1mm}
  \begin{picture}(0,0)
     \put(50,20){\makebox(0,0){$\Delta'$}}
     \put(48,42){\makebox(0,0){$(n)$}}
     \put(48,9){\makebox(0,0){$(-n)$}}
     \put(36,5){\makebox(0,0){$\HS n$}}
     \put(96,42){\makebox(0,0){$(n-1)$}}
     \put(96,9){\makebox(0,0){$(-n)$}}
     \put(145,42){\makebox(0,0){$(n-1)$}}
     \put(145,9){\makebox(0,0){$(-n+1)$}}
     \put(133,5){\makebox(0,0){$\HS{n-1}$}}
  \end{picture}
  \caption{The transition $n\to n-1$.}
  \label{fig:t2}
\end{figure}
\fi

As in section \ref{ss:si}, we use $\Delta'$ to denote the part of
discriminant left over after we subtract the contribution from the two
curves of ${\rm II}^*$ fibres. In the first diagram in figure
\ref{fig:t2} we show locally how $\Delta'$ loops around a collision
between it and a line, $C_\infty$, of ${\rm II}^*$ fibres together
with the class, $f$, that passes though this point. Now when we do the
blow-up by switching on a scalar in a tensor multiplet we go to the
middle diagram. The exceptional divisor is a line of self-intersection
$-1$. The line that was in the class $f$ also becomes a
$(-1)$-line. The middle diagram of figure \ref{fig:t2} is obviously
symmetric and we may blow-down the latter $(-1)$-curve to push the
loop onto the bottom line of ${\rm II}^*$ fibres as shown in the last
diagram.

The effect of this is to change $C_\infty$ into a line of
self-intersection $n-1$ and $C_0$ into a line of self-intersection
$-n+1$. Thus we have turned $\HS n$ into $\HS{n-1}$. It is also clear
that we have moved the small instanton from one of the $E_8$'s into
the other $E_8$. Thus we may connect up all our theories which are
elliptic fibrations over $\HS n$. Note that the use of tensor moduli
means that we do not expect a perturbative interpretation of this
process in the heterotic string language.

\subsection[The $\spnh$ heterotic string]
  {The $\Spin(32)/\mbox{\bigbbbfont Z}_2$ heterotic string}  \label{ss:so32}

Now that we can content ourselves with the knowledge that we know how to
build a type IIA dual to the generic $E_8\times E_8$ heterotic string
on a K3 surface we turn our attention to the $\spnh$ heterotic
string. Whereas the topology of the $E_8\times E_8$ bundle required
specifying how the 24 instantons were divided between the two $E_8$'s,
the $\spnh$ heterotic string is quite different. In this case the
gauge group is not simply-connected and the topology of the bundle is
not simply specified by $c_2$.

It will be important to recall some fundamentals of the construction
of the $\spnh$ heterotic string from \cite{GHMR:let}. The 16 extra
right-movers of the heterotic string are compactified on an even
self-dual lattice of definite signature. There are two such lattices,
which we denote $\Gamma_8\oplus\Gamma_8$ and $\Gamma_{16}$. The former
is two copies of the root lattice of $E_8\times E_8$ with which we
have become well-acquainted in these talks. The second lattice is the
``Barnes-Wall'' lattice \cite{CS:sphere}. This may be constructed by
supplementing the root lattice of $\so(32)$ by the weights of one of
its spinors. Such spinor weights are never of length squared 2 are so
do not give massless states. Thus, as far as massless states are
concerned, the lattice is the root lattice of $\SO(32)$ and the string
states fill out the adjoint representation. Massive representations
may fill out spinor representations for one of the spinors and but we
never have representation in the vector representation or the other
spinor representation. The gauge group can be viewed as a $\Z_2$
quotient of $\Spin(32)$ which does not admit vector
representations. Hence $\pi_1$ of our gauge group is $\Z_2$. When we
try to build $\spnh$-bundles the
situation is very similar to real vector bundles over $M$ where the fact that
$\pi_1(\SO(d)) \cong \Z_2$ leads to the notion of the second
``Stiefel-Whitney'' class, $w_2$, of a bundle as an element of
$H^2(M,\Z_2)$. Here we have a similar object characterizing the
topology of the bundle which we denote $\tilde w_2$. If $\tilde
w_2\neq0$ then bundles with fibre in the vector representation are
obstructed just as spinor representations are obstructed for non-spin
bundles with $w_2\neq0$. See \cite{BLPSSW:so32} for a detailed
account of this. 

Thus, rather than being classified by how 24 is split between second
Chern class, which was the case for the $E_8\times E_8$ string, the
topological class of an $\spnh$ heterotic string compactified over a
K3 surface is characterized by $\tilde w_2\in H^2(S,\Z_2)$. Whereas
elements $\tilde w_2$ are in one-to-one correspondence with the
homotopy classes of $\spnh$-bundles on a fixed (marked) K3 surface, in
our moduli space we are also allowed to vary the moduli of the K3
surface. Thus two elements of $H^2(S,\Z_2)$ should be considered to be
equivalent if they can be mapped to each other by a diffeomorphism of the
K3 surface. That is,
\begin{equation}
  \tilde w_2
    \in\frac{\Gamma_{3,19}/2\Gamma_{3,19}}{\GO^+(\Gamma_{3,19})}.
\end{equation}
It was shown in \cite{BLPSSW:so32} that there are
only three possibilities:
\begin{enumerate}
  \item $\tilde w_2=0$,
  \item $\tilde w_2\neq0$ and $\tilde w_2.\tilde w_2=0\pmod4$,
  \item $\tilde w_2\neq0$ and $\tilde w_2.\tilde w_2=2\pmod4$.
\end{enumerate}

We will focus on the case $\tilde w_2=0$ in this section. This will allow us
to use the same arguments as in the last section about shrinking
instantons down to retrieve the entire primordial gauge group. If
$\tilde w_2$ were not zero, the topology of the bundle would
obstruct the existence of arbitrary point-like instantons at smooth points in
the K3 surface. See \cite{BLPSSW:so32} for an account
of this.

We have seen
already that if a perturbatively understood heterotic string is
dual to a limit of a type IIA string on \CY\ threefold, $X$, then $X$
must be an elliptic fibration with a section over $\HS n$, where $0\leq n\leq
12$. Thus, if our duality picture is going to continue working for any
$\spnh$ heterotic string then we must have already encountered it in
disguise as the $E_8\times E_8$ heterotic string for a particular
$n$. 

The statement that the $\spnh$ heterotic string compactified on a K3
surface is the same thing as an $E_8\times E_8$ heterotic string
compactified on a K3 surface should not come as a surprise as the same
thing has been known to be true for toroidal compactifications for
some time \cite{N:torus,Gins:torus}. The identification of the dilaton
as the size of the base of $X$ as a K3 fibration has nothing to do
with whether we deal with the $E_8\times E_8$ or the $\spnh$ string
and so the duality between these theories cannot effect the string
coupling. Thus there must be some T-duality statement that connects
these two theories. This may be highly nontrivial however, as it may
mix up the notion of what constitutes the base of the bundle over K3
and what constitutes the fibre. A construction of such a T-duality at
a special point in moduli space was given in \cite{BLPSSW:so32}.

Now let us return to the issue concerning the extra states in the
$\Gamma_{16}$ lattice not contained in the root lattice of $\so(32)$.
What is the dual analogue of these extra massive states coming from
the spinor of $\so(32)$? Let us think in terms of the type IIA string
compactified on a K3 surface versus the $\spnh$ heterotic string
compactified on a 4-torus as in section \ref{ss:enh}. We may switch
off all the Wilson lines of the heterotic string compactification by
rotating the space-like 4-plane, $\Pi$, so that $\Gamma_{4,20}\cap
\Pi^\perp\cong\Gamma_{16}$. Now view this as the ``fibre'' of the type
IIA string compactified on $X$ versus the heterotic string
compactified on K3${}\times T^2$. This restores the full $\so(32)$
gauge symmetry and so can be thought of as shrinking down all 24
instantons. As discussed in section \ref{ss:N2enh} we may now vary the
$T^2$ and the Wilson lines by varying the 2-plane, $\mho$, in
$\Upsilon\otimes_\Z\R\cong\R^{2,\rho}$, where $\Upsilon$ is the
quantum Picard lattice of the generic fibre of $X$ as a K3
fibration. Thus we see that
$\Upsilon\cong\Gamma_{2,2}\oplus\Gamma_{16}$. In other words, {\em the
Picard lattice of the generic K3 fibre of $X$ is $\Gamma_{1,1}\oplus
\Gamma_{16}\cong\Gamma_{1,17}$ and is therefore self-dual.}

Let the limit of a type IIA string compactified on $X_0$ be dual to
the $\spnh$ heterotic string compactified on a K3 surface (times a
2-torus of large area) with $\tilde w_2=0$ and all the instantons
shrunk down and let
$X$ be the blow-up of $X_0$. This theory will have an $\so(32)$ gauge symmetry.
We know the following:
\begin{enumerate}
 \item $X$ is a K3 fibration and an elliptic fibration with a section
over a Hirzebruch surface.
 \item $X_0$ contains a curve of singularities of type $D_{16}$.
 \item The generic K3 fibre of $X$ has a self-dual Picard lattice (of
rank 18).
\end{enumerate}
Let us construct $X$.

Table \ref{tab:Mir} tells is that the base, $\Theta$, of $X$ as an
elliptic fibration will contain a curve of ${\rm I\pII}^*_{12}$
fibres. We may put this curve along $C_0$, the isolated section of
$\HS n$. A generic fibre of $X$ as a K3 fibration will be a K3 surface
built as an elliptic fibration with a ${\rm I\pII}^*_{12}$ fibre
(and, generically, 6 ${\rm I}_1$ fibres). Let us denote this K3
surface as $S_t$. What is $\Pic(S_t)$?

Let $\sigma$ denote the section of $S_t$ as an elliptic fibration
``at infinity'' guaranteed by the Weierstrass form. Let $R$ be the
sublattice of $\Pic(S_t)$ generated by the irreducible curves within
the fibres not intersecting $\sigma$. Let $\Phi$ be the set of
sections. One may show \cite{MirPer:ell}
\begin{equation}
  \disc(R) = |\Phi|^2\disc(\Pic(S_t)),
\end{equation}
where $\disc$ denotes the ``discriminant'' of a lattice, i.e., the
determinant of the inner product on the generators.

In our case, $R$ is generated by a set of rational curves forming the
Dynkin diagram for $D_{16}$. Thus, $\disc(R)$ is the determinant of
the Cartan matrix of $D_{16}$ which is 4. In order for $\Pic(S_t)$ to
be unimodular we see that we require exactly two sections. Writing an
elliptic fibration with two sections in Weierstrass form is easy. We
are guaranteed one section at infinity. Put the other section
along $(y=0, x=p(s,t))$. Thus the general Weierstrass form with two
sections is
\begin{equation}
  y^2 = (x-p(s,t))(x^2+p(s,t)x+q(s,t)),
\end{equation}
where
\begin{equation}
  \eqalign{a(s,t) &= q(s,t) - p(s,t)^2\cr
  b(s,t) &= -p(s,t)q(s,t).\cr}
\end{equation}
From (\ref{eq:ABD}) the divisors $P$ and $Q$, given by the zeros of
$p$ and $q$, 
are in the class $4C_0+(4+2n)f$ and $8C_0+(8+4n)f$ respectively. The
discriminant is 
\begin{equation}
  \eqalign{\delta&=4a^3+27b^2\cr
  &=(q+2p^2)^2(4q-p^2).\cr}
\end{equation}
The fact that the discriminant factorizes will have some profound
consequences. In terms of divisor classes let us write $\Delta=2M_1
+M_2$, where $M_1$ is the divisor given by $q+2p^2$ and $M_2$
corresponds to $4q-p^2$.

We know that $2M_1+M_2$ contains $18C_0$ from the ${\rm I\pII}^*_{12}$
fibres. Using the fact that $a$ and $b$ vanish to an order no greater
than 2 and 3 respectively along $C_0$ and the fact that $M_1$ and
$M_2$ must contain a nonnegative number of $C_0$ and $f$, one may show
that $M_1$ contains $8C_0$ and $M_2$ contains $2C_0$ as their
contributions towards the ${\rm I\pII}^*_{12}$ fibres. What remains is
\begin{equation}
\eqalign{M_1'&= (8+4n)f\cr
  M_2'&= 6C_0+(8+4n)f.\cr}
	\label{eq:8+4n}
\end{equation}
Putting $M_2'.C_0\geq0$ fixes $n\leq4$. There are no other constraints.

We have not yet achieved our goal in determining what $n$ is for the
$\spnh$ heterotic string. All we know is that $n\leq4$. Note however
that along the $8+4n$ zeros of $M_1$ we have a double zero of
$\Delta$. Thus, generically we have $8+4n$ parallel lines along the
$f$ direction of ${\rm I}_2$
fibres. Thus means the gauge algebra acquires an extra $\su(2)^{8+4n}$
factor. The necessary appearance of an extra gauge symmetry can
ultimately be traced to the way the determinant factorized.
We will denote this as $\sp(1)^{8+4n}$ to fit in with later analysis.
It is very striking how different this behaviour is from the
$E_8\times E_8$ analysis of the last section. In the latter case we
acquired 24 massless tensors when the instantons were shrunk down to
zero size, whereas for the $\spnh$ heterotic string we acquire new
massless gauge fields.

The natural thing to do would be to identify each $\sp(1)$ with a
small instanton. To do this fixes $n=4$ as one can have 24 small
instantons, at least in the case that the underlying K3 surface is
smooth and $\tilde w_2=0$. Thus we arrive at the following
\begin{prop}
  The $\spnh$ heterotic string compactified smoothly on a K3 surface with
$\tilde w_2=0$ is dual
to (a limit of) the type IIA string compactified on an elliptic
fibration over $\HS 4$.
\end{prop}
This proposition depends on the assumptions governing proposition
\ref{prop:K3f} and the somewhat schematic way we associated each of the
$\sp(1)$ factors to a small instanton to show $n=4$.

Another way \cite{MV:F} to argue $n=4$ is that the generic point in
moduli space 
of $n=4$ theories has gauge symmetry $\so(8)$ and then compare to the
analysis in section
\ref{ss:E8}. We can try to break as much of $\so(32)$ as possible by
finding the largest subalgebra of $\so(32)$ consistent with
(\ref{eq:c2lim}).
Since $\tilde w_2=0$ we may consider vector representations in which case
$\so(24)$ is the largest such algebra with $c_2=24$ and this
breaks $\so(32)$ down to $\so(8)$. A glance at table \ref{tab:ggg}
then confirms that $n=4$.

Actually, the analysis of the $\spnh$ heterotic string was first
done by Witten \cite{W:small-i}, where the appearance of the $\sp(1)$ factors
from small instantons was argued directly. Unfortunately this involves
methods beyond the scope of these talks. It is worth contrasting
Witten's method with the above. It is a remarkable achievement of string
duality that the same answer results.

\iffigs
\begin{figure}[t]
  \centerline{\epsfxsize=13cm\epsfbox{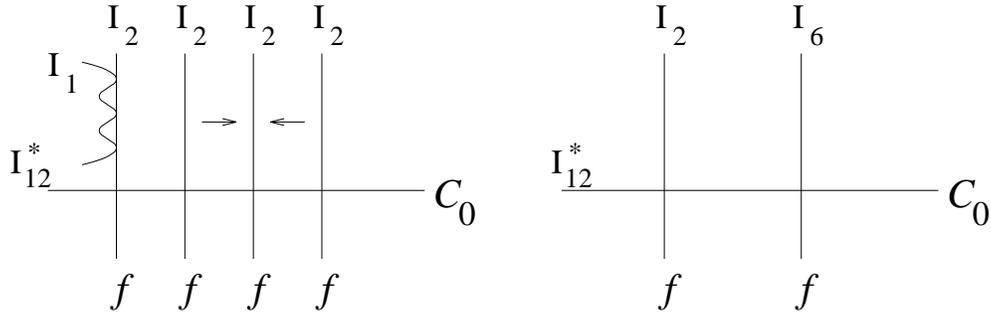}}
  \caption{Three $\so(32)$ point-like instantons coalescing to form $\sp(3)$.}
  \label{fig:coal}
\end{figure}
\fi

We may consider a specialization of the moduli to bring together some
of the components of $M_1'$ along $f$. Suppose we bring $k$ of the 24
lines together. This will produce an ${\rm I}_{2k}$ line of fibres. As usual
we have to worry about monodromy to get the full gauge group. In this
case the transverse collision between the ${\rm I}_{2k}$ line and the
${\rm I\pII}^*_{12}$ line produces 
monodromy acting on the ${\rm I}_{2k}$ fibres to
give an $\sp(k)$ gauge symmetry.\footnote{Actually, since the $f$
curves are topologically spheres, there had better be more than one
point within them around which there is nontrivial monodromy. There
are three other points within each $f$ curve where there is monodromy
provided by non-transverse collisions of $f$ with what is left of the
discriminant. We show this on one of the $f$ curves in figure
\ref{fig:coal}. We omit this on the other $f$ curves to make the
figure more readable.}
This corresponds to $k$ of the 24
point-like instantons {\em coalescing\/} \cite{W:small-i}. We show an
example of this in figure \ref{fig:coal}. As an extreme case, all 24
instantons may be pushed to the same point. This results in a gauge
algebra of $\so(32)\oplus\sp(24)$.

As we said above, it is interesting to contrast the behaviour of the point-like
$E_8$ instantons with the point-like $\so(32)$ instantons. The former
produced massless tensor multiplets whereas the latter produce extra
massless vector multiplets to enhance the gauge group. Of course,
since both theories are meant to live in the same moduli space for
$n=4$, it should be possible to continuously deform one type of
point-like instanton into another. This may well require a deformation
of the base K3 surface as well as the bundle. The T-duality analysis
of \cite{BLPSSW:so32} should provide a good starting point for such a
description. 

\subsection{Discussion of the heterotic string}  \label{ss:dh}

We have analyzed both the $E_8\times E_8$ and $\spnh$ heterotic
string compactified on a K3 surface in terms of a dual type IIA string
compactified on an elliptically fibred \CY\ threefold. In both cases
we discovered curious nonperturbative behaviour associated to
point-like instantons. 

It is worth noting that in the case of the $\spnh$ string we appear
to have acquired a free lunch concerning the analysis of the gauge
group. To see this note the following. In old perturbative string
theory a type IIA string compactified on a \CY\ manifold never has a
nonabelian gauge group. Similarly a heterotic string may have a
subgroup of the original $E_8\times E_8$ or $\spnh$ as its gauge
group together with perhaps a little more from massless strings. This
latter can only have rank up to a certain value limited by the central
charge of the corresponding conformal field theory (see, for example,
\cite{AL:ubiq} for a discussion of this). In the case of point-like
instantons we have claimed above to have found gauge algebras as large
as $\so(32)\oplus\sp(24)$. Such a gauge group must be nonperturbative
from both the type IIA {\em and\/} heterotic point of view.

When discussing duality one would normally claim that its power lies
in its ability to relate nonperturbative aspects of one theory to
perturbative aspects of another. This then allows the nonperturbative
quantities to be calculated. In
the above we appear to have discovered things about a subject that was
nonperturbative in both theories at the same time! How did we do this?
The answer is that we first related perturbative nonabelian gauge
groups in the heterotic string to curves of orbifold singularities in
the type IIA picture. To do this the curves of singularities were
always formed by each K3 fibre of $X$, as a K3 fibration, acquiring a
singular point. There is no reason why a curve of singularities need
be in this form --- it may be completely contained in a bad fibre and
not seen by the generic fibre at all. Why should the type IIA string
intrinsically care about the fibration structure? Assuming it does
not, this latter kind of curve should have just as much right to
produce an enhanced gauge group as those we could understand
perturbatively from the heterotic string.
This is exactly the type of curve that gives the
$\sp(24)$ gauge symmetry which is now nonperturbative in terms of the
heterotic string.

The example of heterotic-heterotic duality studied in section
\ref{ss:hh} gives a clear picture of how nonperturbative effects are
viewed in this way. Begin with a perturbative gauge group in the
heterotic string. This corresponds to a curve of singularities in the
type IIA's K3 fibration passing through each generic
fibre. Heterotic-heterotic duality corresponds to exchanging the two
$\P^1$'s in the base of the \CY\ viewed as an elliptic fibration. This
gives another K3 fibration but now the curve of singularities lies
totally within a bad fibre. Thus it has become a nonperturbative
gauge group for the dual heterotic string.

An issue which is very interesting but we do not have time to discuss
here concerns the appearance of extra massless hypermultiplets at
special points in the moduli space. 
This question appears to be rather straight forward in the case of
hypermultiplets in the adjoint representation, as shown in
\cite{KMP:enhg}. The more difficult question of other representations
has been analyzed in
\cite{FHSV:N=2,MV:F2,AG:sp32,BKV:enhg,KV:hyp}. It is essential to do this
analysis to complete the picture of possible phase transitions in terms
of Higgs transitions in the heterotic string. As usual, in discussions
about duality, it also hints at previously unsuspected relationships
in algebraic geometry.

Another very important issue we have not mentioned so far concerns
anomalies. The heterotic string compactified on a K3 surface produces
a chiral theory with potential anomalies. This puts constraints on
the numbers of allowed massless supermultiplets (see \cite{Sch:a6} for
a discussion of this). One constraint may
be reduced to the condition \cite{AW:grav,GSW:a6,Sg:6d}
\begin{equation}
  273-29n_T-n_H+n_V=0,
	\label{eq:6d-anom}
\end{equation}
where $n_T$, $n_H$ and $n_V$ count the number of massless tensors,
hypermultiplets, and vectors respectively. The reason we have been able
to ignore this seemingly important constraint is that, assuming one
does the geometry of elliptic fibrations correctly, it always appears
to be obeyed. At present this appears to be another string miracle!

\begin{difficult}
  As an example consider the following. Compactify the $E_8\times E_8$
heterotic string, with $c_2$ split as $12+n$ and $12-n$ between the
two $E_8$ factors, on a K3 surface so that the unbroken gauge symmetry is
precisely $E_8$. This means that the corresponding elliptic threefold,
$X$, has a curve of ${\rm II}^*$ fibres which we may assume lies along
$C_0$ in the Hirzebruch surface $\HS n$. We will calculate $n_T$,
$n_H$, and $n_V$. What is left of the divisors $A$, $B$, and $\Delta$
after subtracting the contribution from the curve of ${\rm II}^*$
fibres is given by
\begin{equation}
  \eqalign{A'&=4C_0+(8+4n)f\cr
  B' &= 7C_0+(12+6n)f\cr
  \Delta'&=14C_0+(24+12n)f.\cr}
\end{equation}
$B'$ collides with $C_0$ generically $12-n$ times. Each of these
points corresponds to a point-like instanton required to produce the
$E_8$ gauge group. Each such point must be blown up within the base to
produce a \CY\ threefold. Thus we have $12-n$ massless tensor
multiplets in addition to the one from the six-dimensional dilaton.
Since the gauge group is $E_8$, we have 248 massless vectors
furnishing the adjoint representation.

To count the massless hypermultiplets we require $h^{2,1}(X)$. It is
relatively simple to compute $h^{1,1}(X)$ (assuming there is a finite
number of sections) as this is given by 2, from
the Hirzebruch surface, plus $12-n$ from the
blow-ups within the base, plus 1 from the generic fibre, plus 8 from
${\rm II}^*$ fibres. That is, $h^{1,1}(X)=23-n$. Now $h^{2,1}$ may be
determined from the Euler characteristic of $X$,
$\chi(X)=2(h^{1,1}-h^{2,1})$. To find this, recall 
that the Euler characteristic of a smooth bundle is given by the
product of the Euler characteristic of the base multiplied by the Euler
characteristic of the fibre. Thanks to the nice way Euler
characteristics behave under surgery, we may thus apply this rule to
each part of fibration separately. The Euler characteristic of any
fibre is given by $N$ in table \ref{tab:Mir}.

Over most of the base, the fibre is a smooth elliptic curve which has
Euler characteristic zero. Thus only the degenerate fibres contribute
to our calculation. We have a curve of ${\rm II}^*$ fibres over $C_0$
which contributes $10\times 2$. The rest of the contribution comes
from $\Delta'$. Let us determine the geometry of
$\Delta'$. Firstly we know that, prior to blowing up the base,
$\Delta'$ has $12-n$ double points as seen in the upper part of figure
\ref{fig:curly}. Secondly, whenever $A'$ and $B'$ collide, which
happens at $A'.B'=24n+104$ points, $\Delta'$ will have a cusp (assuming
everything is generic). Na\"\i vely, the Euler characteristic of
$\Delta'$ can be given by the adjun
\begin{equation}
  -\Delta'.(\Delta'+K) = -596-130n.
\end{equation}
However, each cusp will increase this by value by 2 and each double
point by 1. In addition, when we do the blow-up of the base, the double
points will be resolved and an additional 1 must be added for each
double point. Thus
\begin{equation}
  \eqalign{
  \chi(\Delta') &= -596-130n+2(24n+104)+2(12-n)\cr
    &= -364-84n.}
\end{equation}

Over most of the points in $\Delta'$ the fibre will be type ${\rm I}_1$ but
there will be type II fibres over each cusp. Thus the Euler
characteristic of $X$ is given by
\begin{equation}
  \eqalign{\chi(X) &= 10.2+1.(-364-84n-(24n+104))+2.(24n+104)\cr
   &= -240-60n,}
\end{equation}
and so $h^{2,1}(X) = 143+29n$. To obtain the number of hypermultiplets we
add one to this figure from the four-dimensional dilaton in the type
IIA string. In summary, we have
\begin{equation}
  \eqalign{n_T &= 13-n\cr
     n_H &= 144+29n\cr
     n_V &= 248.\cr}
\end{equation}
It can be seen that this satisfies (\ref{eq:6d-anom}). See
\cite{Sch:a6,MV:F2} for more examples.
\end{difficult}

In section \ref{ss:so32} we analyzed the $\spnh$ heterotic string
compactified on a K3 surface with $\tilde w_2=0$. What about the case
$\tilde w_2\neq0$? An example of this was studied in
\cite{BLPSSW:so32} based on the construction of Gimon and Polchinski
\cite{GP:open}. This uses
open string models which we have not discussed here. In this case,
$\tilde w_2$ is dual to $\ff12\sum_{i=1}^{16} C_i$, where $\{C_i\}$
are sixteen  
disjoint $(-2)$-curves. The fact that such a class lies in
$H^2(S,\Z)$ can be seen in \cite{BPV:}. Such a $\tilde w_2$ is
conjectured to be dual to the case of the $E_8\times E_8$ heterotic
string with $n=0$. 
It would be interesting to analyze this from
the elliptic fibration point if view presented here. One 
indication that things will work is that the algebra $\sp(8)$
appears naturally in the nonperturbative group in the $n=0$ case from
the analysis 
following (\ref{eq:8+4n}). This agrees with the model of
Gimon and Polchinski. The problem is that we have the
full $\so(32)$ present 
as a gauge symmetry which does not appear in Gimon and Polchinski's
model. One can also show that massless tensor multiplets appear from
the type IIA approach.
These issues are resolved in \cite{me:sppt}.

An interesting issue we had no time to pursue is that concerning what
happens as we take the coupling of the heterotic string to be strong.
All of the above analysis was done for a weakly-coupled
heterotic string. It has been noted that one can expect some kind of
phase transition to occur as the coupling reaches a particular value
\cite{DMW:hh}. This can be analyzed in the context of elliptic
fibrations \cite{MV:F2}.

As concluding remarks to the discussion concerning the heterotic
string on a K3 surface we should note that this analysis is far from
complete. In the case of the type IIA and type IIB string we were able
to give a fairly complete picture of the entire moduli space. For the
heterotic string we have been able to study a few components of the
moduli space and a few points at extremal transitions. A first
requirement to study the full moduli space will be a classification of
elliptic fibrations. A more immediate shortcoming in our discussion is
that we do not yet have an explicit map between the moduli of the
elliptically fibered \CY\ manifold and the K3 surface and bundle on
which the heterotic string is compactified. This must be understood
before we can really answer questions such as what happens when
point-like instantons collide with singularities of the K3 surface. 

%%%%%%%%%%%%%%%%%%%%%%%%%%%%%%%%%%%%%%%%%%%%%%%%%%%%%%%%%%%%%%%%%%%

\section*{Acknowledgements}

I would like to thank my collaborators M.~Gross and D.~Morrison for
their contribution to much of the work discussed above. I would also
like to thank them for answering a huge number of questions concerning
algebraic geometry many of which I should have looked up in a book
first. 
I thank J.~Louis, with whom I have also collaborated on some of
the work covered above.
It is a pleasure to thank O.~Aharony, S.~Kachru, S.~Katz, E.~Silverstein, and
H.~Tye for useful conversations and of course B.~Greene and 
K.~T.~Mahanthappa for organizing TASI 96.
The author was supported by an NSF grant while at Cornell University,
where part of these lecture notes were written,
and is supported in part by DOE grant DE-FG02-96ER40959 at Rutgers
University. 

%\bibliographystyle{my-phys}
%\bibliography{string}

\end{document}